\documentclass[iop]{emulateapj}
\usepackage[citebordercolor={0 .5 .5}]{hyperref}

\newcommand{\beq}{\begin{equation}}
\newcommand{\eeq}{\end{equation}}
\newcommand{\be}{\begin{equation}}
\newcommand{\ee}{\end{equation}}
\newcommand{\bea}{\begin{eqnarray}}
\newcommand{\eea}{\end{eqnarray}}
\newcommand{\bdi}{\begin{displaymath}}
\newcommand{\edi}{\end{displaymath}}

\newcommand{\kt}{k_{\theta}}
\newcommand{\invarcmin}{\, \rm arcmin^{-1}}

\newcommand{\rmicron}{$\,\micro$m}

\def\lsim{\,\lower2truept\hbox{${<\atop\hbox{\raise4truept\hbox{$\sim$}}}$}\,}
\def\gsim{\,\lower2truept\hbox{${>\atop\hbox{\raise4truept\hbox{$\sim$}}}$}\,}
\newcommand{\rmnum}[1]{\romannumeral #1}
\newcommand{\Rmnum}[1]{\expandafter\@slowromancap\romannumeral #1@}

\usepackage{amssymb,amsmath}
\usepackage{natbib}  
\usepackage[mediumspace,Gray, squaren]{SIunits} 

\shorttitle{Cosmic Infrared Background Anisotropies from HerMES}
\shortauthors{Viero et al.}

\begin{document}
\title{HerMES: Cosmic Infrared Background Anisotropies and the Clustering of Dusty Star-Forming Galaxies$^{\dagger}$}

\author{M.P.~Viero\altaffilmark{1,$\ddagger$},
L.~Wang\altaffilmark{2,3},
M.~Zemcov\altaffilmark{1,4},
G.~Addison\altaffilmark{5},
A.~Amblard\altaffilmark{6},
V.~Arumugam\altaffilmark{7},
H.~Aussel\altaffilmark{8},
M.~B{\'e}thermin\altaffilmark{8,9},
J.~Bock\altaffilmark{1,4},
A.~Boselli\altaffilmark{10},
V.~Buat\altaffilmark{10},
D.~Burgarella\altaffilmark{10},
C.M.~Casey\altaffilmark{11},
D.L.~Clements\altaffilmark{12},
A.~Conley\altaffilmark{13},
L.~Conversi\altaffilmark{14},
A.~Cooray\altaffilmark{15,1},
G.~De Zotti\altaffilmark{16},
C.D.~Dowell\altaffilmark{1,4},
D.~Farrah\altaffilmark{17,3},
A.~Franceschini\altaffilmark{18},
J.~Glenn\altaffilmark{19,13},
M.~Griffin\altaffilmark{20},
E.~Hatziminaoglou\altaffilmark{21},
S.~Heinis\altaffilmark{10},
E.~Ibar\altaffilmark{22},
R.J.~Ivison\altaffilmark{23,7},
G.~Lagache\altaffilmark{9},
L.~Levenson\altaffilmark{1,4},
L.~Marchetti\altaffilmark{18},
G.~Marsden\altaffilmark{24},
H.T.~Nguyen\altaffilmark{4,1},
B.~O'Halloran\altaffilmark{12},
S.J.~Oliver\altaffilmark{3},
A.~Omont\altaffilmark{25},
M.J.~Page\altaffilmark{26},
A.~Papageorgiou\altaffilmark{20},
C.P.~Pearson\altaffilmark{27,28},
I.~P{\'e}rez-Fournon\altaffilmark{29,30},
M.~Pohlen\altaffilmark{20},
D.~Rigopoulou\altaffilmark{27,5},
I.G.~Roseboom\altaffilmark{3,7},
M.~Rowan-Robinson\altaffilmark{12},
B.~Schulz\altaffilmark{1,31},
D.~Scott\altaffilmark{24},
N.~Seymour\altaffilmark{32},
D.L.~Shupe\altaffilmark{1,31},
A.J.~Smith\altaffilmark{3},
M.~Symeonidis\altaffilmark{26},
M.~Vaccari\altaffilmark{18,33},
I.~Valtchanov\altaffilmark{14},
J.D.~Vieira\altaffilmark{1},
J.~Wardlow\altaffilmark{15},
C.K.~Xu\altaffilmark{1,31}}
\altaffiltext{$\ddagger$}{Email: marco.viero@caltech.edu}
\altaffiltext{$\dagger$}{Herschel is an ESA space observatory with science instruments provided by European-led Principal Investigator consortia and with important participation from NASA.}
\altaffiltext{1}{California Institute of Technology, 1200 E. California Blvd., Pasadena, CA 91125}
\altaffiltext{2}{Institute for Computational Cosmology, Department of Physics, University of Durham, South Road, Durham, DH1 3LE, UK}
\altaffiltext{3}{Astronomy Centre, Dept. of Physics \& Astronomy, University of Sussex, Brighton BN1 9QH, UK}
\altaffiltext{4}{Jet Propulsion Laboratory, 4800 Oak Grove Drive, Pasadena, CA 91109}
\altaffiltext{5}{Department of Astrophysics, Denys Wilkinson Building, University of Oxford, Keble Road, Oxford OX1 3RH, UK}
\altaffiltext{6}{NASA, Ames Research Center, Moffett Field, CA 94035}
\altaffiltext{7}{Institute for Astronomy, University of Edinburgh, Royal Observatory, Blackford Hill, Edinburgh EH9 3HJ, UK}
\altaffiltext{8}{Laboratoire AIM-Paris-Saclay, CEA/DSM/Irfu - CNRS - Universit\'e Paris Diderot, CE-Saclay, pt courrier 131, F-91191 Gif-sur-Yvette, France}
\altaffiltext{9}{Institut d'Astrophysique Spatiale (IAS), b\^atiment 121, Universit\'e Paris-Sud 11 and CNRS (UMR 8617), 91405 Orsay, France}
\altaffiltext{10}{Laboratoire d'Astrophysique de Marseille - LAM, Universit\'e d'Aix-Marseille \& CNRS, UMR7326, 38 rue F. Joliot-Curie, 13388 Marseille Cedex 13, France}
\altaffiltext{11}{Institute for Astronomy, University of Hawaii, 2680 Woodlawn Drive, Honolulu, HI 96822}
\altaffiltext{12}{Astrophysics Group, Imperial College London, Blackett Laboratory, Prince Consort Road, London SW7 2AZ, UK}
\altaffiltext{13}{Center for Astrophysics and Space Astronomy 389-UCB, University of Colorado, Boulder, CO 80309}
\altaffiltext{14}{Herschel Science Centre, European Space Astronomy Centre, Villanueva de la Ca\~nada, 28691 Madrid, Spain}
\altaffiltext{15}{Dept. of Physics \& Astronomy, University of California, Irvine, CA 92697}
\altaffiltext{16}{INAF - Osservatorio Astronomico di Padova, Vicolo dell'Osservatorio 5, I-35122 Padova, Italy.}
\altaffiltext{17}{Department of Physics, Virginia Tech, Blacksburg, VA 24061}
\altaffiltext{18}{Dipartimento di Astronomia, Universit\`{a} di Padova, vicolo Osservatorio, 3, 35122 Padova, Italy}
\altaffiltext{19}{Dept. of Astrophysical and Planetary Sciences, CASA 389-UCB, University of Colorado, Boulder, CO 80309}
\altaffiltext{20}{School of Physics and Astronomy, Cardiff University, Queens Buildings, The Parade, Cardiff CF24 3AA, UK}
\altaffiltext{21}{ESO, Karl-Schwarzschild-Str. 2, 85748 Garching bei M\"unchen, Germany}
\altaffiltext{22}{Pontificia Universidad Cat\'olica de Chile, Departamento de Astronom\'ia y Astrof\'isica, Vicu\~na Mackenna 4860, Casilla 306, Santiago 22, Chile}
\altaffiltext{23}{UK Astronomy Technology Centre, Royal Observatory, Blackford Hill, Edinburgh EH9 3HJ, UK}
\altaffiltext{24}{Department of Physics \& Astronomy, University of British Columbia, 6224 Agricultural Road, Vancouver, BC V6T~1Z1, Canada}
\altaffiltext{25}{Institut d'Astrophysique de Paris, UMR 7095, CNRS, UPMC Univ. Paris 06, 98bis boulevard Arago, F-75014 Paris, France}
\altaffiltext{26}{Mullard Space Science Laboratory, University College London, Holmbury St. Mary, Dorking, Surrey RH5 6NT, UK}
\altaffiltext{27}{RAL Space, Rutherford Appleton Laboratory, Chilton, Didcot, Oxfordshire OX11 0QX, UK}
\altaffiltext{28}{Department of Physical Sciences, The Open University, Milton Keynes MK7 6AA,UK}
\altaffiltext{29}{Instituto de Astrof{\'\i}sica de Canarias (IAC), E-38200 La Laguna, Tenerife, Spain}
\altaffiltext{30}{Departamento de Astrof{\'\i}sica, Universidad de La Laguna (ULL), E-38205 La Laguna, Tenerife, Spain}
\altaffiltext{31}{Infrared Processing and Analysis Center, MS 100-22, California Institute of Technology, JPL, Pasadena, CA 91125}
\altaffiltext{32}{CSIRO Astronomy \& Space Science, PO Box 76, Epping, NSW 1710, Australia}
\altaffiltext{33}{Astrophysics Group, Physics Department, University of the Western Cape, Private Bag X17, 7535, Bellville, Cape Town, South Africa}

\begin{abstract}
We present measurements of the auto- and cross-frequency power spectra of the cosmic infrared background (CIB) at 250, 350, and 500\rmicron\ (1200, 860, and 600\,GHz) from observations totaling $\sim 70\, \rm deg^2$ made with the SPIRE instrument aboard the \emph{Herschel Space Observatory}.  
We measure a fractional anisotropy $\delta I / I = 14 \pm 4$\%,  detecting signatures arising from the clustering of dusty star-forming galaxies in both the linear (2-halo) and non-linear (1-halo) regimes; and that the transition from the 2- to 1-halo terms, below which power originates predominantly from multiple galaxies within dark matter halos, occurs at $k_{\theta} \sim 0.10$--$0.12\invarcmin$ ($\ell \sim 2160$--2380), from 250 to 500\rmicron.
New to this paper is clear evidence of a dependence of the Poisson and 1-halo power on the flux-cut level of masked sources --- suggesting that some fraction of the more luminous sources occupy more massive halos as satellites, or are possibly close pairs.   
We measure the cross-correlation power spectra between bands, finding that bands which are farthest apart are the least correlated, as well as hints of a reduction in the correlation between bands when resolved sources are more aggressively masked.  
In the second part of the paper we attempt to interpret the measurements in the framework of the halo model.   
With the aim of fitting simultaneously with one model the power spectra, number counts, and absolute CIB level in all bands, we find that this is achievable by invoking a luminosity-mass relationship, such that the luminosity-to-mass ratio peaks at a particular halo mass scale and declines towards lower and higher mass halos.  
Our best-fit model finds that the halo mass which is most efficient at hosting star formation in the redshift range of peak star-forming activity, $z\sim 1-3$, is ${\rm log}(M_{\rm peak}/\rm M_{\odot}) \sim 12.1\pm 0.5$, and that the minimum halo mass to host infrared galaxies is ${\rm log}(M_{\rm min}/\rm M_{\odot}) \sim 10.1\pm 0.6$.   

\end{abstract}
%
\section{Introduction}
\label{sec:intro}
Star formation is well traced by dust, which absorbs the UV/optical light produced by young stars in actively star-forming regions and re-emits the energy in the far-infrared/submillimeter \citep[FIR/submm; e.g.,][]{savage1979}.  
Roughly half of all starlight ever produced has been reprocessed by dusty star-forming galaxies \citep[DSFGs; e.g.,][]{hauser2001,dole2006}, and this emission is responsible for the ubiquitous cosmic infrared background \citep[CIB;][]{puget1996,fixsen1998}.  
The mechanisms responsible for the presence or absence of star formation are partially dependent on the local environment (e.g., major mergers: \citealt[][]{narayanan2010}; condensation or cold accretion: \citealt[][]{dekel2009}, photoionization heating, supernovae, active galactic nuclei, and virial shocks: \citealt[][]{birnboim2003, granato2004, bower2006}).
Thus, the specifics of the galaxy distribution --- which can be determined statistically to high precision by measuring their clustering properties --- inform the relationship of star formation and dark matter density, and are valuable inputs for models of galaxy formation.     
However, measuring the clustering of DSFGs has historically proven difficult to do.    

Owing to the relatively large point spread functions (PSF's) of ground-, balloon-, and space-based submillimeter observatories, coupled with very steep source counts, 
maps at these wavelengths are dominated by confusion noise.  For the 250\rmicron\  channel on \emph{Herschel}, for example, this means that no matter how deeply you observe a field, without some sort of spatial deconvolution at best only $\sim 15\%$ of the flux density will be resolved into individually detected galaxies \citep{oliver2010}. 
Add to that the fact that the redshift distribution of DSFGs is relatively broad \citep[e.g.,][]{casey2012b,chapman2005,bethermin2012b}, clustering measurements of resolved sources have consequently had limited success  \citep[e.g.,][]{blain2004, scott2006, weiss2009}, and somewhat contradictory results \citep[e.g.,][]{cooray2010,maddox2010}.  

The remaining intensities in the maps appear as fluctuations, or anisotropies, in the CIB.  
Contained in CIB anisotropies (or CIBA) is the clustering pattern, integrated over luminosity and redshift, of \emph{all}  DSFGs --- including  those too faint to be resolved. 
And analogous to the two-point function typically used to estimate the clustering of resolved galaxies, 
 the power spectrum of these intensity fluctuations is a probe of the clustering properties of 
 those galaxies \citep[e.g.,][]{bond1984,scott1999,knox2001,negrello2007}.  
Initial power spectrum measurements from \emph{Spitzer} \citep{grossan2007, lagache2007}, BLAST \citep{viero2009, hajian2012}, ACT \citep{dunkley2011}, and SPT \citep{hall2010} found a signal in excess of Poisson noise 
originating from the clustering of DSFGs, but were limited to measuring the galaxy bias in the linear regime, rather than their distribution within dark matter halos. 
Subsequent measurements from \emph{Herschel}/SPIRE \citep{amblard2011}, and \emph{Planck} \citep{lagache2011} were able to isolate the linear and non-linear clustering signals, but the two groups found that their measurements agreed only after correcting for multiple systematics.  

Power spectra can be interpreted with modeling frameworks in much the same way as is done for two-point function measurements of resolved sources.  Among the most commonly adopted models are so-called \lq\lq halo models\rq\rq\ \citep[e.g.,][]{seljak2000, cooray2002}, which use halo occupation distributions \citep[HODs; e.g.,][]{peacock2000a,scoccimarro2001} 
to statistically assign galaxies to dark matter halos in order to re-create observed clustering measurements.  
Halo models have been adopted to interpret CIBA spectra 
from BLAST \citep{viero2009}, \emph{Herschel}/SPIRE \citep{amblard2011, penin2012b, xia2012}, and \emph{Planck} \citep{lagache2011, penin2012b, shang2012, xia2012}, with varying success.   
 
Precisely measuring the CIBA power spectra 
and decoding the information contained within them is a rapidly growing field, and it is also the focus of this paper.  
First and foremost, we aim to advance the field by providing state-of-the-art 
 measurements of the auto- and cross-frequency power spectra of CIB anisotropies at 250, 350, and 500\rmicron, spanning angular scales $0.01 \le k_{\theta} \lsim 2\invarcmin$ (or $350 \lsim \ell \lsim 45{,}000$) (\S~\ref{sec:results}).  
With the addition of more than four times the area, we extend the efforts of \citet{amblard2011} --- who definitively resolved a signature of non-linear clustering on small scales --- 
by illustrating how the strength of the non-linear clustering signal depends strongly on the flux-cut level of masked sources (\S~\ref{sec:clustering}). 
We improve on the efforts of BLAST \citep{viero2009, hajian2012} by measuring the cross-frequency power spectra and estimate the level of correlation between bands (\S~\ref{sec:bandband}).   

We then attempt to interpret our measurements with a series of halo models, whose common feature is to tie the luminosities of sources to their host halo masses  (\S~\ref{sec:formalism}), but which differ by their treatment of the spectral energy distribution (SED) of galaxy emission.  
Our models fit the auto and cross-frequency power spectra in each band, and measured number counts of sources, simultaneously, thereby introducing a new level of sophistication to the body of existing halo models in the literature. 
When required, we adopt the concordance model, a flat $\Lambda$CDM cosmology with $\Omega_{\rm M} = 0.274$, $\Omega_{\Lambda}  = 0.726$, $H_0 = 70.5\, \rm km\, s^{-1}\, Mpc^{-1}$, and $\sigma_8 = 0.81$ \citep{komatsu2011}.

\newpage
\section{Data}
\label{sec:data}
The primary data set for this work comes from the Herschel Multi-tiered Extragalactic Survey\footnote[1]{{\tt http://hermes.sussex.ac.uk}} \citep[HerMES;][]{oliver2012}, a guaranteed time (GT) key project of the \emph{Herschel Space Observatory} \citep{pilbratt2010}. 
We use submillimeter maps observed with the SPIRE instrument \citep{griffin2010} at 250, 350, and 500\rmicron.  
We also use reprocessed 100\rmicron\ \emph{IRAS} \citep{neugebauer1984} maps 
 in order to quantify the contribution to the power spectra from Galactic cirrus (see \S~\ref{sec:cirrus}).   
Each of the data sets is described in detail below.    
 \subsection{HerMES/SPIRE}
\label{sec:spire}
 HerMES fields are organized, according to area and depth, into levels 1 through 7, with level 1 maps being the smallest and deepest ($\sim 310\, \rm arcmin^2$), and level 7 maps the widest and shallowest \citep[$\sim 270\, \rm deg^2$;][]{oliver2012}. 

This study focuses on a subset of the level 5 and 6 fields, totaling $\sim 70\, \rm deg^2$, 
chosen for their large area and uniformity, and because they have a manageable level of Galactic cirrus contamination.  
 The fields used for this study, and a summary of their properties, are given in Table~\ref{tab:specs}.  Combined, they represent an increase of more than four times the area of the initial HerMES study \citep{amblard2011}.    
 The largest of the HerMES fields, the HerMES Large-Mode Survey ({\sc HeLMS}) --- which was designed specifically to measure the power spectrum on large angular scales ---  is still in preparation, and will be the subject of a future study.   
Maps will be made available to the public through {\sc HeDaM}\footnote[2]{{\tt http://hedam.oamp.fr/HerMES/}} \citep{roehlly2011} as a part of data release 2 (DR2).

\begin{table}[t!]
\scriptsize
\vspace{2.5mm}
 \centering
 \caption{Map Properties of the HerMES Fields} 
   \begin{tabular}{l||c|c|c|c}
   \hline
Field  Name & Area & 1$\sigma$ Noise  & Repeats & Scan Speed\\
  $ $  & (deg$^2$) &  (MJy sr$^{-1}$) & $ No. $ & $ $  \\
\hline
 {\sc bootes} & 11.3 & 1.11,\,0.61,\,0.29 & 9 & Parallel \\
{\sc cdfs-swire} & 12.2 & 1.00,\,0.57,\,0.27 & 5/20 & Parallel/Fast \\
{\sc elais-s1}  & 8.6 & 1.12,\,0.63,\,0.31 & 8 & Parallel \\
{\sc lockman-swire}  & 15.2 & 1.08,\,0.60,\,0.29 & 4/20 & Parallel/Fast \\
{\sc xmm-lss} & 21.6  & 1.15,\,0.68,\,0.52 & 8 & Parallel \\
   \hline
  \end{tabular}
 \tablecomments{Total noise (1$\sigma$ including confusion) are given at 250, 350, and 500\rmicron, respectively.  Repeats are defined as the number of times a field has been observed in \emph{two} orthogonal passes, thus one repeat equals two passes.  
{\sc cdfs-swire} and {\sc lockman-swire} were observed partially in Parallel mode, and partly in Fast mode. The scan speed of the telescope is either $20\, \rm arcsec\, s^{-1}$ (Parallel) or $60\, \rm arcsec\, s^{-1}$ (Fast).
 }
 \label{tab:specs}
\end{table}

The data obtained from the Herschel Science Archive were processed with a combination of standard ESA software and a customized software package SMAP.   
The maps themselves were then made using an updated version of SMAP/SHIM \citep{levenson2010}, an iterative map-maker designed to optimally separate large-scale noise from signal.  
SMAP differs from HIPE \citep{ott2010} in three fundamental ways which are relevant for power spectrum studies.  First, the standard scan-by-scan temperature drift correction module within HIPE is overridden in favor of a custom correction algorithm which stitches together all of the time-ordered data (or timestreams), allowing us to fit to and remove a much longer noise mode.  Further, the standard processing is modified such that a \lq\lq sigma-kappa\rq\rq\ deglitcher is used instead of a wavelet deglitcher, to improve performance in large blank fields.    
Lastly, imperfections from thermistor jumps, the \lq\lq cooler burp\rq\rq\ effect, and residual glitches, are removed manually  before map construction.
Detailed descriptions of the updates to the SMAP pipeline of \citet{levenson2010} are presented in Appendix \ref{sec:smap}.

Following \citet{amblard2011}, we make maps with 10 iterations, fewer than SMAP's default of 20,  
in order to minimize the time needed to measure the transfer functions (\S~\ref{sec:tf}) and uncertainties (\S~\ref{sec:errors}) with Monte-Carlo simulations.   

Additionally, timestream data are divided into two halves and unique \lq\lq jack-knife\rq\rq\ map-pairs are made.  These map-pairs are those ultimately used for estimating power spectra (\S~\ref{sec:mask}).  
SMAP maps are natively made with pixel sizes of 6, 8.33, and 12\arcsec, which is motivated by the beam size (sampling them by $\sim 1/3\, \rm FWHM$).  
But because the cross-frequency power spectra calculations needs maps of equivalent pixel sizes, three additional sets of map-pairs are made, 
with the extra sets having custom pixel sizes so that the cross-frequency power spectra can be performed at the pixel resolution native to the maps with the larger instrumental beams.  In other words, the maps used to calculate the $250\times 350$\rmicron\ spectra have identical 8.33\arcsec\ pixels; and those used for $250\times 500$ and $350\times 500$\rmicron\ have 12\arcsec\ pixels.
    
\subsection{\emph{IRAS}/IRIS}
\label{sec:iras}
At 100\rmicron, we use the Improved Reprocessing of the \emph{IRAS} Survey\footnote[3]{\tt{http://www.cita.utoronto.ca/\textasciitilde{}mamd/IRIS/IrisDownload.html}} \citep*[IRIS;][]{miv2005},  a data set which corrects the original plates for calibration, zero level and striping problems.  
The resulting full width at half maximum (FWHM) resolution and noise level are $4.3\pm0.2\arcmin$, and $0.06\pm0.02\, \rm MJy\, sr^{-1}$, respectively, and the gain uncertainty is 13.5\%.  
Data are available for up to three independent observations, or HCONs, although two of our fields were only observed twice.  For fields in which tiles intersect, maps can be stitched together with custom software provided on their site.\footnote[4]{\tt{http://www.cita.utoronto.ca/\textasciitilde{}mamd/IRIS/data/irispro.tar}}
\begin{figure}
\vspace{-5mm}
\hspace{-12mm}
\includegraphics[width=0.565\textwidth]{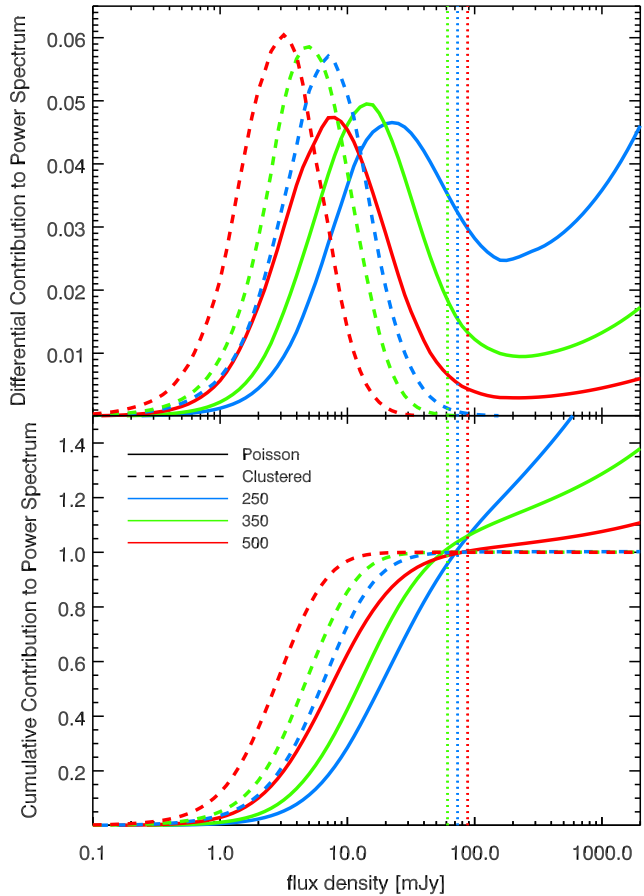}
\vspace{-8mm}\caption{
Differential (top panel) and cumulative (bottom panel) contributions to the Poisson (solid lines) and clustering (dashed lines) power from sources of different flux densities, estimated from the \citet{bethermin2011} model for illustrative purposes.   
{\bf Top panel:} the Poisson curves are determined by the normalization of $S^3 dN/dS$; and the clustering curves of $S^4 (dN/dS)^2$.  
By multiplying by an additional power of $S$, the peak values represent where the contribution to the integral per logarithmic interval are maximum.  
The integral under the curves for $S_{\rm cut} < 73$, 61, and 88\,mJy at 250, 350, and 500\rmicron, respectively (dotted vertical lines) --- which represent the 5$\sigma$ source detection threshold for a map made with five repeats --- are set equal to unity.  
The median values of sub-100\,mJy local maxima are (Poisson) 22, 14, and 8\,mJy, and (CIB) 7, 5, and 3\,mJy, at 250, 350, and 500\rmicron, respectively.  {\bf Bottom panel:} cumulative contribution to the power spectra normalized to unity at the detection threshold.  It is evident from this that the Poisson level at 250\rmicron, and to a lesser extent at 350\rmicron, are very sensitive to the masking level of resolved sources, while the 500\rmicron\ channel is fairly insensitive to source masking.  The clustering signal, on the other hand, is largely insensitive to masking.  Note that the clustering curves are estimates of linear clustering power, and do not include non-linear effects which may be more sensitive to source masking.  }
\label{fig:Poisson}
\end{figure}

\section{Power spectrum of CIB Anisotropies} 
\label{sec:method}
The cosmic infrared background at submillimeter wavelengths is dominated by emission from dusty star-forming galaxies, while other potential sources of signal, like the cosmic microwave background (CMB),  Sunyaev-Zel\rq dovich (SZ) effect, GHz-peaking radio galaxies or quasars, and intergalactic dust, are subdominant and can be safely ignored.  Anisotropies arise from galaxy 
over-densities (i.e., galaxy clustering) which appear as background fluctuations.  These anisotropies can be described by their power spectrum, and are made up of the following contributions:
\begin{equation}
P_{k_{\theta}}  =  P_{k_{\theta}}^{\rm shot} +P_{k_{\theta}}^{\rm clust}   +   P_{k_{\theta}}^{\rm fore} + N^{\rm inst}.
\label{eq:ps}
\end{equation}
Here $P^{\rm shot}_{k_{\theta}}$ is Poisson (or shot) noise,  $P^{\rm clust}_{k_{\theta}} $ is the power resulting from the clustering of galaxies; i.e., the excess above Poisson, $P^{\rm fore}_{k_{\theta}}$ is the noise from foregrounds, and $N^{\rm inst} $ is the instrumental noise. The foreground noise term could in principle include Galactic cirrus, free-free, synchrotron, and zodiacal emission, but in practice all but the cirrus term are negligible at SPIRE's wavelengths \citep[e.g.,][]{hajian2012}.

The Poisson noise component arises from the discrete sampling of the background, and as such is decoupled from the clustering term.  
For sources with a distribution of flux densities $dN/dS_{\nu}$ the effective Poisson level is
\begin{equation}
P^{\rm shot}_{\nu} = \int_0^{S_{\rm cut}} S_{\nu}^2\frac{dN}{dS_{\nu}}(S_{\nu}) dS_{\nu}.
\label{eqn:shot}
\end{equation}   
While the clustered power for the same sources can be estimated as roughly the three dimensional power spectrum of the galaxy number density field weighted by the square of the redshift distribution of the cumulative flux, $(dS_{\nu}/dz)^2$.  

The contribution to the Poisson noise and clustered power from galaxies with different flux densities are illustrated as solid and dashed lines in Figure~\ref{fig:Poisson}, respectively.  The peak contribution to the Poisson noise is from galaxies with $S_{\nu} \approx 22,$ 14, and $8\, \rm mJy$ at 250, 350, and 500\rmicron; while the peak contribution to the clustering power comes from fainter (higher-$z$) sources, with $S\approx 7,$ 5, and $3\, \rm mJy$ at 250, 350, and 500\rmicron. 
Shown as dotted vertical lines at $\sim 90$, 75, and 60\,mJy are the 5$\sigma$ limits of resolved sources in maps of equivalent depth \citep{nguyen2010}.
Poisson noise in the power spectrum is flat (in units of Jy$^2\, \rm sr^{-1}$), behaving as a level of white noise which can be reduced by masking brighter sources.  
Note that masking sources is more effective at reducing the Poisson level at 250\rmicron\ than it is at 350 or 500\rmicron.

\subsection{Estimating the Power Spectra: Masking, Filtering, and Transfer Functions}
\label{sec:mask}
The intensity in a given SPIRE map, $I_{\rm map}$, can be approximated as
\begin{equation}
I_{\rm map}=\left( T\otimes \left[ I_{\rm sky}\otimes B + N\right] \right)W, 
\label{eqn:one}
\end{equation}  
where  $I_{\rm sky}$ is the sky signal we wish to recover, $T$ is the transfer function of the map-maker, $B$ is the instrumental beam, $N$ is the noise, and $W$ is the window function, which includes the masking of map edges and of bright sources.  We use $\otimes$ to represent a convolution in real-space.  
The instrumental noise, $N$, is made up of white noise, which dominates on angular scales  $\kt \gsim 0.2\invarcmin$, and $1/f$ noise.  
We note that as $T$ has structure in 2D, and as the true beam may vary slightly across the map (particularly for bigger maps).  These corrections are small enough that Equation~\ref{eqn:one} remains a reasonable approximation.

In the auto-correlation of a map (i.e., the auto-power spectrum of a single map), all of the power present ---  which includes both signal and noise --- is correlated, while in the cross-correlation of jack-knife map-pairs, in principle the only signal correlated between them  is the sky signal. 
Since we are interested in recovering the sky signal, we estimate the power spectrum from the cross-correlation of jack-knife map-pairs (discussed in \S~\ref{sec:spire}).  
In practice, some correlated noise could exist between maps, particularly on large scales, as $1/f$.  
To minimize this, we use map-pairs constructed by dividing the timestreams in half by time, which ensures that the maps are made from data taken at time intervals corresponding to very large-scales.  Note, this would \emph{not} be true if, say, the data were split into those from even and odd detectors, since the same large-scale noise would be present in both maps.  The remaining $1/f$ results from serendipitous alignment of large-scale noise  \citep[see Appendix~B of][]{hajian2012}, i.e., independent noise clumps in either map that happen to line up.  Also note that excessive high-pass filtering of the TODs before constructing maps would extragalactic large-scale signal along with unwanted signal, and is thus not employed.   

The one-dimensional power spectrum is the azimuthal average of the (nearly isotropic) two-dimensional power spectrum of map-pairs in $k$-space.  In order to recover the true power spectrum of the sky, the cross-spectrum of the map-pairs must be corrected for the transfer function, masking, and instrumental beam.  We now describe these corrections in detail.  
\begin{table}
 \centering
  \caption{Number of masked sources}
   \begin{tabular}{l||r|r|r|r}
   \hline
 Obs ID     & 300\,mJy  & 200\,mJy  & 100\,mJy & 50\,mJy    \\
\hline
\hline
$ $        & 18 & 38 & 180 & 1567 \\
{\sc bootes} &  4  & 9 & 46 & 723 \\
$ $        & 1 & 2 & 13 & 248\\
\hline
$ $            & 9   & 23 & 165  & 1569 \\
{\sc cdfs-swire} &  3 & 6 & 30 & 725 \\
$ $            & 1   &  1 & 6 & 141\\
   \hline
$ $          & 9 & 23 & 104 & 1073 \\
{\sc elais-s1} & 1  & 3 & 22 & 437 \\
$ $          & 1 & 1 & 3 & 146\\
\hline
$ $                      & 16 & 42 & 221 & 2043 \\
{\sc lockman-swire} &  4  & 9 & 57 & 930 \\
$ $                       & 1 & 3 & 10 & 220 \\
\hline
$ $            & 23 & 66 & 368 & 2973 \\
{\sc xmm-lss} &  5   & 13 & 85 & 1170 \\
$ $            & 1 &   2  & 16 & 437\\
   \hline
  \end{tabular}
\tablecomments{Number of sources masked at 250, 350, and 500\rmicron\ are given from top to bottom in each panel.   
 Not accounted for in the table are the extended sources masked in each field, of which there are 3, 2, 2, 4, and 4, respectively.  A 50\,mJy cut amounts to masking approximately 1.4, 1.2, and 0.5\% of the pixels at 250, 350, and 500\rmicron, respectively.}
 \label{tab:masked_sources}
\end{table}
\subsubsection{Masking and the Mode-Coupling Matrix}
Before calculating the power spectra, the maps are multiplied with windows whose values equal unity in the clean parts of the maps and taper to zero with a Gaussian profile (90\arcsec\ FWHM) at the edges.  
Note that jack-knife maps typically do not cover identical patches of sky because the orientation of the telescope between observations changes, so that the windows of the two jack-knife map-pairs are different.   

Next, because we are interested in the behavior of the power spectrum with the masking level of resolved point sources, we create a series of masks for each field and band to mask sources whose flux densities  are greater than 50, 100, 200, and 300\,mJy, in addition to extended sources.  
Extended sources are exceptional objects, e.g., local IRAS galaxies, which exceed 400\,mJy but can be as bright as 1{,}500\,mJy at 250\rmicron. 
Mask positions are determined from catalogs of resolved sources which we construct.  The source identification code first high-pass filters the maps in Fourier space to remove Galactic cirrus and other large-scale power, then convolves the maps by the instrumental beam, and finally measures all peaks with signal-to-noise greater than 3$\sigma$.   
The total number of masked sources in each field and band are tabulated in Table~\ref{tab:masked_sources}.  
Note that this method could potentially suffer from Eddington bias \citep[e.g.,][]{chapin2009}, a phenomenon where instrumental white noise systematically boosts faint sources above the detection limit.   
To ensure that this is not a significant problem, we compare the total number of sources above the cut to cumulative number counts from \citet{glenn2010} and find that they are consistent.  

Each source is masked by circles of $1.1\times \rm FWHM$ in diameter, or 19.1, 27.7, and 40.3\arcsec\ at 250, 350, and 500\rmicron, respectively --- chosen to cover the full first lobe of the beam, though we check that the exact size of the mask has a negligible effect on the spectra.  
For the data that concerns this paper, unique masks are made for each field and each band (i.e., one at 250, one at 350, and one at 500\rmicron), so that only sources above the given cut in that band are masked.  This means that when calculating the cross-frequency power spectra, not all of the sources masked at e.g., 250\rmicron\ will also be masked at 350\rmicron, and vice-versa.  However, we additionally calculate an alternative set of spectra where we mask in all bands the sources identified at 250\rmicron, 
 (i.e., the same mask at 250, 350, and 500\rmicron), 
and the power spectrum pipeline is rerun.  Plots and tables for this alternate masking scheme are presented in Appendix~\ref{sec:alt}, where we also show that the spectra at longer wavelengths are less sensitive to the level of source masking than at shorter wavelengths. 
Finally, we note that this method differs from that of \citet{amblard2011}, who instead masked all pixels above 50\,mJy, as well as all neighboring pixels; 
the motivation being that a catalog-based masking scheme is better able to distinguish sources from spurious noise.  
The total number of masked sources in each field and band are tabulated in Table~\ref{tab:masked_sources}.  
\begin{figure}[t!]
\hspace{-10mm}
\includegraphics[width=0.55\textwidth]{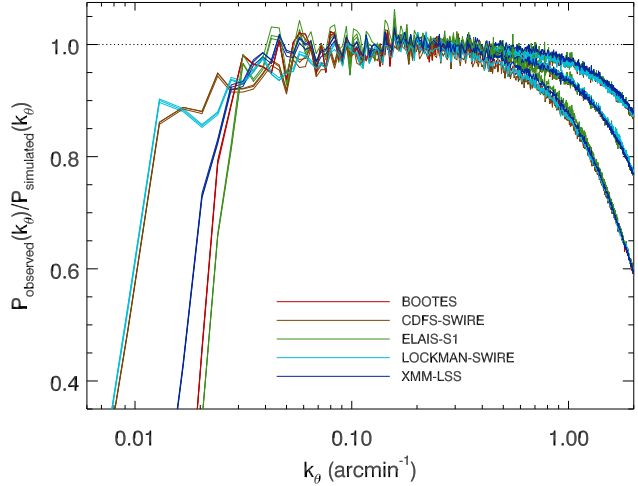}
\caption{Transfer functions of the map-maker, averaged in 2D, for each of our fields.  For any given field, the transfer functions on large scales are indistinguishable between bands, while on small scales, the transfer functions converge to that of the pixel windows, which are band dependent.  The shape of the transfer function depends largely on: the scan speed, which determines on what scales the $1/f$ noise is projected onto the timestreams; and scan lengths, which determines the order of polynomial which is removed from the timestreams in the SMAP pipeline. 
Thus, maps which are smaller or which were observed with slower scans are attenuated on smaller angular scales.  
} 
\label{fig:tf}
\end{figure}
 
Masking in map-space can result in mode-coupling in Fourier space, which can bias the power spectrum.  The coupling kernel, or mode-coupling matrix \citep{hivon2002}, in the flat sky approximation is  
\begin{equation}
M_{kk^{\prime} } = \displaystyle\sum\limits_{\theta_k}\displaystyle\sum\limits_{\theta_{k^{\prime}}}\left| w_{kk^{\prime} }\right|^2/N(\theta_k),
\label{eqn:mkk}
\end{equation}
where  $\left| w_{kk^{\prime} }\right|^2$ 
is the auto-power spectrum of the mask, and $N(\theta_k)$ is the number of modes in annulus of radius $k$.  
Since in our case the masks of the map-pairs are not identical, Equation~\ref{eqn:mkk} is generalized for different masks by replacing $\left| w_{kk^{\prime} }\right|^2 $ with the cross-power spectrum of the masks, $\langle w^A_{kk^{\prime} } w^{^\ast B}_{kk^{\prime} } \rangle$ \citep[see][]{tristram2005}. 

The mode-coupling matrix must be inverted in the final step in order to recover the de-coupled power spectrum.   
A unique mode-coupling matrix is calculated for each auto- and cross-frequency power spectrum, and for each flux cut, per field.  
Additionally, each $M_{kk^{\prime}}$ is tested on 1000 simulated maps with steep input spectra and found to be unbiased.

\begin{figure}[t!]
\centering
\hspace{-10mm}
\includegraphics[width=0.51\textwidth]{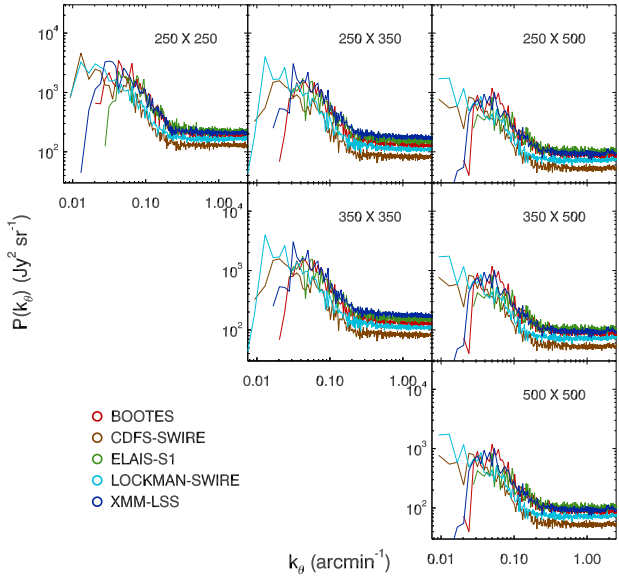}
\caption{Noise levels calculated from the power spectrum of the difference map of jack-knife map-pairs. Not shown but consistent with these are noise curves estimated as the auto- minus cross-power spectra of the maps.
White noise dominates the spectra on scales $k_{\theta} \gsim 0.25\, \rm arcmin^{-1}$, while $1/f$ noise is prominent on larger angular scales.    
As expected, deeper maps have lower white noise than shallower maps.  
The turnover on the largest scales, which is map-dependent, reflects the high-pass filtering by the map-maker.  
Note that the ordinate (y-axis) differs from that of the following figures presenting power spectra, as the signal is nearly two orders of magnitude greater than the noise.  
} 
\label{fig:noise_ps}
\end{figure}

\subsubsection{Transfer Function}
\label{sec:tf}
The large-scale correlated noise of SPIRE is extremely low \citep[e.g.,][]{pascale2011}. 
As a result, 
only minimal high-pass filtering is required to make well-behaved maps, and the resulting power spectrum can be measured 
out to relatively large scales ($k_{\theta} \gsim 0.01\,\invarcmin$).  
This filtering is quantified by the transfer function of the map-maker, $T$, which must be accounted for in the final spectra.  

We measure $T$ with a Monte-Carlo simulation whose steps are: \rmnum{1}) running the SMAP map-making pipeline on simulated observations of input maps with known power spectra resembling that of clustered DSFGs; \rmnum{2}) calculating the power spectrum of the output map with a pipeline identical to that used for the real data, including all masking, Fourier space filtering, and mode-coupling corrections;  and iii) computing the average of the ratio of the output power spectrum to the known input spectrum.  
Transfer functions are calculated from 100 simulations of each field and wavelength, and are shown in Figure~\ref{fig:tf}.  
We check that the spectra in step (\rmnum{1}) are not sensitive to the steepness of the input spectra, and that the transfer functions have converged, with a mean error of $\sim 1\%$.  

As anticipated, all three bands converge on large-scales for each field, meaning that the same filtering is performed in each band.  
And on small scales ($k_{\theta}>0.3\invarcmin$) all fields converge to the same three curves, which are the pixel window functions of the three bands.  
The angular scales on which the high-pass filtering occurs in each field is related to the average speeds and lengths of the maps scans: the former determines the scale in which $1/f$ noise is projected onto the timestreams; while the latter dictates the order of the polynomial removed from the timestreams.  
As described later in \S~\ref{sec:combine}, the final spectra are weighted combinations of those in each field which are attenuated by less than 50\%, corresponding to $k_{\theta} \ge 0.021$, 0.009, 0.023, 0.015, and $0.009\invarcmin$, for {\sc bootes}, {\sc cdfs-swire}, {\sc elais-s1}, {\sc lockman-swire}, and {\sc xmm-lss}, respectively. 
Maps scanned with longer and faster scans have less attenuation on large scales, which is why the maps observed in fast-scan mode ({\sc cdfs-swire} and {\sc lockman-swire}) are also those which best measure the largest scales. 
Note, the excess power introduced on large scales by earlier versions of SMAP \citep[][]{levenson2010, amblard2011} is no longer present.  

\subsubsection{Instrument Beam}
\label{sec:psf}
The instrumental PSF (or beam) attenuates power on scales smaller than $\sim 0.25$--$0.5\invarcmin$, depending on the band.  
This window function can be corrected by dividing the power spectrum of the map by the power spectrum of the beam.
The instrumental beam is measured from maps of Neptune --- a source which to SPIRE is effectively point-like (angular size $\lsim 2.5\arcsec$).

The beam power spectra are estimated in the following way.  All pixels beyond a 10$\times \rm FWHM$ radius from the peak  are masked, due to uneven coverage and excessive noise.  We check that the dependence of the beam spectra on the choice of the radius is small, with any differences contributing to the systematic uncertainties.  Furthermore, point sources in the background above 30\,mJy at 250\rmicron, which are subdominant but contribute to the noise, are masked in all bands.  Again, we check that the level of this masking makes a negligible difference, but account for each of these differences as part of the error budget.     
Thus, uncertainties in the beam power spectra measurements are largely systematic.  These uncertainties couple to the uncertainties in the estimate of the power spectra, and are accounted for in the Monte Carlo procedure described in \S~\ref{sec:errors}.

\begin{table*}[t!]
 \centering
 \caption{Galactic cirrus properties in each field}
 \begin{tabular}{l||c|c|c|c|c|c}
 \hline
Obs ID & $\alpha_{\rm c} $ & $P_{0,100}$ & $P_{0,250}$ & $P_{0,350}$ & $P_{0,500}$ & $T$ \\
$ $ & $ $ & $\rm (Jy^2/sr)$ & $\rm (Jy^2/sr)$ & $\rm (Jy^2/sr)$ & $\rm (Jy^2/sr)$ & $ (K) $  \\
 \hline
{\sc bootes } &  $ -3.52 \pm    0.41 $ &  $ 7.22 \times 10^{5} $  &  --- &  --- &  --- &  --- \\ 
{\sc elais-s1 } &  $ -3.75 \pm    0.08 $ &  $ 3.10 \times 10^{5} $  &  --- &  --- &  --- &  --- \\ 
{\sc lockman-swire } &  $ -3.66 \pm    0.05 $ &  $ 2.8 \times 10^{5} $  &  $ 4.5 \times 10^{5} $  &  $ 2.23 \times 10^{5} $  &  $ 9.22 \times 10^{4} $  &  $ 17.2 \pm 1.2 $  \\  
{\sc xmm-lss } &  $ -2.96 \pm    0.17 $ &  $ 7.31 \times 10^{5} $  &  $ 1.1 \times 10^{6} $  &  $ 5.24 \times 10^{5} $  &  $ 2.9 \times 10^{5} $  &  $ 18.5 \pm 0.7 $  \\  
{\sc cdfs-swire } &  $ -3.93 \pm    0.06 $ &  $ 7.36 \times 10^{5} $  &  $ 6.66 \times 10^{5} $  &  $ 3.22 \times 10^{5} $  &  $ 1.22 \times 10^{5} $  &  $ 20.4 \pm 1.4 $  \\  
\hline
\end{tabular}
\tablecomments{Best-fit variables from Equation~\ref{eqn:cirrus} are Column 2: the index $\alpha_{\rm c}$;  Column  3--6: the amplitudes in each band $P_0$; and Column 7: the temperature $T$ with $\beta = 1.8$.  Long dashes represent fits which were unconstrained by data because filtering in those maps was too aggressive to recover the large scales where the power from cirrus would be present.  
}
\label{tab:cirrus}
\end{table*}

\subsection{Estimating Uncertainties}
\label{sec:errors}
Present in each map-pair is correlated signal from the sky, and both correlated and uncorrelated noise.  Three terms contribute to the uncertainties in the power spectrum: a non-Gaussian term due to the Poisson distributed compact sources; sample variance in the signal due to limited sky coverage; and the noise.  In this order, the  
variance of the cross-spectra of maps $\rm A\times B$ can be written as
\begin{eqnarray}
\sigma^2(\hat{P}_{b}^{A\times B}) & = & \frac{\sigma^2_{\rm P}}{f_{\rm sky}} + \frac{2}{n_b}\left(  \hat{P}_{b}^{A\times B}  \right)^2  \nonumber \\ 
							& + & \frac{\hat{P}_{b}(\hat{N}_{b}^{A} + \hat{N}_{b}^{B})+\hat{N}_{b}^{A}\hat{N}_{b}^{B}}{n_b},                                                             
\label{eqn:uncertainties}
\end{eqnarray}
where $\hat{P}_{b}$ is the mean cross-spectrum of map-pairs, $\hat{N}_{b}$ is the average noise power spectrum of the map, $n_b$ is the number of Fourier modes measured in bin $b$,  and $f_{\rm sky}$ is the observed area divided by the solid angle of the full sky.  The first term, $\sigma^2_{\rm P}$ 
is given by the non-Gaussian part of the four-point function \citep[as described in e.g.,][]{acquaviva2008,hajian2012}, and is particularly sensitive to the flux cut of the masked sources.  

Shown in Figure~\ref{fig:noise_ps} are the noise levels calculated from the power spectrum of the difference map of jack-knife map-pairs, which are consistent with the difference between the auto- and cross-power spectra of the maps (not shown).  The noise behavior demonstrates the impressive performance and stability of the SPIRE instrument, with white noise in most cases nearly two orders of magnitude below the power from the sky signal.  
The noise spectra turn over on large scales due to the filtering performed by the map-maker, which is related to the length and speed of the scans.  

Uncertainties in the estimates of the power spectra are derived from Monte Carlo simulations of the pipeline on realistically simulated sky-maps.  The maps include sources correlated between bands (i.e., the same sources appear in all three maps, but with different flux densities), which are necessary for estimating uncertainties in the cross-frequency power spectra, as well as both $1/f$ and white noise, and Galactic cirrus.  
Also included in the Monte Carlo simulations are systematic uncertainties arising from the beam and transfer function corrections, such that for each iteration, the beam and transfer function corrections are perturbed by the appropriate amount.  

The ensemble of estimated output power spectra are used to measure, ${\bf V}$, the covariance matrix  
\begin{equation}
{\bf V}_{bb^{\prime}}=\left \langle \left(P_b -  \tilde{P}_b \right)  \left(P_{b^{\prime}} - \tilde{P}_{b^{\prime}}\right) \right \rangle_{{\rm MC}}, 
\end{equation}
where the tilde denotes the mean over every iteration in bin $b$.  
The resulting errors are 
\begin{equation}
\sigma_{P_{b}^{\rm map}}=\sqrt{{\bf V}_{bb}}.
\label{eqn:diagonal}
\end{equation}
The non-Gaussian term emerges from the simulations as an offset in ${\bf V}$.   
We check that this level is realistic by comparing it to the level of the four-point function estimated directly from data, with appropriate masking, following \citet{fowler2010}, which we found to be between 5 and 10\% of the total error in the Poisson dominated regime, depending on flux cut of masked sources: more aggressive source masking results in a smaller non-Gaussian term.  

In addition, there are $\sim 8 \%$ systematic errors due to absolute calibration uncertainty, of which $\lsim 1 \%$ is due to beam area uncertainty, as described in Appendix~\ref{sec:calibration}.  Though they are accounted for when model fitting, they are not included in the reported error bars.

\begin{figure*}[t!]
\centering
\includegraphics[width=1.0\textwidth]{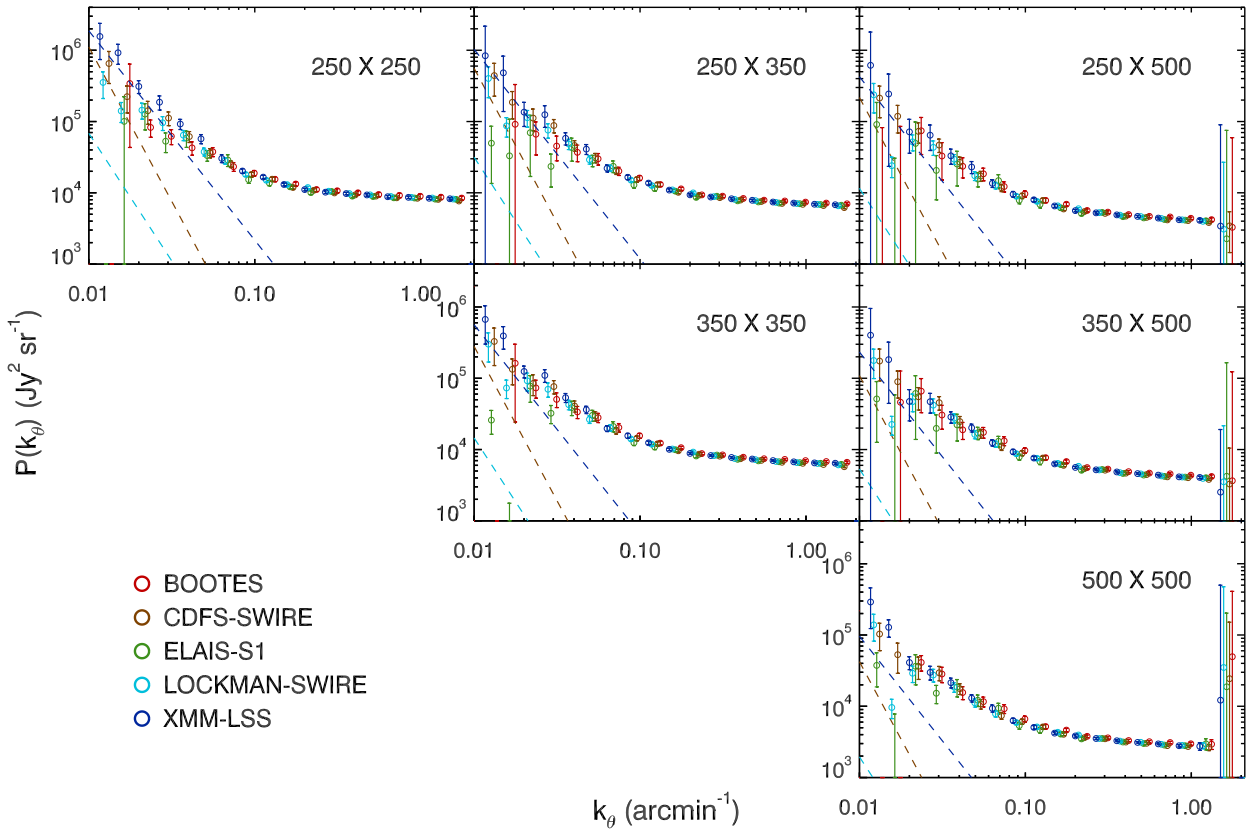}
\caption{Auto- and cross-frequency power spectra of the sky (circles with error bars) before correcting for Galactic cirrus (dashed lines) for all fields after masking sources with flux densities greater than 100\,mJy.  Data points are shifted horizontally for clarity.  Error bars are derived for each field from Monte Carlo simulations, as described in \S~\ref{sec:errors}.   
All fields agree within the errors on scales where cirrus contamination is not significant ($\gsim 0.06\invarcmin$), and the variance in the shot noise levels is consistent with expectations derived from simulations.  Maps which are smaller or which were observed with slower scans are attenuated on smaller angular scales, which is a reflection of the pattern also see in the transfer functions (Figure~\ref{fig:tf}). } 
\label{fig:raw_ps}
\end{figure*}

\subsection{Galactic Cirrus}
\label{sec:cirrus}

The most significant foreground for the extragalactic power spectrum is that from Galactic cirrus, which can dominate the signal on scales greater than $\sim 30\arcmin$.  \citet{gautier1992} showed that the power spectrum of Galactic cirrus can be well approximated by a power law 
\begin{equation}
P_{k_{\theta}}^{\rm cirrus}=P_0\left( \frac{k}{k_0} \right)^{\alpha_{\rm c}},
\label{eqn:cirrus}
\end{equation}
whose amplitude, $P_0$, normalized at $k_0=0.01\invarcmin$, may vary from field to field, but whose index $\alpha_{\rm c} \approx -3.0$.  
More recently, studies of various fields using BLAST and SPIRE data have found a much wider range of the index, e.g., $\alpha_{\rm c} \sim -2.4$ \citep{bracco2011}, $-2.6$ \citep{roy2010}, $-2.8$ \citep{martin2010, miv2010}, $-2.9$ \citep{lagache2007}, and even ranging from $-2.5$ to $-3.6$, depending on the field \citep{miv2007}.   
Thus, rather than assume a power law with $\alpha = -3$ \citep[e.g.,][]{viero2009}, we proceed by treating each field independently.  

In diffuse cirrus regions \citep[i.e., column density, $N_{\rm HI} < 2 \times 10^{20}\, \rm  cm^{-2}$ and brightness temperature, $T_{\rm b} < 12\,$K; e.g.,][]{lockman2005,gillmon2006}, H\,{\sc i} is a good tracer of dust, and the dust-to-gas ratio can be measured from the slope of the pixel-pixel scatter plot.  
Cirrus contamination can then be \lq\lq cleaned\rq\rq\ by scaling the H\,{\sc i} maps by the dust-to-gas ratio and subtracting them directly from the dust maps in question.  This approach has been used quite successfully by e.g., \citet[][]{lagache2011} or \citet[][]{penin2012a}, on maps which have high fidelity on large angular scales.  
Unfortunately, this technique is made complicated for SMAP-made SPIRE maps because the filtering of scales larger than $\gsim 20\, \rm arcmin$ attenuates the very structure that the differencing with H\,{\sc i} is meant to remove.  Though H\,{\sc i} maps can be filtered with the SMAP simulator to attenuate large scales, the remaining structures are faint with respect to the noise, and thus difficult to regress with SPIRE maps.    

We instead adopt an approach similar to that used by \citet{lagache2007}, \citet{viero2009}, and \citet{amblard2011}, with some additional modifications.  
Diffuse Galactic cirrus emits as a modified blackbody proportional to $\nu^{\beta}B({\nu})$, where $B({\nu})$ is the Planck function and $\beta$ is the emissivity index.
Typically, it has a temperature of $\sim 18\,\rm K$ and $\beta \sim 1.8$ \citep[e.g.,][]{bracco2011} resulting in a SED which peaks at $\sim 170$\rmicron\ \citep[e.g.,][]{martin2012}.  At 100\rmicron, cirrus emission has roughly the same amplitude as at 250\rmicron, but unlike in SMAP/SPIRE maps, the favorable large-scale properties of the IRIS maps make it possible to accurately measure the power spectra out to scales of $\sim 4^{\circ}$.  
Thus, assuming that the Galactic cirrus power spectrum is well described by a power law \citep{roy2010}, we use the 100\rmicron\ power spectra, calculated from IRIS maps with sizes identical to their SPIRE counterparts, and with sources above 500\,mJy masked, to estimate the best-fit to the cirrus spectra in each field.  
Note that although larger size regions would better constrain the large-scale spectra, we intentionally use the exact same regions because these fields were chosen specifically because they were special places in the sky with low Galactic cirrus, and thus the spectra inside and the spectra surrounding the field are unlikely to be the same.   
Uncertainties in the power spectra are estimated analytically following \citet{fowler2010}.  To distinguish between the power originating from cirrus and that from clustered galaxies, we include an estimate of the linear power (i.e., 2-halo) term constrained by the measured galaxy spectra of \citet{penin2012a}.  Also, we adopt the \citet{bethermin2011} model to fix the Poisson level, which is unconstrained by data because of the 4\arcmin\ \emph{IRAS} beam, although we note that on these angular scales the exact choice for the Poisson level has a negligible effect on the fit.  

Next, assuming that the linear power spectrum from clustered DSFGs is independent of field, 
and after masking all sources above 300\,mJy in the SPIRE bands, we estimate the contribution to the SPIRE spectra from cirrus
 by fitting the 100\rmicron\  and SPIRE auto-power spectra of all five fields simultaneously with: a Poisson term;
 a 1- and 2-halo clustered galaxy terms; and a temperature (with fixed $\beta = 1.8$) which sets the band-to-band amplitudes, $P_0$.  
 
Lastly, uncertainties are estimated with a Monte Carlo simulation where the slope and amplitudes of the best-fit power law at 100\rmicron\ are perturbed by an amount dictated by their errors, and the cirrus estimate pipeline described above is rerun with those values fixed. 
Results are given in Table~\ref{tab:cirrus}.  Note, gain uncertainties in the IRAS maps are not accounted for in the fit, as they would only act to increase the error of the best-fit temperature, but not the uncertainty in the best-fit power law.  
Also note that many of the indices are steeper than those of most previous analyses of cirrus power spectra, which may be due to the fact that these regions are specifically chosen as windows through the cirrus, and not representative of the mean.  And although they are steep, they remain consistent with results at shorter wavelengths from \citet{bazell1988}, or the extreme end of spectra found by \citet{miv2007}.  Nevertheless, we check that fixing the slope of spectra in each field to $\alpha = -3$ does not significantly alter the correction, and indeed find that the resulting cirrus-corrected data fall within the uncertainties.

We find that the method provides good constraints for the {\sc lockman-swire}, {\sc cdfs-swire} and {\sc xmm-lss} fields, while in the {\sc bootes} and {\sc elais-s1} fields,  because of the aggressive filtering in the SPIRE maps, 
the measured spectra only probe scales in which the cirrus contribution is negligible.     
Consequently, when later combining the spectra, the largest scale bins are constrained using a subset of the maps.  


\begin{figure}[t!]
\centering
\hspace{-10mm}
\includegraphics[width=0.5275\textwidth]{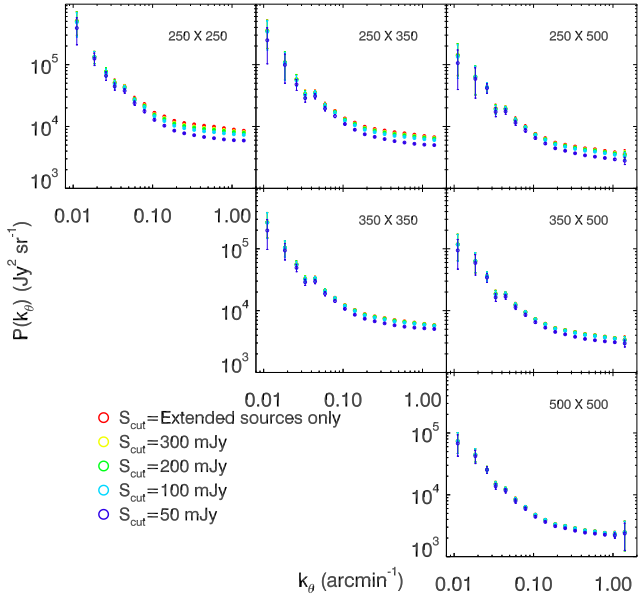} 
\caption{Spectra combined from the different fields for different levels of resolved source masking, plotted as circles with error bars.  The best estimates of the cirrus spectra in each field (\S~\ref{sec:cirrus}) are removed before combining.  All data and uncertainties are tabulated in Table~\ref{tab:combined_spectra}.
} 
\label{fig:ps_vs_flux_cut}
\end{figure}
\subsection{Re-binning and Combining Spectra}
\label{sec:combine}

Individual spectra in each field, for each band and flux cut, are first re-binned following \citet{amblard2011} into logarithmic intervals with width equal to $\Delta k_{\theta}/k_{\theta} = 0.25$ for $k_{\theta} \ge 0.033$ ($\Delta \ell/\ell = 720$), and linearly for larger scales, with bin-widths of $\Delta k_{\theta} = 7.41 \times 10^{-3}\invarcmin$ ($\Delta \ell=160$).  
These bin-width values are chosen to ensure that, with the exception of correlations introduced on small scales from Poisson errors, the off-diagonals in the covariance matrix are always less than $\sim 10\%$. 
The re-binned uncertainties are given by
\begin{equation}
\sigma_{P_b^{\rm sky}} ^2=\frac{ 1}{ \displaystyle\sum\limits_{i,j} \left({\bf V}_{i,j}^{-1}\right)}, 
\label{eqn:sigma}
\end{equation}
where $i,j$ span the entries of bin $b$ and 
$\left({\bf V}_{i,j}^{-1}\right)$ is the inverse of the subset  of the covariance matrix ${\bf V}$ calculated from simulations (\S~\ref{sec:errors}).

Next, the best estimates of the cirrus power spectra for each field (estimated in \S~\ref{sec:cirrus} and reported in Table~\ref{tab:cirrus}) are subtracted in order to recover the power spectra of extragalactic sources, $P_{k_{\theta}}^{\rm exgal}$.  Uncertainties in the cirrus estimate are propagated into the final uncertainties assuming that the errors are uncorrelated so that
\begin{equation}
\sigma_{P_b^{\rm exgal}} ^2=\sigma_{P_b^{\rm sky}} ^2 + \sigma_{P_b^{\rm cirrus}} ^2.
\label{eqn:final_sigma}
\end{equation}
\begin{figure}[t!]
\vspace{-2.6mm}
\hspace{-5mm}
\includegraphics[width=0.53\textwidth]{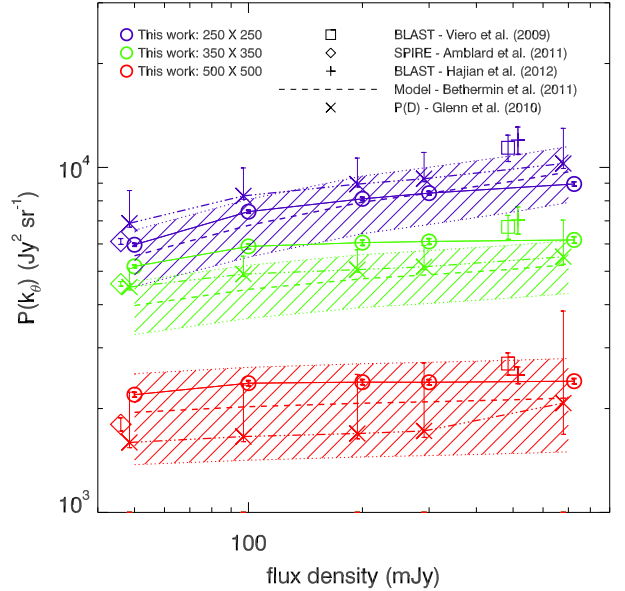}
\vspace{-2.0mm}
\caption{Poisson noise level vs.\@ flux cut of masked sources.  Best-estimates of the auto-frequency spectra values are shown as open circles and tabulated in Table~\ref{tab:poisson}.   It is worth noting that the data points between bands are not independent.  
Estimates of the Poisson level derived from SPIRE $P(D)$ source counts \citep{glenn2010} are shown as exes with asymmetric error bars, whose sizes are functions of the uncertain upper limits on the faint end, and that the fields studied were relatively small.  
Dashed lines and shaded regions represent the best estimate and 1$\sigma$ uncertainties of the \citet{bethermin2011} model, with which we find good agreement at all but 350\rmicron, which is underpredicted by approximately 1$\sigma$.  
Also shown are measured Poisson levels from BLAST (squares: \citealt[][]{viero2009}; crosses: \citealt[][]{hajian2012}), and SPIRE \citep[diamonds:][]{amblard2011}.   
} 
\vspace{2.5mm}
\label{fig:shot}
\end{figure}

\begin{figure*}
\hspace{-7.0mm}
\includegraphics[width=1.05\textwidth]{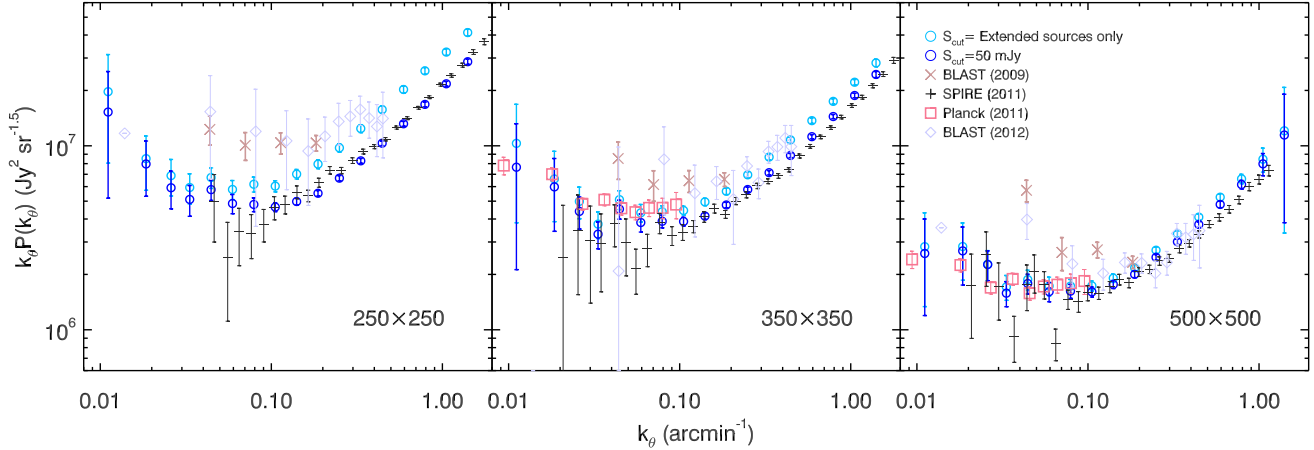} 

\caption{Comparison of our cirrus-corrected, combined data to published measurements.  
Data are plotted as $k_{\theta}P(k_{\theta})$ in order to reduce the dynamic range of the plotted clustering signal,
 and thus better visualize the differences between the measurements.  
Furthermore, in order to adequately compare to the wide range of source masking found in the literature, we present the two 
masking extremes of our analysis: spectra with sources greater than 50\,mJy masked (dark blue circles), and those with only extended sources having been masked (light blue circles).  
Previous SPIRE measurements  from \citet{amblard2011}, which should be compared to the dark blue circles, are shown as black crosses.  The remaining data and curves should be compared to the light blue circles.  
Note that to help aid the comparison, error bars include systematics due to calibration and beam uncertainties.  
They are: 
BLAST data from \citet{viero2009} (brown exes) and \citet{hajian2012} (lavender diamonds); and \emph{Planck} data from 
\citet{lagache2011} (red squares) at 350 and 500\rmicron. Note, \emph{Planck} data 
are color corrected to account for their different passbands by multiplying the 350 and 500\rmicron\ data by factors of 0.99, and 1.30, respectively, and adjusted to the most current calibration by dividing them by 1.14 and 1.30, respectively \citep[see][]{planck_viii}.} 
\label{fig:published_data}
\end{figure*}

Finally,  data in each band and flux cut are combined for the five fields, $f$, following \citet{lagache2011}, where
\begin{equation}
P_b^{\rm combined}= \displaystyle\sum\limits^5_{f=1} W^f_b \times P_b^{f,\rm exgal},  
\label{eqn:combine}
\end{equation}
and $W^f_b$ is the weight of each field and bin,
\begin{equation}
W^f_b= \frac{\sigma^{-2}_{P^{f,\rm exgal}_b}}{\displaystyle\sum\limits^5_{f=1} \sigma_{P^{f,\rm exgal}_b}^{-2}},  
\label{eqn:combine_weight}
\end{equation}
which assumes that fields are far enough apart to be uncorrelated. 
Note that for each field, spectra at angular scales where the transfer function falls below 0.5 are omitted in the combined fit.

\section{Results}
\label{sec:results}
\subsection{Total Sky Spectra}
\label{sec:total}
We measure signals in excess of Poisson noise in all auto- and cross-frequency power spectra, in each field.  This excess signal originates from the clustering of DSFGs, and to varying degrees from Galactic cirrus on large scales.  
Total power spectra of the five fields (which includes power from Galactic cirrus and Poisson noise) with sources $\ge 100\, \rm mJy$ masked are shown in Figure~\ref{fig:raw_ps}.  
Spectra with different levels of source masking behave similarly, and are thus here omitted for clarity.    
Also shown are dashed lines representing the best estimate of the Galactic cirrus, approximated as power laws.   
The spectra from each field agree within errors on angular scales where the power from Galactic cirrus is subdominant ($\gsim 0.06\invarcmin$).    

\begin{figure*}[t!]
\hspace{-7.0mm}
\includegraphics[width=1.05\textwidth]{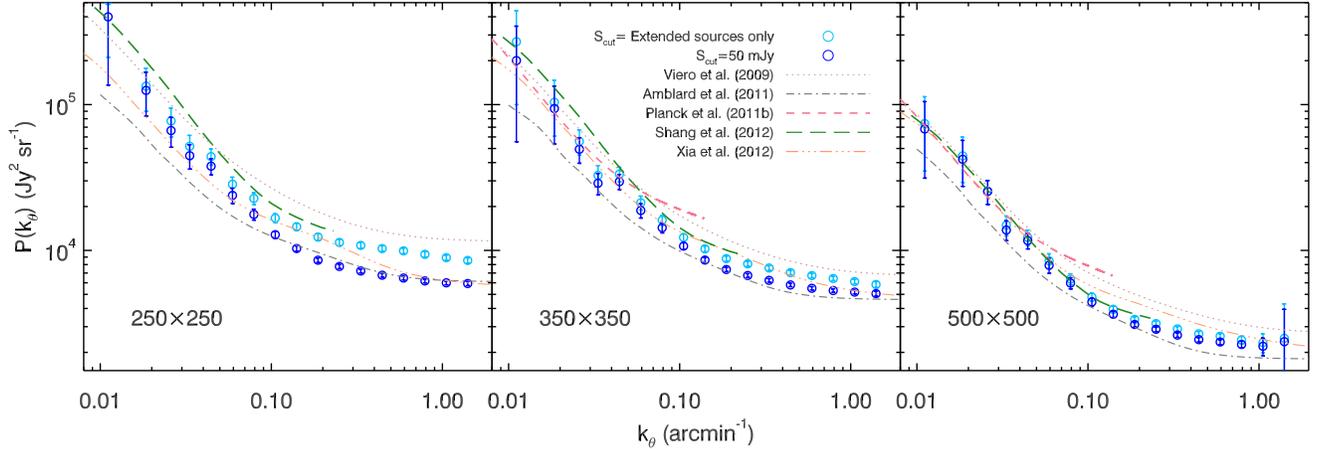} 
\caption{Comparison of our cirrus-corrected, combined data to published models.  
Best-fit halo model to previously published SPIRE data \citep{amblard2011} are shown as grey dot-dashed lines, and should be compared to the dark blue circles.  
The remaining  curves should be compared to the light blue circles.  
Note, with the exception of the BLAST model \citep{viero2009}, all of the following were originally fit to the 857 and 545\,GHz channels of \emph{Planck}, and have thus been color-corrected by 0.99 and 1.30 at 350 and 500\rmicron, respectively.  
They are from: 
 \citet[][brown dotted lines]{viero2009}; 
\citet[][red dashed lines]{lagache2011} at 350 and 500\rmicron; case 0 of \citet[][green dashed lines]{shang2012}; and \citet[][orange three-dot-dashed lines]{xia2012}.  } 
\label{fig:published_models}
\end{figure*}
\subsection{Combined Extragalactic Sky Spectra}
\label{sec:combine}
The cirrus-subtracted power spectra in each field, which are combined using Equation~\ref{eqn:combine}, are presented in Figure~\ref{fig:ps_vs_flux_cut} and tabulated in Appendix~\ref{sec:tables}, Table~\ref{tab:combined_spectra}.  As expected, the spectra in each panel converge for increasing angular scales as the contribution from Poisson noise becomes subdominant.  
Consistent with expectations from Figure~\ref{fig:Poisson}, the flux cut has a significant effect on the Poisson level at 250\rmicron, and a nearly negligible effect at 500\rmicron.  

Poisson levels are determined through a simultaneous fit to the combined spectra of the Poisson and clustered galaxy terms with templates adopted from the \citet[][]{viero2009} halo model, and are shown as a function of flux cut of masked sources in Figure~\ref{fig:shot}.  
We note that this estimate is subject to systematic uncertainties due to the mild degeneracy of the Poisson and 1-halo terms, more so for spectra with fewer masked sources, and that we account for those uncertainties in the estimate.  
As anticipated, shorter wavelengths are significantly more affected by removal of the brightest sources (see Figure~\ref{fig:Poisson}).  Previous measurements from \citet{amblard2011} with 50\,mJy sources masked are in relatively good agreement, with 250\rmicron\ higher by $\sim 5$\%, and 350 and 500\rmicron\ lower by $\sim 7$\% and $ 17$\%, respectively.    
The BLAST measurements, which masked sources with flux densities greater than 500\,mJy  \citep{viero2009,hajian2012} appear to be higher than our values by $\sim 26$\%, $12$\%, and $11$\% at 250, 350, and 500\rmicron, respectively.  

Also plotted are estimates of the Poisson level derived from the \citet{glenn2010} P(D) number counts using Equation~\ref{eqn:shot}.  We find that the number count predictions over-estimate our values by $\sim 16\%$ at 250\rmicron, and under-estimate our measured values by $\sim 12$ and 16\% at 350 and 500\rmicron, respectively.   It should be noted that differences between different cuts in a single band are correlated, and that the shot noise levels fall within the calibration uncertainties.  

Finally, we compare to the model predictions of \citet{bethermin2011}, a phenomenological model which to-date is the best at reproducing the observed number counts from 15\rmicron\ to 1.1\,mm. 
We find that the model is in very good agreement with the data at all flux-cut levels and at all but 350\rmicron, which underpredicts the data by approximately 1$\sigma$.

\subsection{Comparison to Published Measurements}
\label{sec:compare_to_data}

In Figure~\ref{fig:published_data} we plot our combined auto-frequency power spectra along with a selection of recently published CIBA measurements.  
We show the two masking extremes of our data: those in which all sources greater than 50\,mJy were masked (dark blue open circles); and those where only extended sources were masked (light blue open circles).  
We do this in order to adequately compare with the wide range of masking in the literature.       

Shown as black crosses are the SPIRE auto-frequency power spectra of $\sim 15\, \rm deg^2$ from \citet{amblard2011} in which pixels greater than 50\,mJy were masked, so that they should be compared to our dark blue circles.   
On larger scales (e.g., $k_{\theta} \lsim  0.08\invarcmin$), particularly at shorter wavelengths, their spectra suffer from overcorrection of Galactic cirrus contamination \citep[see discussion in][]{lagache2011}.  On smaller scales we find that our spectra differ by factors of $\sim 0.91\pm 0.01$, $1.09\pm 0.01$, and $1.09\pm 0.01$. 
Note that the calibration of the science demonstration phase (SDP) maps used in \citet{amblard2011} differed by 1.02, 1.05, and 0.94, at 250, 350, and 500\rmicron, respectively, but that those corrections were not applied here.  
These calibration differences, combined with the offsets resulting from the new estimate of the beam (Appendix~\ref{sec:calibration}) may partially account for this difference.  
Also note that although the error bars on their data are comparable to ours at small angular scales, correlations due to the non-Gaussian term (first term in Equation~\ref{eqn:uncertainties}) were not included when they re-binned into log bins, thus artificially deflating their errors.  

Shown as red squares at 350 and 500\rmicron\ are results from the \citet[][]{lagache2011}, which should be compared to our light blue circles.  
Note, comparisons of the two spectra must be made with caution, bearing in mind that the flux density of masked sources in \emph{Planck} is much higher (710 and 540\,mJy at 350 and 550\rmicron).  In addition, because the passbands of the two instruments are not the same, \emph{Planck} data at 857 and 545\,GHz (350 and 550\rmicron) are color corrected by multiplying them by factors of 0.99, and 1.30, respectively. 

We find possible inconsistencies between the two sets of measurements.  
On scales $k_{\theta} \lsim  0.04\invarcmin$, the \emph{Planck} spectra appear to be offset high by factors of $1.26\pm 0.06$ and $1.40\pm 0.06$ at 350 and 500\rmicron, respectively;  
while on smaller angular scales there is an apparent excess of power in the \emph{Planck} data, though it should be noted that point-by-point, the 350\rmicron\ values do agree within errors.  
Potential explanations for this discrepancy include calibration, excess Poisson, or reconstruction systematic errors in the \emph{Planck} beam.  We discuss these scenarios in more detail in \S~\ref{sec:planck}.  

\subsection{Comparison to Published Models}
\label{sec:compare_to_models}
In Figure~\ref{fig:published_models} we plot our combined auto-frequency power spectra next to a selection of published halo models.  
As in Figure~\ref{fig:published_data}, we show the two masking extremes of our data in order to adequately compare with the wide range of masking in the literature.       

\begin{figure*}[t!]
\centering
\includegraphics[width=0.99\textwidth]{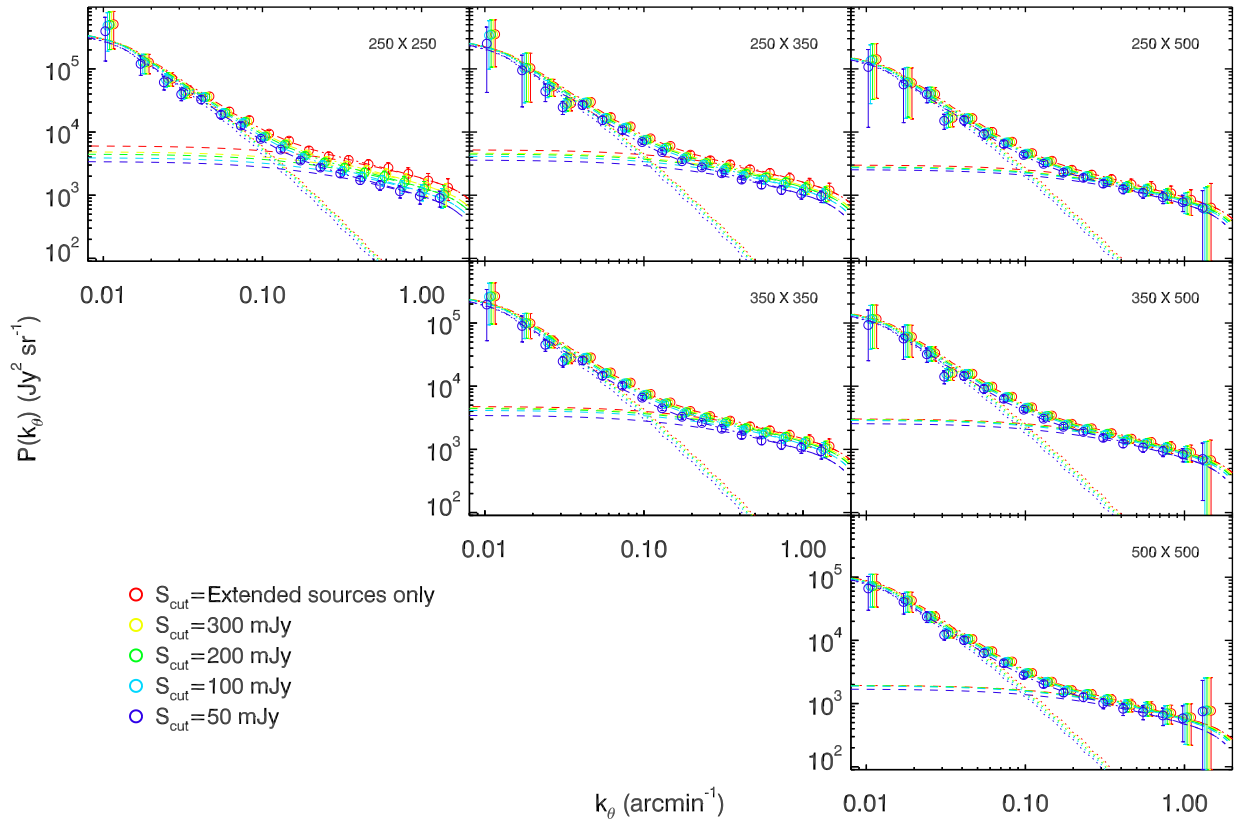}
\caption{Combined clustering spectra vs.\@ flux-cut level of masked sources.  Spectra are shifted horizontally for visual clarity.  
Overlaid are best-fit templates to the data from our Model 3 (\S~\ref{sec:intro2}). 
If resolved sources contributed solely to the Poisson noise component of the spectra then these points would lie on top of one another.  Instead, there is a clear reduction of 1-halo power with masking level at 250\rmicron, suggesting that some fraction of bright DSFGs are either close pairs, or reside, as satellites, in more massive dark matter halos. } 
\label{fig:clust_ps_vs_flux_cut2}
\end{figure*}

The \citet{viero2009} models (dotted brown lines), which were fit to BLAST data with sources greater than 500\,mJy masked, and appeared to be a good match to the \citet{lagache2011} data, here do a poor job of describing the SPIRE measurements, overestimating the power on scales greater than $\sim 40\arcsec$.  

The \citet{amblard2011} models, which assumed a masking level of 50\,mJy and were fit to each band individually, are shown as grey dot-dashed lines.   Similar to the differences in the data, the large scale power is underestimated due to the over-correction for the contribution from Galactic cirrus; while on small scales, the small differences may be due to difference in the calibration. 

The halo models from the \citet{lagache2011} are shown as red dashed lines at 350 and 500\rmicron.  They assumed a masking level consistent with the masking level of \emph{Planck} data, as do all following models.  
Their published data must be corrected for calibration by dividing them by factors of 1.14 and 1.30 at 350 and 500\rmicron, respectively \citep{planck_viii}, after which 
they agree very well with our data. 

The \citet{xia2012} halo model, shown as orange three-dot-dashed lines in Figure~\ref{fig:published_models}, was fit to \emph{Planck} and \emph{corrected} SPIRE data from \citet[][\S~5.3]{lagache2011}. 
It adopts a description for the source population from \citet{lapi2011} \citep[an update of][]{granato2004}.  It appears to be consistent with the overall amplitude of the data, but has a bump of excess emission at around $k_{\theta} \sim 0.1$--0.03 that the data do not show.  
This evolutionary model assumes that the steep part of the source counts at submillimeter wavelengths is dominated by massive, proto-spheroidal galaxies in the process of forming most of their stars.   
The model also includes small contributions from late-type and starburst galaxies.   
Notable in this model is that the redshift distribution of the emission peaks at slightly higher redshifts, broadly around $z\sim 1.7$--2.2, increasing with increasing wavelength; this is in distinction to other models which are strongly peaked at $z\sim 1$.  
It was fit to data from \emph{Herschel}/SPIRE \citep{amblard2011}, \emph{Planck}/HFI \citet{lagache2011}, SPT \citep{shirokoff2011} and ACT \citep{dunkley2011}, and as it is physically based, it is much more constrained than the phenomenological models used by of e.g., \citet{viero2009} or \citet{shang2012}.  
That two populations are represented is evidenced by the clear feature at $\sim 0.2\invarcmin$, a feature which is not apparent in the data, suggesting an overestimate of the contribution of late-type galaxies to the total spectrum.  

Lastly, shown as green dashed lines is the \citet{shang2012} model, whose main feature is to implement a luminosity-mass ($L-M$) relation, such that more massive halos host more luminous sources.  
Though the model was fit primarily to \emph{Planck} data, it appears to fit our spectra at 500\rmicron\ quite well.   The fit is less good at 250 and 350\rmicron; though the shape is in good agreement, the curves are high by $\sim 30\%$.  Despite this success, the model has some points of concern.  In particular, it underpredicts the contribution of lower redshift sources to the CIB  \citep[e.g., they are significantly below the lower limits measured from stacking 24\rmicron\ selected sources;][]{jauzac2011}, and also find an uncharacteristically large $\beta$ (discussed further in \S~\ref{sec:hm_discussion}).  

Finally, we highlight the feature that appears in the data at $k_{\theta} \sim 0.03$--$0.04\invarcmin$, particularly in all auto- and cross-frequency spectra, but predominantly at 500\rmicron.   A similar feature was visible in the SPIRE data of \citet{amblard2011} as well.  It is not clear if this is a real feature in the sky, which would be unexpected and is unlikely, or due to noise present in the data.  Similar features do not appear in the transfer function, or in the simulations of the pipeline which tested for potential biases.  
\begin{figure}[t!]
\centering
\vspace{-2mm}
\hspace{-4mm}
\includegraphics[width=0.49\textwidth]{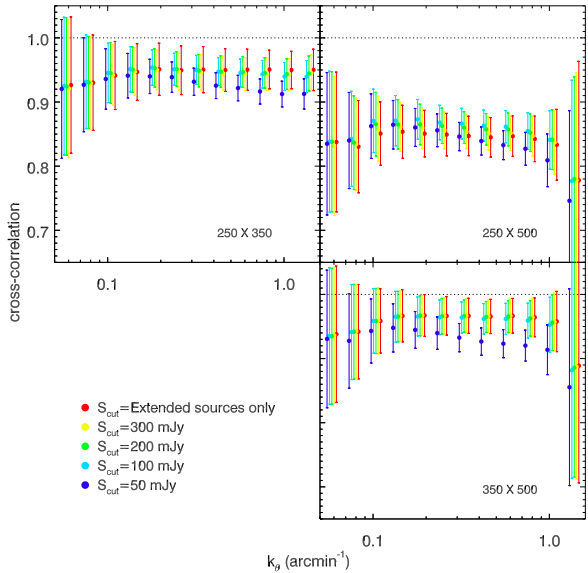}
\caption{Cross-correlation power spectra.  
Spectra are shifted horizontally for visual clarity. 
For two identical maps, or two maps in which all the sources are at the same redshift, this measurement would have unit amplitude, which is represented by a dotted line.   We find that for unmasked maps, the cross-correlation is approximately $0.95\pm0.04$, $0.86\pm0.04$, and $0.95\pm0.03$, for $250\times 350$, $250\times 500$, and $350\times 500$, respectively.  For maps with sources greater than 50\,mJy masked, this cross-correlation is reduced, and appears to weaken further with decreasing angular scale. } 
\label{fig:cc}
\end{figure}
\subsubsection{Clustered Galaxy Power Spectra}
\label{sec:clustering}
Ultimately, we are interested in the power spectra of clustered DSFGs, which we estimate by removing the Poisson noise from the cirrus-subtracted, combined power spectra of Figure~\ref{fig:ps_vs_flux_cut}.  Results are shown in Figure~\ref{fig:clust_ps_vs_flux_cut2}.    
The fraction anisotropy, calculated as
\begin{equation} 
\delta I/I = \sqrt{2\pi k_{\theta}^2 P(k_{\theta})}/I_{\rm CIB}, 
\end{equation}
where $I_{\rm CIB}$ is the overall amplitude amplitude of the CIB, measured to be $0.71\pm0.17$, $0.59\pm 0.14$, and $0.38\pm 0.10\, \rm MJy sr^{-1}$ at 250, 350, and 500\rmicron, respectively \citep[e.g.,][]{lagache2000,marsden2009}, and $k_{\theta}$ is converted to sr.  We find $\delta I/I =14\pm 4\%$, consistent with findings from \citet{viero2009}.  

We fit a simple power law to the clustering spectra over the range $k_{\theta} = 0.01$--$1.4\invarcmin$, finding a mild change in the slope with changes in the masking level.  Specifically, from most ($S_{\rm cut} > 50\, \rm mJy$) to least aggressively masked (only extended sources), we find slopes of $-1.60\pm 0.05$ to  $-1.50\pm 0.07$ at 250\rmicron; $-1.52\pm 0.05$ to $-1.36\pm 0.06$ at 350\rmicron;  and $-1.52\pm 0.06$ to $-1.47\pm 0.06$ at 500\rmicron.  The $\chi^2$ of these fits for 15 degrees of freedom (d.o.f.) are 7, 10 and 3 (reduced $\chi^2 \sim 0.5$, 0.7, and 0.3) at 250, 350, and 500\rmicron, respectively.  

The next notable feature is the reduction of 1-halo power with flux cut of masked sources, particularly at 250\rmicron, which are shown as dashed lines in Figure~\ref{fig:clust_ps_vs_flux_cut2}, whereas the 2-halo power, shown as dotted lines, remain relatively unchanged.  
We demonstrate in \S~\ref{sec:one_halo} how this result can be interpreted as more luminous sources residing in more massive halos, motivating the use of a model later in the paper in which a luminosity-mass relationship is invoked \citep[e.g.,][]{sheth2005,skibba2006, shang2012}.  We show that attempting to account for the reduction of power entirely with the Poisson term leads to significant tension in the fit.  Though the reduction of power is much less significant at 500\rmicron, we remind the reader that there are far fewer sources at each given flux-cut level at 500\rmicron\ than at 250 or 350\rmicron\ (see Table~\ref{tab:masked_sources}).  
The capability to mask fainter sources reliably at 500\rmicron\ would require either maps with higher angular resolution (i.e., less confusion  noise) or a way to probe deeper into the confusion using ancillary data \citep[e.g., XID;][]{roseboom2010}.  
\subsection{Cross-Correlation Power Spectra}  
\label{sec:bandband}
The cross-correlation power spectrum is defined as 
\begin{equation}
C_{\rm A\times B} = \frac{P_{k_{\theta}}^{\rm A\times B}}{\sqrt{P_{k_{\theta}}^{\rm A}\cdot P_{k_{\theta}}^{\rm B}}},
\label{eqn:cc}
\end{equation}
i.e., the ratio of the cross-frequency power spectra to the geometric mean of the two auto-frequency power spectra.  
Identical maps would thus have a cross-correlation of unit amplitude, as would maps containing sources located at identical redshifts and with identical colors.  
Departures from unity would be an indication that sources are not all at the same redshift, or that their colors (or average temperatures) are variable, and the strength and shape of the cross-correlation signal would depend on the level of correlation between maps.   
Consequently, the cross-correlation provides strong constraints for source population models.  

We show the cross-correlations as functions of the flux cut of masked sources in Figure~\ref{fig:cc}.  The measurement becomes very uncertain at angular scales $k_{\theta} \le 0.1\invarcmin$.  At larger $k_{\theta}$, with the exception of the 50\,mJy cut, we find similar levels of correlation for all levels of masking, which can be approximated as horizontal lines at $0.95\pm0.04$, $0.86\pm0.04$, and $0.95\pm0.03$ for $\rm A\times B = 250\times 350$, $250\times 500$, and $350\times 500$, respectively.   
Cross-correlations of maps with sources masked at 50\,mJy appear to be less correlated, with hints of a reduction in the correlation with increasing $k_{\theta}$, which as first predicted by \citet{knox2001}, would be an indication that longer wavelengths are more sensitive to higher-$z$.     

These results compare favorably with the \citet{lagache2011} measurements of the cross-correlation, who found  0.89 and 0.91 for two different fields at $350\times 550$\rmicron, and are also consistent within errors with the cross-correlations measured by \citet{hajian2012}. 

\section{Halo model interpretation of CIB anisotropy measurements}
\label{sec:intro2}
The angular power spectrum of intensity fluctuations, $P_{\nu\nu^{\prime}}(k_{\theta})$, is obtained using Limber's approximation \citep{limber1953}, which is valid on small angular scales ($2\pi k_{\theta}\gsim 10$).  The projection of the flux-weighted spatial power spectrum is  
\begin{equation}
P_{\nu\nu^{\prime}}(k_{\theta}) = \int \frac{dz}{\chi^2} \frac{dz}{d\chi} P_{\nu\nu^{\prime}} \left(k=\frac{2\pi k_{\theta}}{\chi(z)}, z\right) \frac{dS_{\nu}}{dz} \frac{dS_{\nu^{\prime}}}{dz},
\end{equation}
where $\chi (z)$ is the comoving distance to redshift $z$, and $dS_{\nu}/dz$ is the redshift distribution of the cumulative flux.  
The redshift range used here is $0<z<4$, from which most of the CIB is emitted \citep[e.g.,][]{bethermin2012c}.   
For sources with flux densities $S_{\nu} \leq S_{\rm cut}$ 
\begin{equation}
\frac{dS_{\nu}}{dz} (z) = \int_0^{S_{\rm cut}} S_{\nu} \frac{d^2 N}{dS_{\nu} dz} (S_{\nu}, z) dS_{\nu},
\end{equation}
and the differential number counts are related to the epoch-dependent comoving luminosity function, $dn/dL(L, z)$, through
\begin{equation}
\frac{d N}{dS_{\nu}} = \int dz \chi^2 \frac{d\chi}{dz} dn/dL[L(S_{\nu}, z), z] .
\end{equation}

\subsection{Halo Model Formalism}
\label{sec:formalism}

The power spectrum of CIB anisotropies in the halo model formalism is written as the sum of three terms: the linear (or 2-halo) term, which accounts for pairs of galaxies in separate halos and dominates the spectrum on large scales;  the non-linear (or 1-halo) term, which describes pairs of galaxies residing in the same halo and is the dominant term on small scales;  and the Poisson (or shot) noise term: 
\begin{equation}
P_{\nu\nu^{\prime}} (k, z) = P_{\nu\nu^{\prime}}^{\rm 1h} (k, z) + P_{\nu\nu^{\prime}}^{\rm 2h} (k, z) + P_{\nu\nu^{\prime}}^{\rm shot} (k, z).
\label{eqn:ps4}
\end{equation}
Common to most halo models is that a distinction is made between central and satellite galaxies, with $N^{\rm gal}=N^{\rm cen}+N^{\rm sat}$.  
All halos above a minimum mass $M_{\rm min}$ host a galaxy at their center, 
\begin{equation}
N^{\rm cen}(M) = 
\begin{cases} 
0  & M < M_{\rm min}, \\
1  & M \ge M_{\rm min}, 
\end{cases}
\label{eqn:ncen}
\end{equation}
while any additional galaxies in the same halo would be designated as satellites which trace the dark matter density profile \citep[e.g.,][]{zheng2005}.  Halos host satellites when their mass exceeds the pivot mass $M_1$ (also known as $M_{\rm sat}$ in the literature), and the number of satellites is an exponential function of halo mass:
\begin{equation}
N^{\rm sat}(M)=\left( \frac{M}{M_1} \right)^{\alpha}.
\label{eqn:hon}
\end{equation}

In earlier halo models, galaxies were assumed to contribute equally to the emissivity density \citep[e.g.,][]{viero2009,amblard2011,lagache2011}.  Therefore, assuming that the CIB originates from galaxies, spatial variations in the specific emission coefficient $j_{\nu}$ directly trace fluctuations in the galaxy number density
\begin{equation}
\delta j_{\nu}/\bar{j}_{\nu} = \delta n^{\rm gal} / \bar{n}^{\rm gal}.
\label{eqn:j}
\end{equation}
The linear, 2-halo term dominates on large scales and is given by the clustering of galaxies in separate dark matter halos:
\begin{equation}
P_{\nu\nu^{\prime}}^{\rm 2h} (k, z) =  \frac{1}{\bar{j}_{\nu}\bar{j}_{\nu}^{\prime}}  P_{\rm lin}(k, z) D_{\nu}(k, z) D_{\nu}^{\prime}(k, z),
\label{eqn:2hA}
\end{equation}
with 
\begin{eqnarray}
D_{\nu} (k, z) & = & \int dM \frac{dN}{dM} (z) b(M, z) u_{\rm gal}(k, z, M) \nonumber \\
&&\times [ N_{\nu}^{\rm  cen}(M, z) + N_{\nu}^{\rm sat}(M, z)],
\label{eqn:2hB}
\end{eqnarray} 
where $P_{\rm lin} (k, z)$ is the linear dark matter power spectrum, $b(M, z)$ is the linear large-scale bias, and $u_{\rm gal}(k, z,M)$ is 
the normalized Fourier transform of the galaxy density distribution within a halo, which is assumed to equal the dark matter density profile, i.e.,  $u_{\rm gal}(k, z, M) =  u_{\rm DM}(k, z, M)$.  
The non-linear, 1-halo term dominates on small scales, and is written as
\begin{eqnarray}
P_{\nu\nu^{\prime}}^{\rm 1h} (k, z) & =  &\frac{1}{ \bar{j}_{\nu} {\bar{j}_{\nu}^{\prime}}} \int dM \frac{dN}{dM} (z)\nonumber\\
& & \times \Big{[}N_{\nu}^{\rm cen}(M,z) N_{\nu^{\prime}}^{\rm sat}(M,z) u_{\rm gal}(k, z , M) +\nonumber \\
& & N_{\nu^{\prime}}^{\rm cen}(M,z) N_{\nu}^{\rm sat}(M,z) u_{\rm gal}(k, z, M) +\nonumber \\
& & N_{\nu^{\prime}}^{\rm sat}(M, z) N_{\nu}^{\rm sat}(M, z) u^2_{\rm gal}(k, z, M) \Big{]}, 
\label{eqn:1h}
\end{eqnarray}
\citep{cooray2002}, where $dN/dM$ is the halo mass function.   

However, conceptually it is wrong to assume that galaxies of different luminosities have equal weight in contributing to the power spectrum of the intensity fluctuations. A consequence of this assumption is that the excess signal on small angular scales from galaxies in massive halos can only be reproduced by having more satellite galaxies,   
leading to previous estimates of $\alpha$ which exceed predictions for sub-halo indices from semi-analytic models, \citep[$\alpha \le 1$: e.g.,][]{gao2004,hansen2009} and a significant overabundance of satellites. 
For example, previous halo model fits to SPIRE data from \citet{amblard2011} found $\alpha \sim 1.7$ at 250\rmicron, and $\sim 1.8$ at 350 and 500\rmicron, albeit with large errors.  

A way to overcome this excess satellite problem, while still producing enough 1-halo power, is to have a model with fewer but more luminous satellites and weight galaxies by their luminosities.  
One such model is that of \citet{shang2012}, which invokes a luminosity-mass ($L-M$) relation to tie the emissivity from galaxies to their host halo masses \citep[also see e.g.,][]{yang2003,vale2004}.  
The advantage of this model is that in principle one can predict the abundance as well as the clustering of galaxies observed at different frequency bands simultaneously, while in previous halo models it was impossible to predict the power spectrum across different frequency bands at the same time. As we will now show, this new formalism is similar to previous ones 
up until the number of central and satellite galaxies are substituted for luminosity weighted quantities. 

\subsection{Luminosity Weighted Halo Model}
\label{sec:lm_model}
Hereafter, we follow the formalism of \citet{shang2012}, with some modifications.  Novel to our implementation is the simultaneous fitting to the power spectra in all three bands, and to the number counts of sources above about $0.1\, \rm mJy$ \citep{glenn2010}.   
In this implementation of the halo model, the mean comoving specific emission coefficient  is
\begin{equation} 
\bar{j}_{\nu} (z) = \int dL \frac{dn}{dL}(L,z) \frac{L_{\nu}[(1+z)\nu]}{4\pi},
\label{meanspecificemission}
\end{equation}
where $L$ is the luminosity and $dn/dL$ is the luminosity function of  DSFGs.  What this means is that, unlike the earlier models, which assume that all galaxies contribute equally to the emissivity density (i.e., have the same luminosity), here the emissivities of galaxies in a given halo, by way of their luminosities,  depend on the redshift, halo mass and frequency:
\begin{equation}
L_{\nu}[(1+z)\nu] = L_0 (1+z)^{\eta} \Sigma(M)\Theta \left[ \left(1+z \right) \nu \right],
\label{eqn:lm_eqn}
\end{equation}
where $L_0$ is the overall normalization factor, $\eta$ describes the redshift evolution, $\Sigma(M)$ describes the relation between infrared luminosity and halo mass (the $L-M$ relation), and $\Theta(v)$ describes the shape of the infrared SED.     Note that here \emph{M} represents both the mass of the main halo and the infall mass of the subhalo.  

Furthermore, the effective, luminosity-weighted, number of central and satellite galaxies is 
\begin{equation}
f_{\nu}^{\rm cen} = N^{\rm cen}  \frac{L_{(1+z)\nu}^{\rm cen}(M, z)}{4\pi},
\end{equation}
\begin{equation}
f_{\nu}^{\rm sat} = \int dm \frac{dn}{dm} (M, z) \frac{L_{(1+z)\nu}^{\rm sat}(m, z)}{4\pi},   
\end{equation}
where $dn/dm (M,z)$ is the subhalo mass function of the main halo whose mass is $M$.

The terms in Equations~\ref{eqn:2hB} and \ref{eqn:1h} remain the same, except the number of central and satellite galaxies are substituted with their luminosity weighted counterparts: such that $D_{\nu}$ in the 2-halo term (Equation~\ref{eqn:2hB}) becomes
\begin{eqnarray}
D_{\nu} (k, z)  & =  & \int dM \frac{dN}{dM} (z) b(M, z) u_{\rm gal}(k, z, M) \nonumber\\
&& \times [f_{\nu}^{\rm cen}(M, z) +  f_{\nu}^{\rm sat}(M, z) ], 
\label{eqn:2hC}
\end{eqnarray}
and the 1-halo term (Equation~\ref{eqn:1h}) becomes
\begin{eqnarray}
P_{\nu\nu^{\prime}}^{\rm 1h} (k, z) & =  &\frac{1}{\bar{j}_{\nu}  \bar{j}_{\nu^{\prime}}}\int dM \frac{dN}{dM} (z) \nonumber\\ 
& & \times \Big{[} f_{\nu^{\prime}}^{\rm cen}(M, z) f_{\nu}^{\rm sat}(M, z) u_{\rm gal}(k, z, M) + \nonumber \\
& &  f_{\nu}^{\rm cen}(M, z) f_{\nu^{\prime}}^{\rm sat}(M, z) u_{\rm gal}(k, z, M) +\nonumber \\
& &f_{\nu^{\prime}}^{\rm sat}(M, z) f_{\nu}^{\rm sat}(M, z)u^2_{\rm gal}(k, z, M) \Big{]}.
\label{eqn:1hC}
\end{eqnarray}

We define halos here as overdense regions whose mean density is 200 
times the mean background density of the Universe according to the spherical collapse model, and we adopt the density profile of \citet*[][ or NFW]{nfw1997} with the concentration parameter of \citet{bullock2001}, and the fitting function of \citet{tinker2008} for the halo mass function and its associated prescription for the halo bias \citep*{tinker2010b}. For the subhalo mass function, we use the fitting function of \citet*{tinker2010c}.  
We will now describe each of the terms in Equation~\ref{eqn:lm_eqn} in more detail.  Note that using instead the concentration parameter of \citet{duffy2008} leads to very little change in the final best-fit parameters.  
\begin{figure*}[t!]
\centering
\hspace{-10mm}
\includegraphics[width=1.05\textwidth]{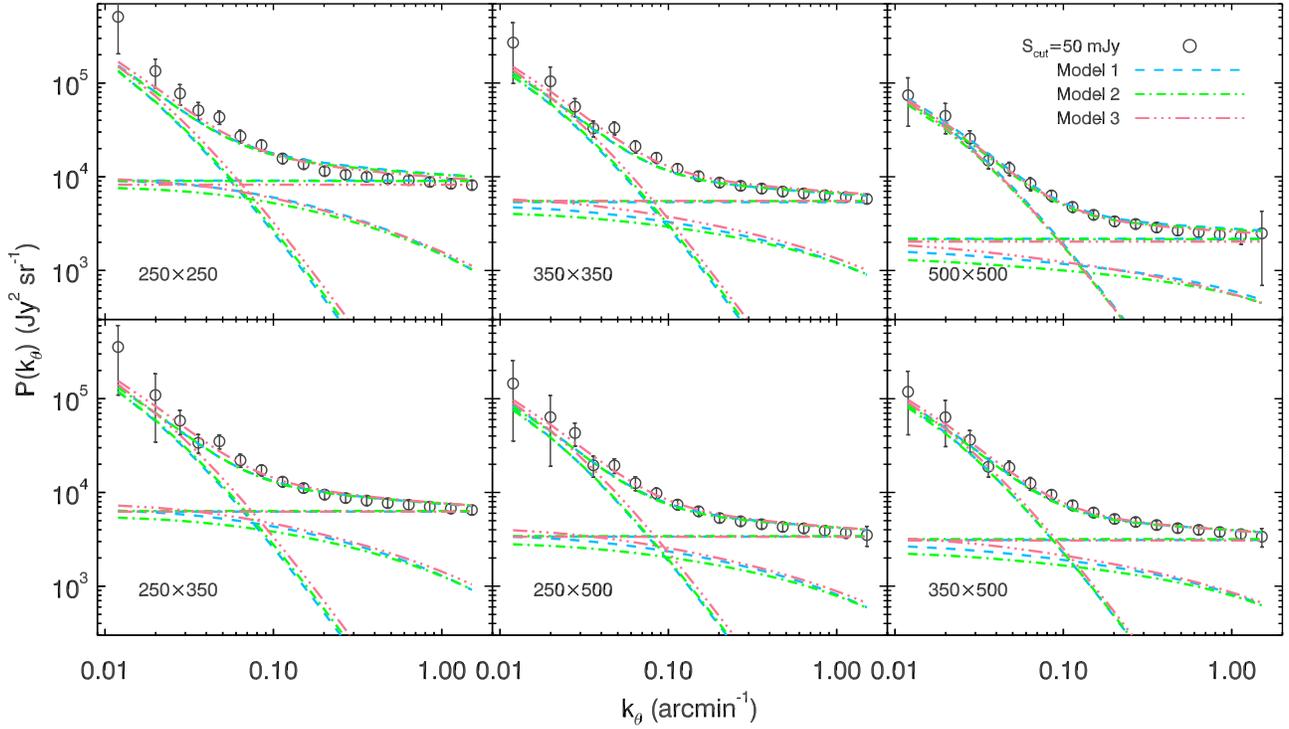}

\vspace{-5mm}
\caption{
Best-fit halo models fit simultaneously to cirrus-subtracted, combined auto- and cross-frequency power spectra (with sources greater than $S_{\rm cut} = 50\, \rm mJy$ masked) and to the P(D) number counts from \citet{glenn2010}. 
Spectra are fit with three terms: Poisson (horizontal lines); 2-halo (steep lines dominant at low $k_{\theta}$); and 1-halo (less steep and contributing at all $k_{\theta}$).  The sum of the three terms are also plotted.
} 
\label{fig:halo_model_fit}
\end{figure*}
\begin{figure}[t!]
\centering
\hspace{-10mm}
\includegraphics[width=0.525\textwidth]{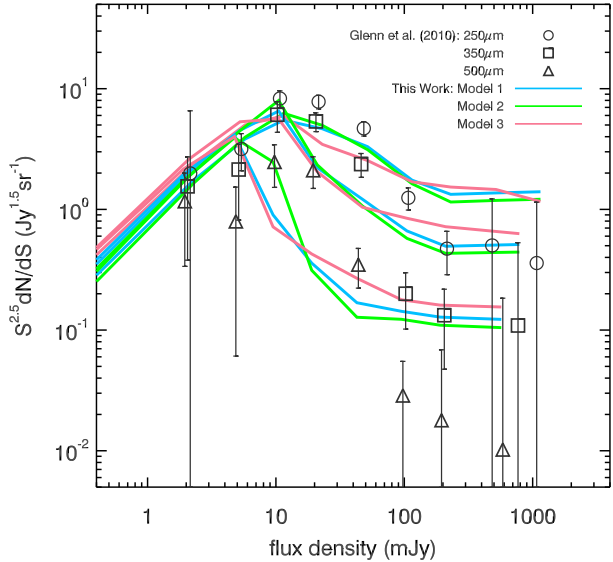}
\caption{
Euclidean normalized differential number counts from \citet[][shown as circles, squares and triangles at 250, 350, and 500\rmicron, respectively]{glenn2010} along with best-fit curves from the three models.  These curves were found by simultaneously fitting to spectra (shown in Figure~\ref{fig:halo_model_fit}) as well as to these counts.  
This fit reveals a possible tension between the modeling of the counts and the modeling of the clustering terms.  
} 
\label{fig:counts}
\end{figure}
\subsubsection{$(1+z)^{\eta}$: The Luminosity Evolution }
\label{sec:hm_z}
The luminosity evolution in this model is motivated by the known increase of specific star-formation rate (sSFR) with redshift \citep[e.g.,][]{elbaz2007, oliver2010b, karim2011,noeske2007, sargent2012, wang2012}, and the fact that SFRs and infrared luminosities are correlated for DSFGs \citep[e.g.,][]{kennicutt1998}.  
The exact form of the evolution is still not clear: though measurements find a rapid rise followed by plateau at $z\gsim 2$  \citep[e.g.,][]{stark2009, gonzalez2010},  semi-analytic models have difficultly reproducing  observations without invoking a number of ad-hoc modifications to the standard physical recipes \citep*[e.g.][]{weinmann2011}.  
Yet, without a convincing alternative, we proceed motivated by observations, letting $\eta$ be a free parameter over $0<z<2$ and set $\eta=0$ at $z\ge 2$. 

\subsubsection{$\Sigma(M)$: The $L-M$ relation}
\label{sec:hm_lm}
Observationally it is clear that some halos are more efficient than others at hosting star formation \citep[e.g.,][]{bethermin2012a, wang2012}, and that the halo mass of most efficient star formation evolves with redshift \citep[i.e., downsizing; e.g.,][]{cowie1996, bundy2006}.  It is also clear that star formation in halos is suppressed by several plausible mechanisms at the high mass \citep[e.g., accreting black holes][]{birnboim2003, keres2005} and low mass  \citep*[e.g., feedback from supernovae, photoionization heating][]{dekel1986,thoul1996} extremes. 
Thus, following \citet{shang2012}, we assume that the $L-M$ relation, $\Sigma(m)$, can be parameterized by a simple log-normal distribution
\begin{equation}
\Sigma(m) = m \frac{1}{\sqrt{2\pi \sigma^2_{L/m}}} \exp\left[-\frac{(\log m - \log M_{\rm peak})^2}{2\sigma^2_{L/m}}\right],
\end{equation}
where $M_{\rm peak}$ describes the peak of the specific IR emissivity per unit mass, and $\sigma^2_{L/m}$ describes the range of halo masses in which galaxies producing IR emission reside.  
The minimum halo mass to host a galaxy, $M_{\rm min}$, is left as a free parameter, but we place a lower limit on it such that $L=0$ at $M<M_{\rm min}$.

Note that we have implicitly assumed that the shape of the relation between halo mass and infrared luminosity is redshift-independent  and identical for both central and satellite galaxies. 
Equation~\ref{meanspecificemission} can then be recast as 
\begin{eqnarray}
\bar{j}_{\nu} (z) &=& \int dM \frac{dN}{dM}(z) \frac{1}{4\pi} \Big{[} N^{\rm cen}L_{(1+z)\nu}^{\rm cen} \nonumber \\
& &+ \int dm \frac{dn}{dm} (M, z) L_{(1+z)\nu}^{\rm sat} \Big{]},
\end{eqnarray}
where $m$ is the subhalo mass at the time of accretion \citep[e.g.,][]{wetzel2010,shang2012} and $dn/dm$ is the subhalo mass function in a host halo of mass $M$ at a given redshift.  

\subsubsection{$\Theta(\nu)$: The Model SED}
\label{sec:hm_sed}
Following \citet{hall2010} and \citet{shang2012}, we begin with the simplest model SED, a single modified blackbody:  
\begin{equation}
\Theta(\nu) \propto \nu^{\beta} B(\nu,T_{\rm d}),
\end{equation}
where $B(\nu,T_{\rm d})$ is the blackbody spectrum (or Planck function), with effective dust temperature $T_{\rm d}$,
and $\beta$ is the emissivity index.  
Both $T_{\rm d}$ and $\beta$ are free variables with no redshift evolution. 
We refer to this as Model 1. 

Next, in attempting to address the growing observational evidence for evolving temperature with redshift \citep[e.g.,][]{pascale2009, amblard2010, viero2013}, we introduce an additional parameter, $T_z$, such that $T_{\rm dust} \propto (1+z)^{T_z}$.  
Note the $\beta$ parameter remains a free variable without redshift evolution in this model, which  
we refer to as Model 2.  

Lastly,  motivated by the findings of e.g., \citet{dunne2001} or \citet{elbaz2011} --- who found that a typical dusty star-forming galaxy spectrum is better fit by a linear combination of two SEDs: i) those from \emph{hotter} ($T_{\rm d, warm}\sim 50\, \rm K$), star-forming regions, and ii) those from \emph{colder} ($T_{\rm d, cold}\sim 20\, \rm K$) regions of diffuse ISM --- we adopt a two-component SED.  
The ratio of the masses in the two components is defined as $\xi = $log($N_c/N_w$), and is independent of  redshift.  
Here $\beta$ is redshift independent and fixed to equal 2.  We refer to this model as Model 3.  

\subsubsection{Markov Chain Monte Carlo}
\label{sec:mcmc}

We make use of Markov Chain Monte Carlo (MCMC) methods to derive the posterior probability distributions for all parameters by fitting to the P(D) number counts \citep{glenn2010} and all auto- and cross-frequency power spectra at 250, 350 and 500\rmicron\ simultaneously.  
Further, fits are performed simultaneously to spectra with sources above 50 and 300\,mJy masked. 
Finally, the absolute CIB as measured by \citep[$10.4\pm 2.3$, $6.5\pm 1.6$, $2.6\pm0.6\, \rm nW\,m^{-2}\,sr^{-1}$]{lagache2000} provide additional constraints.       

Models 1, 2 and 3 consist of seven, eight and nine free parameters, respectively.  
All models include:
one for the low mass halo cutoff ($M_{\rm min}$); 
two for the $L-M$ relation ($M_{\rm peak}$ and $\sigma_{L/m}$);  
one for the redshift evolution ($\eta$); 
and an overall normalization ($L_0)$.    
Models 1 and 2 have two parameters for the SED ($T_{\rm d}$ and $\beta$); while   
Model 3 has 2 SED temperature parameters ($T_{\rm cold}$ and $T_{\rm warm}$) and   
a parameter describing the ratio of the masses in the two components ($\xi$). 
Lastly, Models 2 and 3 have parameters to describe the evolution of the dust temperature with redshift ($T_z$).


\subsection{Halo Model Results}
\label{sec:hm_results}

\begin{table*}[t!]
\footnotesize
\centering
\caption{Model 1: Best-fit parameters and corresponding correlation matrix. }
\begin{tabular}{l|ccccccccc}
Parameter & log($M_{\rm min}$) & log($M_{\rm peak}$) & $T$ & $\beta$ & $\sigma^2_{L/m}$ & log($L_0$) & $\eta$ \\ 
\hline
log($M_{\rm min}$) & $9.8 \pm 0.5 $  & 0.16 & -0.09 & 0.06 & -0.15 & -0.23 & 0.23 \\ 
log($M_{\rm peak}$) &  ---  & $12.2 \pm 0.5 $  & -0.11 & 0.16 & -1.00 & -0.42 & 0.49 \\ 
$T$ &  ---  &  ---  & $23.1 \pm 1.3 $  & -0.97 & 0.10 & 0.74 & -0.29 \\ 
$\beta$ &  ---  &  ---  &  ---  & $1.4 \pm 0.1 $  & -0.16 & -0.62 & 0.25 \\ 
$\sigma^2_{L/m}$ &  ---  &  ---  &  ---  &  ---  & $0.4 \pm 0.0 $  & 0.41 & -0.50 \\ 
log($L_0$) &  ---  &  ---  &  ---  &  ---  &  ---  & $-1.7 \pm 0.1 $  & -0.74 \\ 
$\eta$ &  ---  &  ---  &  ---  &  ---  &  ---  &  ---  & $2.0 \pm 0.1 $  \\ 
\end{tabular}
\tablecomments{Best-fit parameters and correlations between them in Model 1.  Off-diagonal values of $+1$ or $-1$ mean that the parameters are highly correlated or anti-correlated, respectively, while values near 0 mean that they are independent of one another.}
\label{tab:correlations1}
\end{table*}
\begin{table*}[t!]
\footnotesize
\centering
\caption{Model 2: Best-fit parameters and corresponding correlation matrix. }
\begin{tabular}{l|ccccccccc}
Parameter & log($M_{\rm min}$ & log($M_{\rm peak}$) & $T$ & $T_z$ & $\beta$ & $\sigma^2_{L/m}$ & log($L_0$) & $\eta$ \\ 
\hline
log($M_{\rm min}$ & $10.1 \pm 0.5 $  & -0.02 & 0.20 & -0.27 & -0.10 & 0.02 & 0.26 & -0.25 \\ 
log($M_{\rm peak}$) &  ---  & $12.3 \pm 0.5 $  & 0.21 & -0.01 & -0.23 & -1.00 & -0.18 & 0.20 \\ 
$T$ &  ---  &  ---  & $20.7 \pm 1.2 $  & -0.66 & -0.92 & -0.21 & 0.76 & -0.56 \\ 
$T_z$ &  ---  &  ---  &  ---  & $0.2 \pm 0.0 $  & 0.38 & 0.01 & -0.81 & 0.89 \\ 
$\beta$ &  ---  &  ---  &  ---  &  ---  & $1.6 \pm 0.1 $  & 0.23 & -0.53 & 0.31 \\ 
$\sigma^2_{L/m}$ &  ---  &  ---  &  ---  &  ---  &  ---  & $0.3 \pm 0.0 $  & 0.18 & -0.20 \\ 
log($L_0$) &  ---  &  ---  &  ---  &  ---  &  ---  &  ---  & $-1.8 \pm 0.1 $  & -0.90 \\ 
$\eta$ &  ---  &  ---  &  ---  &  ---  &  ---  &  ---  &  ---  & $2.4 \pm 0.1 $  \\ 
\end{tabular}
\tablecomments{Best-fit parameters and correlations between them in Model 2.  Off-diagonal values of $+1$ or $-1$ mean that the parameters are highly correlated or anti-correlated, respectively, while values near 0 mean that they are independent of one another.}
\label{tab:correlations2}
\end{table*}
\begin{table*}[t!]
\footnotesize
\centering
\caption{Model 3: Best-fit parameters and corresponding correlation matrix. }
\begin{tabular}{l|ccccccccc}
Parameter & log($M_{\rm min}$) & log($M_{\rm peak}$) & $T_{\rm warm}$ & $T_{\rm cold}$ & $\xi$ & $T_z$ & $\sigma^2_{L/m}$ & log($L_0$) & $\eta$ \\ 
\hline
log($M_{\rm min}$) & $10.1 \pm 0.6 $  & -0.39 & 0.40 & 0.40 & -0.23 & -0.43 & 0.39 & 0.42 & -0.41 \\ 
log($M_{\rm peak}$) &  ---  & $12.1 \pm 0.5 $  & -0.75 & -0.91 & 0.02 & 0.89 & -1.00 & -0.79 & 0.90 \\ 
$T_{\rm warm}$ &  ---  &  ---  & $26.6 \pm 2.8 $  & 0.80 & -0.05 & -0.90 & 0.75 & 0.94 & -0.90 \\ 
$T_{\rm cold}$ &  ---  &  ---  &  ---  & $14.2 \pm 1.0 $  & 0.05 & -0.93 & 0.90 & 0.80 & -0.92 \\ 
$\xi$ &  ---  &  ---  &  ---  &  ---  & $1.8 \pm 0.1 $  & 0.18 & -0.02 & -0.27 & 0.19 \\ 
$T_z$ &  ---  &  ---  &  ---  &  ---  &  ---  & $0.4 \pm 0.1 $  & -0.88 & -0.95 & 0.99 \\ 
$\sigma^2_{L/m}$ &  ---  &  ---  &  ---  &  ---  &  ---  &  ---  & $0.4 \pm 0.0 $  & 0.79 & -0.89 \\ 
log($L_0$) &  ---  &  ---  &  ---  &  ---  &  ---  &  ---  &  ---  & $-1.9 \pm 0.1 $  & -0.96 \\ 
$\eta$ &  ---  &  ---  &  ---  &  ---  &  ---  &  ---  &  ---  &  ---  & $2.7 \pm 0.2 $  \\ 
\end{tabular}
\tablecomments{Best-fit parameters and correlations between them in Model 3. Here $\xi$ is the ratio of the masses of the cold and warm components.  Off-diagonal values of $+1$ or $-1$ mean that the parameters are highly correlated or anti-correlated, respectively, while values near 0 mean that they are independent of one another.}
\label{tab:correlations3}
\end{table*}

\begin{figure}[t!]
\centering
\hspace{-10mm}
\includegraphics[width=0.525\textwidth]{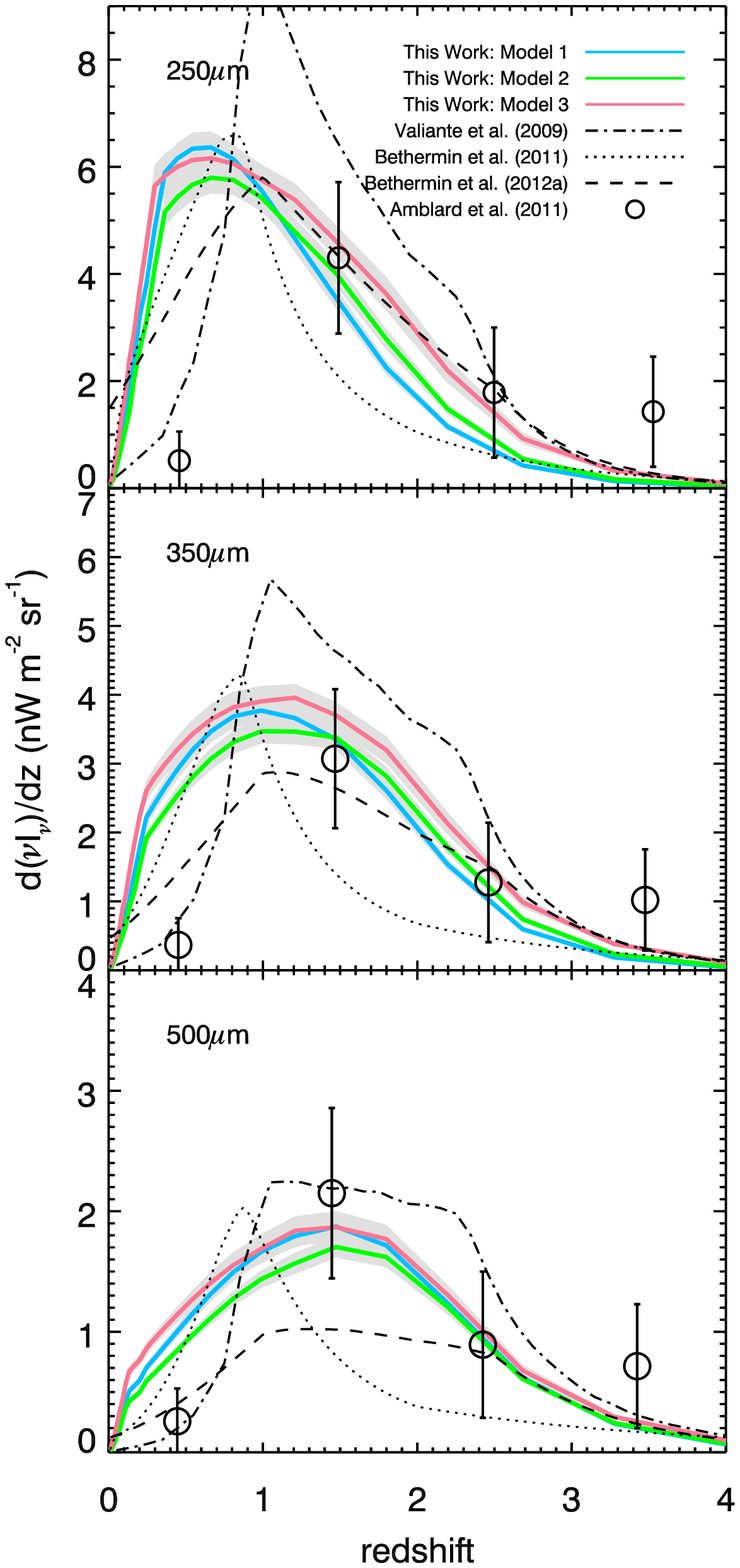}
\caption{Redshift distribution of emission for Models 1, 2, and 3. 
The \citet{amblard2011} points were estimated in a similar manner, finding the best-fit solution to their power spectra with a halo model.  The model predictions of  \citet{bethermin2011} and \citet{bethermin2012c} are shown as dotted and dashed lines, respectively.  Also shown is are model predictions from \citet{valiante2009}.  The models unanimously anticipate more of a contribution from $z \gsim 2$ than our best-fit finds.  
} 
\label{fig:dsdz}
\end{figure}
\begin{figure}[t!]
\centering
\hspace{-10mm}
\includegraphics[width=0.525\textwidth]{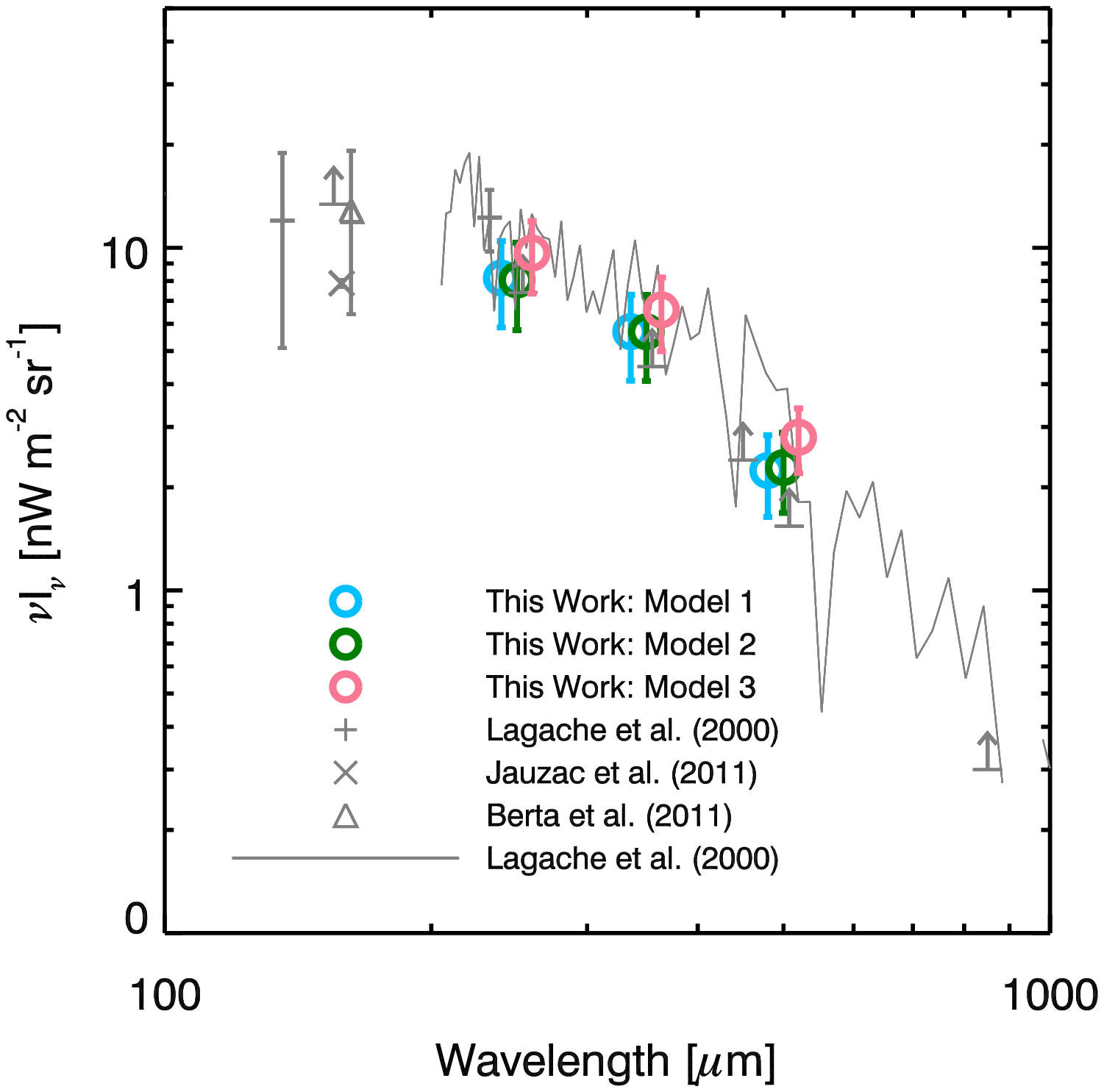}
\caption{The integrated CIB from Models 1, 2, and 3 are plotted as colored circles, 
along with a selection of measurements of the total CIB in grey, including: 
160\rmicron\ with \emph{Herschel}/PACS from \citet[ex]{jauzac2011}  
and \citet[triangle]{berta2011};   
 140 and 240\rmicron\ with WHAM from \citet[crosses]{lagache2000}; 
and $\sim 200$ to 1200\rmicron\ spectra with \emph{COBE}/FIRAS  from \citet[solid line]{lagache2000}.  
Lower limits are from SPIRE at 250, 350, and 500\rmicron\ \citep{bethermin2012b};  
and SCUBA at 450 and 850\rmicron\ \citep{smail2002,serjeant2004}. 
} 
\label{fig:cib}
\end{figure}
The best-fits parameters for Models 1, 2, and 3 are tabulated in Tables~\ref{tab:correlations1}, \ref{tab:correlations2}, and \ref{tab:correlations3}, respectively,  
and shown in Figure~\ref{fig:halo_model_fit}.  
The respective best-fits to the counts are shown in Figure~\ref{fig:counts}.  
The corresponding $\chi^2$ (degrees of freedom)  are 368 (225), 357 (224), and 371 (223), or $\chi_{\rm reduced}^2=1.6$, 1.6, and 1.7, for Models 1, 2, and 3, respectively.  
That the addition of a parameters does not significantly improve the fits is discussed in \S~\ref{sec:hm_discussion}. 

Correlations between parameters are presented in the off-diagonal entries of Tables~\ref{tab:correlations1}, \ref{tab:correlations2}, and \ref{tab:correlations3}  for Models 1, 2, and 3, respectively.  
As pointed out by e.g., \citet{shang2012}, certain pairs of parameters exhibit very high levels of correlation.  In particular, the SED parameters ($\beta$ and \emph{T} in Models 1 and 2, $T_{\rm cold}$ and $T_{\rm warm}$ in Model 3); the luminosity parameters $\eta$ and $L_0$; and the mass parameters $\sigma^2_{L/m}$. 


We find that star formation is most efficient in halos ranging from ${\rm log}(M_{\odot}) = 11.7$ to 12.5, peaking at ${\rm log}(M_{\odot}) \sim 12.1$, which is consistent with several recent results from observations and simulations \citep[e.g.,][]{moster2010, wang2012, behroozi2013}.  
%
In \citet{cooray2010}, where a 
halo model is developed to fit the angular correlation functions of galaxies brighter than 30 mJy,  the minimum halo mass scale is $\log(M_{\rm min}/\rm M_{\odot}) = 12.6, 12.9$ and $13.5$ at 250, 350 and 500\rmicron, respectively. These values are much higher than our best-fit $\log(M_{\rm min}/\rm M_{\odot})=10.1\pm0.6$, which is due to the fact that faint galaxies (around 5\,mJy) dominate the power spectrum of the intensity fluctuations (see Figure~\ref{fig:Poisson}). In \citet{amblard2011}, the minimum halo mass scale is $\log(M_{\rm min}/\rm M_{\odot}) = 11.1, 11.5$ and $11.8$ at 250, 350 and 500\rmicron, respectively, which are higher than our best-fit value.  
Furthermore, the evolution of the dust temperature, characterized by $T_z$, is in very good agreement with stacking measurements found by \citep[e.g.,][]{pascale2009,viero2013}.   

The redshift distribution of the emissivity --- which in previous halo models has been either parametrized or adopted from galaxy population models --- is here an output of the $L-M$ relation, and shown compared to a selection of models \citep{valiante2009,bethermin2011, bethermin2012c} and previous estimates \citep{amblard2011} in Figure~\ref{fig:dsdz}.  

Lastly, we plot the absolute CIB in each band output by our model, along with several measurements from the literature, in Figure~\ref{fig:cib}.  We find that our models are consistent with the fiducial FIRAS values \citep{fixsen1998, lagache2000},  though we note that the uncertainties in the fiducial measurements are of order 30\%.  


\section{Discussion}
\label{sec:interpretation}
We find a clear signature from the clustering of DSFGs in the pattern of CIB anisotropies, with the fidelity to identify linear and non-linear terms.  Notable is how well the spectra are \emph{still} fit by a power law, with $\chi^2 \sim 3$--$10$ for 15 degrees of freedom (reduced $\chi^2 \sim 0.3$--$0.7$) --- a cosmic coincidence first seen in early clustering measurements of resolved galaxies \citep[e.g.,][]{watson2011a}.  
Future measurements on larger scales, which should bracket the expected peak of the linear, 2-halo spectra, will eventually rule out the power law as a viable fit.  
That the halo model is well motivated regardless is evidenced by the change in the power spectrum with changes in the level of source masking: 
while it appears that the 2-halo term is negligibly affected, the 1-halo term is significantly reduced.  We now discuss plausible interpretations of this reduction in power.  

\subsection{The Reduction of 1-Halo Power with Masking}
\label{sec:one_halo}
Since the 1-halo term originates from multiple galaxies occupying the same halo, the reduction of the 1-halo term with more aggressive source masking suggests that  some fraction of the more luminous resolved SPIRE sources are satellites in massive halos or cluster members, though it should be noted that some fraction of bright galaxies could be close pairs, which would make the same imprint on the spectra.   
This is consistent with the interpretation from clustering measurements of resolved SPIRE sources which claims that $\sim 14\%$ of sources with $S_{250} > 30\, \rm mJy$ appear as satellites \citep{cooray2010}.  

One may wonder if the reduction in power can be solely attributed to a reduction in the Poisson level, but  this is unlikely.  We show this by fitting the unmasked 250\rmicron\ spectra with a 2-halo, 1-halo, and Poisson term, and then fitting the same terms to spectra with sources above 50\,mJy masked. The 2-halo term is fixed in both for both levels of masking, while in the latter fit the 1-halo term is first fixed and then allowed to float, in order that the two fits can be compared.  
For a floating 1-halo term, $\chi^2 = 10.0$ for 15 d.o.f.\@ ($\chi_{\rm reduced}^2=0.7$), while for a fixed 1-halo term $\chi^2 = 98.3$ for 16 d.o.f.\@ ($\chi_{\rm reduced}^2=6.1$),  thus ruling out the possibility that the 1-halo term has not been affected by masking.  

That some fraction of more luminous sources are found in satellite halos is also expected from semi-analytic models of DSFGs.  For example, \citet[][]{gonzalez2011} predict 38\% of all DSFGs (defined there as $S \gsim 1\, \rm mJy$ at 850\rmicron) and 24\% of the most luminous sources ($S \gsim 5\, \rm mJy$ at 850\rmicron) are satellites.  
Furthermore, observations of individual groups and clusters universally find that when star-forming galaxies are present they are located on the outskirts of the massive halos \citep[typically defined as the volume between $R_{500}$ and $R_{200}$, e.g.,][]{saintonge2008,tran2009,braglia2011}.  Even more support for this scenario comes from stacking  in the submillimeter at positions of brightest cluster galaxies (BCGs), which shows a bump in emission at $0.8\, \rm Mpc$ from the central galaxy \citep{coppin2011}.  In light of these observations, the reduction of 1-halo power with masking is unsurprising.   

\subsection{Comparison with Planck}
\label{sec:planck}
Released to the archive at about the same time, initial CIBA power spectra from \emph{Herschel}/SPIRE \citep{amblard2011} and \emph{Planck}/HFI \citep{lagache2011} were found to be discrepant by more than 15\%.  The updated, published \citet{lagache2011} paper explored this difference in detail by comparing to power spectra of SPIRE maps that have had no masking applied (rather than the published spectra which masked all pixels greater than $50\, \rm mJy$). They found that a discrepancy still remained, with SPIRE measurements lower by factors of $\sim 1.7$ and $\sim 1.2$ at 857 and 545\,GHz (350 and 550\rmicron) over the angular range $0.02 < k_{\theta}/\rm arcmin < 0.07$ ($400 < \ell < 1500 $).  If we compare our spectra, having only masked extended sources, to the \emph{Planck}/HFI spectra over the same angular range, we find our spectra are also low, but by factors of $\sim 1.3\pm 0.06$ and $1.4\pm 0.06$ at 350 and 500\rmicron, respectively.   

They next proposed a way to resolve the discrepancy.  They showed that the spectra of the two groups could be brought into agreement if: (\rmnum{1}) no Galactic cirrus is removed from the SPIRE data; and (\rmnum{2}) the beam surface area --- which they claim to be overestimated by $\sim 4$ and 9\%  at 350 and 500\rmicron, respectively --- is corrected.   
Indeed, we also find that the previous cirrus values were overestimated; and from a careful estimate of the beam area (described in detail in Appendix~\ref{sec:calibration}), we also find a correction, though equaling $\sim 2$ and 8\%.  
These corrections have brought the overall offset, particularly on scales $k_{\theta} \lsim  0.04\invarcmin$ ($\ell \lsim 900$) into agreement, however, on small scales they have been unable to close the gap entirely.  
The remaining offset on the largest scales can be attributed to systematic calibration uncertainties of the two instruments: 7\% for SPIRE (Appendix~\ref{sec:calibration}), and 7\% for \emph{Planck}/HFI \citep{planck2011a}, which may be a product of the very different calibration strategies of the two instruments; as well as potential systematic uncertainties in from cirrus removal.  
But it is on scales $k_{\theta} \gsim  0.04\invarcmin$ ($\ell \gsim 900$) --- scales on which the contribution from Galactic cirrus is negligible ---  that excess power in the \emph{Planck} curves are still either in tension (350\rmicron) or do not agree (500\rmicron).


Ultimately, since this paper first appeared it was determined that the \emph{Planck} calibration of the 857 and 545\,GHz (350 and 550\rmicron, respectively) channels was off by 7 and 15\%, meaning that to properly compare the data their spectra must be divided by $1.14$ and $1.30$ \citep[see][]{planck_viii}.  This correction --- which has been applied to the data in Figure~\ref{fig:published_data} --- indeed bring the two curves into generally good agreement.  

\subsection{Interpreting Halo Model Results}
\label{sec:hm_discussion}
Our halo models represent a step forward by being the first to fit the auto- and cross-frequency power spectra, number counts, and absolute CIB levels simultaneously.    
The added complexity introduced by these models is justified as simpler halo models have been unable to simultaneously fit published power spectra \citep[e.g.,][]{lagache2011}.  
Yet, considering the wide range of data fit, our models remain relatively simple; e.g., they have a comparable number of parameters 
to the model presented in \citet[][which varied the Poisson level, $M_{\rm min}$, $M_1$, $\alpha$, and 4 $dS/dz$ nodes, but fit each spectra independently and had \emph{no} $L-M$ relation]{amblard2011}.  
However, with reduced $\chi^2$ of $\sim 1.6$, our models cannot formally be claimed to be good fits, and in fact  there are some obvious problems with these models.  

Firstly, the addition of parameters only marginally improves the fits, if at all.  This is because the $\chi^2$ is dominated by the poor fit to the counts, particularly at the bright end (i.e., $S \gsim 20\, \rm mJy$).   Considering that the clustering power is dominated by the sources below that flux density level (Figure~\ref{fig:Poisson}), this tension between the fits to the counts and the fits to the spectra is not terribly surprising.  It may be that this tension is attributable to there being two types of submillimeter-emitting galaxies \citep[i.e., hotter \lq\lq starburst\rq\rq\ type galaxies, and more \lq\lq normal main sequence\rq\rq\ galaxies, e.g.,][]{elbaz2011}.  Though we explore the possibility of two populations in Model 3, we do not decouple their contributions in redshift.  
If  local starbursts are indeed responsible for a separate, lower redshift non-linear term, then our model would struggle to satisfy the linear and non-linear components at all redshifts, as it appears to do.  

Second, there are significant degeneracies, as well as possibly questionable assumptions, built into this model.  
The strong coupling of the SED parameters \emph{T}, $\beta$, and temperature evolution $T_z$, render the resulting best-fit parameters difficult to meaningfully interpret.  
Also, the very strong anti-correlation of luminosity evolution $\eta$, and normalization $L_0$, likewise make interpreting the evolution of the galaxy luminosity or galaxy bias difficult.  
Furthermore, the assumption that the luminosity evolution increases to $z=2$ and then abruptly flattens, though is seen in several observations \citep[e.g.,][]{stark2009, gonzalez2010}, is likely to be extreme and prone to galaxy selection effects \citep[e.g.,][]{weinmann2011}.   Getting this wrong would result in compensating for the discrepant high-$z$ power by the other parameters in unpredictable ways.  
Lastly, the observed quenching of star formation in the cores of the most massive halos \citep[e.g.,][]{cattaneo2006} is not treated by the model, but may have a significant impact on the non-linear component of the power spectrum.

Thus,  our model is far from the final say in the interpretation of CIB anisotropies, as it appears  that the quality of the data demands a model with additional levels of sophistication.   
%
%
Future models can address many of these limitations by carefully implementing observations.  For example, strong constraints on the stellar-mass to halo-mass relationship \citep[e.g.,][]{behroozi2013}, star-formation to halo-mass relationship \citep[e.g.,][]{wang2012}, and infrared luminosity to stellar-mass relationship \citep{viero2013} now exist.  That, combined with knowledge of the quiescent fraction of galaxies with redshift \citep[e.g.,][]{quadri2012}, would result in more constrained models with fewer parameters.   
Furthermore, future models could be extended to fit not only the three SPIRE bands, but also longer wavelength data from e.g., ACT, \emph{Planck} and SPT, as well as at shorter wavelengths from e.g., \emph{Spitzer}, \emph{IRAS} and \emph{WISE}.  
By combining long wavelength multi-band studies of map-based power spectra with discrete object correlations at shorter wavelengths, we should be able to build a much more complete picture of the relationship between stars, star formation, and dark matter halos.  
\section{Summary}
\label{sec:summary}
We have presented the auto- and cross-frequency power spectra of cosmic infrared background anisotropies at 250, 350, and 500\rmicron.  
The background originates from \emph{all} of the dusty star-forming galaxies in the sky; i.e., those which are bright and resolved, as well as those too faint to be resolved.  

We found an unambiguous signature from the clustering of DSFGs in the pattern of the background light and showed that it can be decomposed into linear (or 2-halo) power from galaxies in separate halos, and non-linear (1-halo) power from multiple central and satellite galaxies occupying massive halos.  
We masked resolved sources in stages down to 50\,mJy and found an expected reduction in the level of Poisson noise, as well as a reduction in the 1-halo power.  We interpreted the reduction in 1-halo power as resulting from some fraction of the most luminous sources being satellite galaxies. 
We also measured the cross-correlation of the signal between bands and found that maps with more aggressive masking to be less correlated, as well as hints of a decreasing correlation with decreasing angular scale; which would be indications of decreased correlations between maps for higher-$z$ sources.   

We then attempted to interpret the measurement through the framework of the halo model, building upon and extending the formalism of \citet{shang2012}.  
Our models were able to \emph{simultaneously} fit the auto- and cross-frequency power spectra, as well as measured number counts and absolute CIB levels from the literature.  
We found that, in this framework of these models, the minimum halo mass to host star formation is ${\rm log}(M_{\rm min}/\rm M_{\odot}) \sim 10.1\pm 0.6$, and that star formation is most efficient in a range of halo masses centered around ${\rm log}(M_{\rm peak}/\rm M_{\odot}) \sim 12.1\pm 0.5$ and $\sigma^2_{L/M} \sim 0.4\pm 0.1$, which is in agreement with other estimates from the literature.  

Our measurement has limited power to constrain angular scales $k_{\theta} \lsim 0.2\invarcmin$, due partly to the relatively small areas of the individual fields, but mostly the result of the filtering performed by the SMAP pipeline.   
The situation will improve dramatically with the arrival of {\sc HeLMS}, which was designed to constrain the turnover of the linear term by targeting the largest modes in the sky,  as well as future measurements from  H-ATLAS \citep{eales2010} and  \emph{Planck}.  Add to that cross-frequency correlations over the full range of angular scales will from e.g., $\rm ACT \times SPIRE$ and $\rm SPT \times SPIRE$, and even $\emph{Planck}\times \emph{Planck}$, which will provide powerful new constraints for models of galaxy evolution, the future indeed holds still more breakthroughs

\begin{acknowledgments}

We thank Aur\'elien Benoit-Levy, Kevin Blagrave, Olivier Dor\'e, Duncan Hanson, Amir Hajian,  Peter Martin, Mattia Negrello, Aurelie P\'enin, Anthony Pullen,  and Christian Reichardt for valuable discussions.  
We also thank  Olivier Dor\'e,  Cien Shang, and Jun-Qing Xia for kindly providing their halo model curves.  
Finally, we sincerely thank the referee for helping improve this paper considerably.  

LW acknowledges support from UK's Science and Technology Facilities Council grant ST/F002858/1 and an ERC StG grant (DEGAS-259586).
SPIRE has been developed by a consortium of institutes led
by Cardiff Univ. (UK) and including: Univ. Lethbridge (Canada);
NAOC (China); CEA, LAM (France); IFSI, Univ. Padua (Italy);
IAC (Spain); Stockholm Observatory (Sweden); Imperial College
London, RAL, UCL-MSSL, UKATC, Univ. Sussex (UK); and Caltech,
JPL, NHSC, Univ. Colorado (USA). This development has been
supported by national funding agencies: CSA (Canada); NAOC
(China); CEA, CNES, CNRS (France); ASI (Italy); MCINN (Spain);
SNSB (Sweden); STFC, UKSA (UK); and NASA (USA).

\end{acknowledgments}

\bibliographystyle{apj}
\bibliography{refs.bib} 

\appendix

\section{A. The Updated SMAP Pipeline}
\label{sec:smap}

The reduction and map making algorithms used with HerMES data have
evolved since the description presented in \citet{levenson2010}.  We
review the modifications to the SMAP pipeline which lead to the
DR1 (first data release) HerMES maps in this appendix.  These maps are
available for download from HeDaM.\footnote[5]{{\tt http://hedam.oamp.fr/HerMES/}}

Initial processing for the SMAP pipeline uses the Herschel
Interactive Processing Environment (HIPE).  For HerMES DR1, the
HCSS/HIPE user release version 6.0.3, corresponding to continuous
integration build 6.0.2055, was used (\citealt{ott2006},
\citealt{ott2010}), including calibration tree version
spire\_cal\_6\_1.  The processing script calls the Spire Photometer
Interactive Analysis (SPIA; \citealt{schulz2011}) version 1.2.

In summary, the basic pipeline processing steps that are performed by
HIPE are in order:
\begin{enumerate}

\item Signal jump detection.

\item Common glitch detection.

\item Sigma-Kappa glitch detection.

\item Pointing product generation.

\item Sigma-Kappa glitch repair.

\item Electronics low-pass filter correction.

\item Signal linearization and flux calibration.

\item Bolometer time response correction.

\end{enumerate}
Details of all these steps, and the implementation of each of the
tasks presented below, can be found in the HIPE Owner's Guide\footnote[6]{\tt{http://herschel.esac.esa.int/Docs/DP/HIPE\_4.2.0/hipeowner.pdf}}. 
Below we detail the tasks called and any changes
to their default arguments.

The initial SMAP processing executes a custom HIPE script that
calls the SPIA tasks {\sc spiaLevel0\_5}, {\sc spiaLevel1Repair}, {\sc
  spiaLevel2}, {\sc spiaSaveObs}, and {\sc spiaSaveMaps2Fits}, in that
sequence.  The arguments to {\sc spiaLevel0\_5} are the defaults with
the following exceptions:
\begin{itemize}
\item \textit{waveDeg} is set to ``Inactive'', switching off the
  wavelet deglitcher.
\item \textit{sigKapDeg} is set to ``Active'',  switching on the sigma-kappa deglitcher.
\item \textit{Kappa} is set to ``4'', meaning that glitches will be detected above $4 \sigma$ of the timeline noise. 
\item \textit{LargeGlitchDiscriminatorTimeConstant} is set to ``4'', providing a higher threshold for detecting large glitches. 
\end{itemize}
Note that the task {\sc spiaLevel0\_5} provides only detection of
jumps and glitches through flags so that the original data still can
be inspected later in the processing. The flagged glitches are
repaired and thermistor timelines with jumps are excluded in the task
{\sc spiaLevel1Repair}.

The parameters in {\sc spiaLevel1Repair} are default with the following exceptions:
\begin{itemize}
\item \textit{extend} is set to ``Yes''. This will cut off only half
  of the turn-around datasets after processing one scan, instead of
  the entire turn-around dataset. Because the subsequent scan will
  keep the other half of the turn-around data, the full turn-around
  dataset remains in the Level 1 data, extending the coverage area.
\item \textit{tempDriftCorr} is set to ``Off'', disabling the
  temperature correction based on the signals of the thermistor pixels
  on the bolometer arrays.
\end{itemize}

The parameters in {\sc spiaLevel2} are default with the following exceptions:
\begin{itemize}
\item \textit{displayMap} is set to ``No'', preventing the preview
  images to pop up during processing.
\item \textit{makeBrowseImage} is set to ``No'' to prevent generation
  of browse images irrelevant to this work.
\end{itemize} 
Note that the HIPE Level 2 maps are not used in the SMAP pipeline. 

The newly-generated Level 1 datasets are then saved in local pools by
task {\sc spiaSaveObs}.  These processed time streams are then
exported to FITS files using the task {\sc exportPalToUfDir}; the SMAP
code reads in the time streams at this point.  The code base itself is
written in the Interactive Data Language (IDL; {\tt
  http://www.exelisvis.com/idl/}).  SMAP first applies a customized
set of masks and bad detector lists and appends them to the masks
carried over from HIPE.  These are appended to as required through the
following analysis.

The SPIRE focal planes experience temperature fluctuations which cause
the bolometer signals to drift over time.  These are largely coherent
across the focal plane for each array, and can be large (corresponding
to as much as $50 \,$Jy over $8 \,$h of observation).  The SPIRE focal
planes have sensitive thermistor devices that monitor the temperature
to ${\sim}\,0.5\,\micro$K at the same sample rate as the detectors; in
normal operation, the temperature is stable over 100\,s to
2\,$\micro$K, so though the instantaneous measurement of the drift is
poor, over ${\sim}\,100$\,s scan lengths the signal-to-noise ratio of
the measurement is $> 10$. Since the thermistors experience the same
fluctuations as the detectors, they can be used to remove the
component of the bolometer signal arising from the thermal drift in
the focal plane.  In SMAP, this is achieved by stitching
together all of the astronomical observation requests (AORs) in a contiguous observation of a given field.
Both the bolometer and thermistor signals are low-pass filtered with a
first-order Butterworth filter with a characteristic scale of 1 degree
on the sky.  Because each SPIRE array has two thermistors, and because
the thermistors occasionally experience cosmic ray hits or glitches,
during times when both thermistors have clean signal they are averaged
together to improve the signal-to-noise ratio of the measurement of
the fluctuations.  When one thermistor is masked due to data quality
issues, the other is used for the duration of the mask.  The
reconstructed average thermistor time stream is then fit to each
bolometer in the detector array, and the resulting scaled version of
the thermistor signal is subtracted.  This procedure effectively
removes the component of the signal arising from thermal fluctuations
to the $\sim 10^{-4}$ level.  A consequence of this procedure is that
the mean (after masking) is subtracted from each scan.

Finally, some scans which pass automated quality masking but which
have low-level but visible problems make it into the final maps.  The
maps are inspected, and scans which contribute obviously artificial
structure are masked from the map making.

Once the time streams are completely conditioned, maps can be
constructed.  The SMAP map-maker, SHIM, follows the presentation
in \citet{levenson2010}; we summarize here. Our noise model is
\begin{equation}
S_{ dsj}=g_{\rm d} M(x_{ dsj}, y_{ dsj}) + p_{dsj}+N_{ dsj},
\end{equation}
where $S_{ dsj}$ is the signal for detector $d$, scan $s$, and time
sample $j$, $g_d$ is the detector gain,\footnote[7]{The $g_d$ are in fact
  the deviations from 1.0 of the detector gains already applied by
  HIPE.} $M(x,y)$ is the sky brightness in pixel (x,y), $N_{dsj}$ is
the instrument noise, and $p_{dsj}$ is an order $n$ polynomial
baseline:
\begin{equation}
p_{dsj} = \sum_{l=0}^n a_{ ds}^l \, (t_j)^l.
\end{equation}
The parameters $a^l_{ ds}$ and, optionally, the detector gains, $g_{
  d}$, are iteratively fit to the time stream residuals.
At each iteration $i$ we calculate the residuals:
\begin{equation}
R^{ i}_{ dsj}=S_{ dsj} - \left[ g_{ d}^{ i} M^{ i-1}(x_{ dsj}, y_{ dsj}) + p^i_{dsj}  \right].
\end{equation}  
We first fit each of the $a_{ds}^{l,i}$ by minimizing $\chi^2 = \sum_j
R^{ i}_{ dsj}$ with the $g_d$ held fixed to $g_d^{i-1}$, the values
calculated from the previous iteration. The $g_d^i$ are then fit by
minimizing $\chi^2 = \sum_{sj} R^{ i}_{ dsj}$ with the $a_{ds}^{l,i}$
held fixed. On the first iteration, the sky is assumed to be 0.0 and
the $g_d$ are held fixed to 1.0. The sky map $M^{ i}(x,y)$ is the
weighted mean of all samples falling in each pixel:
\begin{equation}
M^{ i}(x,y) = \frac{ \displaystyle\sum\limits_{dsj \in (x,y)} w^i_{ds}\left( S_{dsj}- p_{ dsj}^i\right)/g_d^i }{\displaystyle\sum\limits_{dsj \in (x,y)} w^i_{ds}},
\end{equation}
where the weights $w^i_{ds}$ are the inverse variance of the timeline residuals,  
\begin{equation}
w^i_{ds} =\left[ \frac{1}{N} \displaystyle\sum\limits_{j=1}^{N} \left( R^i_{dsj} \right)^2 \right]^{-1},
\end{equation}
with $N$ the number of samples in scan $s$. 

The number of iterations and the iteration on which each of the $p_{
  ds}$, $g_{ d}$ and $w^i_{ds}$ are allowed to vary (if any) are
all specified as inputs to the map maker. For the current data
release, DR1, we run for 20 iterations keeping gains fixed to 1.0, and
allow the weights to deviate from 1.0 starting on the 10$^\mathrm{th}$
iteration.

The SMAP map-maker also performs glitch detection.  In addition to the
timestream based sigma-kappa glitch detection from the HIPE
pre-processing mentioned previously, the SMAP map-maker uses an
iterative glitch detection and removal algorithm based on map
information.  Taking advantage of the fact that each pixel in the
final map is sampled by multiple detectors and scans, the SMAP
map-maker builds a model of what each detector should see as a
function of time, including the polynomial baseline.  Timestream
samples which disagree with this model by more than a specified amount
(usually $10\sigma$, where $\sigma$ is computed for each timeline
after masking) are flagged and removed from subsequent map making
iterations.  This procedure is only activated after a fixed number of
iterations (10, by default) in order to allow for the values of $p$ to
settle, and then is applied for all subsequent iterations.  This
approach is particularly well suited for the HerMES data, which have a
large number of scan repeats.

Finally, we apply an absolute astrometry correction to the maps. This
is measured by stacking preliminary maps on \textsl{Spitzer\/} MIPS
24\rmicron\ sources extracted by Vaccari et al.\@ (in prep.) using the SWIRE \citep{lonsdale2003} MIPS
24\rmicron\ data reduction pipeline \citep{shupe2005}. Astrometric registration of MIPS
sources was carried out against 2MASS, returning a mean absolute deviation
of the MIPS-2MASS offset of about 0.5\arcsec in both RA and Dec in all fields.
We first make a \lq\lq quick\rq\rq\ map, running the map-maker
for only 10 iterations, then make individual maps for each AOR using
the parameters determined from the quick map. Each AOR map is stacked
on the 24\rmicron\ catalog and a 2D Gaussian is fit to the resulting
profile. The distance of the center of the fitted Gaussian to the
nominal center of the image is taken as an absolute shift in the
astrometry. These measured offsets are applied to the detector
pointing solutions in subsequent map-making runs. We note that we have
measured the offsets in all three bands independently, but find that
the measured shifts are consistent between bands, and thus apply the
offsets measured at 250\rmicron, where the resolution is highest, to
all three bands. The measured shifts are systematic from AOR to AOR,
and are generally in the range of $1-3$\arcsec.

\section{B. SPIRE Map Calibration}
\label{sec:calibration}

Proper calibration of maps is critical for power spectrum measurements, as any systematic offsets are squared in the power spectrum.    
Here we summarize the calibration and color correction procedures; for a more complete description see the SPIRE Observers Manual\footnote[8]{\tt{http://herschel.esac.esa.int/Docs/SPIRE/html/spire\_om.html}}. 

Since Neptune is very bright, relatively compact (angular size $\lsim 2.5$\arcsec), and can be seen above instrumental noise in the timestreams, SPIRE fluxes are calibrated in the time domain by fitting the point-spread function (or beam profile) to data and setting the peak values to those expected from the \citet{moreno1998} model\footnote[9]{
Tabulations of the Neptune and Uranus brightness temperatures are available from the ESA Herschel Science Centre \tt{ftp://ftp.sciops.esa.int/pub/hsc-calibration}}.  
This measurement is shown to be repeatable at the 2\% level, and the quoted uncertainty in the Neptune model is 5\%, which is conservative and still improving.  As these are systematic uncertainties, the quoted uncertainty in the calibration is thus 7\%. 

In a SPIRE photometer observation, the property of the source that is directly proportional to source power absorbed by the bolometer is the integral over the passband of the flux density weighted by the instrument Relative Spectral Response Function (RSRF).  
Converting from a RSRF-weighted flux density, $\bar{S}_{\rm S}$, to a monochromatic flux density requires the adoption of a standard frequency for the band and some assumption about the shape of the source spectrum. The approach adopted for SPIRE (and PACS) is to assume that the spectrum is a power law across the band defined by the flux density at a standard frequency $\nu_0$, and a spectral index $\alpha_{S_0}$
\begin{equation}
S_{\rm S}(\nu)=S_{\rm S}(\nu_0) \left( \frac{\nu}{\nu_0} \right) ^{\alpha_{S_0}},  
\end{equation}
where $\nu_0$ corresponds to frequency equivalent of the nominal SPIRE wavelengths (i.e., 250, 350 and 500\rmicron), and $\alpha_{S_0}=-1$, so that the source has a spectrum $\nu S(\nu)$ which is flat across the band.  
The monochromatic flux density at frequency $\nu_0$, which is what is output by HIPE, is then
\begin{equation}
S_{\rm S}(\nu_0)=\bar{S}_{\rm S} \left[ \nu^{\alpha_{S_0}}_0 \frac{\int R_{\rm type} (\nu) d\nu}{\int \nu^{\alpha_{S_0}} R_{\rm type} (\nu) d\nu}  \right] = K_{4,\rm type} \bar{S}_{\rm S}, 
\end{equation}
where \lq\lq type\rq\rq\ refers to point or extended source.  
For extended sources, the passband is weighted by an additional $\lambda^{\gamma}$ to account for its width since the beam size increases with increasing wavelength across it.  
The exact value of $\gamma$ is dependent on the optics of the instrument: though nominally it is expected that the beam area would increase as $\lambda^2$, in the limit of a very hard taper (or under-illumination) the illumination on
the primary is proportional to $\lambda$, and the FWHM on the sky is wavelength
independent.  
The SPIRE taper is slightly wider than a pixel with tophat illumination on the primary, meaning that it lies between the two extremes, but closer to nominal.  From the optics model it is found that $\gamma=1.8$.  The SPIRE photometer pipeline is based on a point source, i.e.,  $K_{\rm pip}=K_{\rm 4,P}(\alpha_{S_0})= [1.0119, 1.0094, 1.0073]$ at 250, 350, and 500\rmicron, respectively.

For extended sources whose true spectra differs from a power law with $\alpha_{S_0}=-1$, a color correction, $K_{\rm C, E}=K_{4, \rm E}/K_{\rm pip}$, must be applied, where 
\begin{equation}
K_{4, \rm E}=F_{\rm sky}(\nu_0) \frac{\int R_{\rm E} (\nu) d\nu}{\int F_{\rm sky}(\nu) R_{\rm E} (\nu) d\nu}.
\end{equation}  

We would like to color-correct for the case where $F_{\rm sky}(\nu_0)$  is the infrared background, an extended source which FIRAS showed can described by a modified blackbody with $T=18.5$ and $\beta=0.65$ \citep[][]{puget1996}. 
We estimate $K_{4 \rm E}$ using the SPIRE passbands additionally weighted by $\lambda^2$ and a modified black-body SED, finding $K_{4, \rm E}=[1.0107,1.0022,1.0029]$ at 250, 350, and 500\rmicron, respectively.  We check that these corrections are not sensitive to the approximation made for the FIRAS SED by varying the temperature $\pm 2\, \rm K$, finding a negligible change of $\sim \pm 0.3\%$.  
 These values compare well with the color corrections for extended sources given in Figure 5.11 of the Observers Manual for CIB spectra approximated as power laws across the passbands, with $\alpha_S \approx (0.3, 1.1, 1.6)$.  
In summary, the corrections applied to the maps at 250, 350, and 500\rmicron\ are:
\begin{eqnarray}
K_{\rm FIRAS} & = & K_{\rm C, E} = K_{\rm 4,E}/K_{\rm pip}   \nonumber \\
& = & \left[ 0.9988, 0.9929, 0.9957 \right], 
\end{eqnarray}
and hence, negligible.  

\begin{table}
 \centering
 \caption{Beam Nominal and Effective Areas.}
   \begin{tabular}{l||c|c|c|c}
   \hline
Band  & FWHM & $A_{\rm measured}$ & Correction & $ A_{\rm eff}$ \\
\rmicron &   (arcsec)  & (steradians) &  Factor & (steradians)  \\
\hline
250 & 18.1 & $1.039\times 10^{-8}$ &  1.013 & $1.053\times 10^{-8}$ \\
350 & 25.2 & $1.723\times 10^{-8}$ &  1.004 & $1.730\times 10^{-8}$ \\
500 & 36.6 & $3.707\times 10^{-8}$ &  0.995 & $3.688\times 10^{-8}$  \\
   \hline
  \end{tabular}
 \label{tab:beams}
\end{table}
Finally, since power spectra are performed on maps in surface brightness units of $\rm Jy\, sr^{-1}$, and SPIRE maps are natively produced in units of $\rm Jy\, beam^{-1}$, a conversion factor must be applied to the maps, equal to the inverse of the solid angle of the beams, 
 \begin{equation}
A_{\rm beam}=\int B(\theta,\phi)(\theta,\phi)d\Omega,
\end{equation}
where $ B(\theta,\phi)$ is the normalized beam profile, and $d\Omega$ is the solid angle element in the direction $(\theta,\phi)$
 
The beam solid angles are measured from SMAP generated maps of Neptune with pixel sizes of 2\arcsec, normalized by the peak value.  The area of the SPIRE beams is calculated by summing the Neptune map pixels and multiplying by the pixel area, i.e., $4\arcsec $.
Next, we address the contamination from background galaxies.  We pick a radius, $r_0$, within which to integrate, yielding an integral over an area on the map $A_0$.  We then pick a second area to be an annular ring with $r_0 < r < r_{\rm ID}$, where the inner diameter $r_{\rm ID} = \sqrt{r_0} $, yielding an area equal to the the inner area.  The inner area contains the sum of the response to Neptune and the background galaxies, while the outer annulus is just the sum of from the galaxies. Assuming the statistics of the background do not change, the outer integral can be subtracted from the inner to remove the effect of the background on the beam area. The resulting beam areas given in the second column of Table~\ref{tab:beams}.  
There are systematic uncertainties associated with this calculations.  We estimate the beam integral by repeating the measurement but varying the values of the input parameters, $r_0$ and $r_{\rm ID}$.  
Varying $r_0$ by $\pm 1\arcmin$ from its nominal value results in a fraction of a percent change in the integrals; 
while varying $r_{\rm ID}$ by $\pm 10\%$, which we find dominates the error budget, changes the total area by $< 1\%$.  
Note that these values are specific to SMAP made Neptune maps with the same filtering as was used in the maps used in our study, and as such should not be blindly adopted for just any SPIRE map.  

Lastly, the beam effective area is corrected for the difference in illumination of the passband due to the relative colors of Neptune and the CIB.  Both Neptune and the CIB can be described as modified blackbodies, however, the temperature of Neptune is $\sim 70\, \rm K$, while the CIB is $\sim 18.5\, \rm K$.  To account for this, the beam areas are corrected by the ratio of the integrals of the passbands for extended sources weighted by the two SEDs.  The resulting correction factors are $[1.013,  1.004, 0.995]$ at 250, 350 and 500\rmicron, respectively.  As anticipated, the correction is highest at 250\rmicron, where the SED of the CIB peaks.  We check for potential systematic errors by varying the CIB temperature, and find changes to be at the sub-percent level.   
The final effective beam areas are quoted in the last column of Table~\ref{tab:beams}.  Note that the beam areas used by \citet{amblard2011} were 1.03, 1.77, and $3.99\times 10^{-8}\, \rm steradians$.

\section{C. Conversion to CMB Units}
\label{sec:convert}
The flux density unit of convention for infrared, (sub)millimeter, and radio astronomers is the Jansky, defined as:
\beq
\rm Jy=10^{-26}\, \rm W\, m^{-2}\, Hz,
\eeq
and is obtained by integrating over the solid angle of the source.  For extended sources, the surface brightness is described in Jy per unit solid angle, for example, $\rm Jy\,sr^{-1}$.  Additionally, the power spectrum unit in this convention is given in $\rm Jy^2\, sr^{-1}$.  To convert from $\rm Jy^2\, \rm beam^{-1}$ to $\rm Jy^2\, sr^{-1}$, SPIRE maps must be divided by the area of the beam.  Beam areas are presented in Table~\ref{tab:beams}.   
For more details see Appendix~\ref{sec:calibration} of this paper, or \S~5.2.9 and Table~5.2 of the SPIRE Observers Manual\footnote[10]{\tt{http://herschel.esac.esa.int/Docs/SPIRE/html/spire\_om.html\#x1$-$850005.1}}.

The convention for CMB units is to report a signal as $\delta T_{\rm CMB}$; the deviation from the primordial $2.7255\, \rm K$ blackbody.  To convert from $\rm Jy\, sr^{-1}$ to $\delta T_{\rm CMB}$ in $\mu \rm K$, as a function of frequency:
\begin{eqnarray}
\delta T_{\nu} &=& \left(\frac{\delta B_{\nu}}{\delta T}\right), \\
{\rm where}~~ \frac{\delta B_{\nu}}{\delta T} & = & \frac{2k}{c^2}\left( \frac{kT_{\rm CMB}}{h} \right)^2 \frac{x^2e^x}{(e^x -1)^2}= \frac{98.91~ \rm Jy\, sr^{-1}}{\mu \rm K} \frac{x^2e^x}{(e^x -1)^2}, \\
{\rm and}~~~~~~~ x &=& \frac{h\nu}{k_{\nu}T_{\rm CMB}}=\frac{\nu}{56.79~\rm GHz},
\end{eqnarray}
\citep{fixsen2009}.  %
Because the SPIRE passbands have widths of $\sim 30\%$ \citep{griffin2010}, and because the CMB blackbody at these wavelengths is particularly steep (falling exponentially on the Wien side of the 2.7255 K blackbody), the integral of $\delta B_{\nu}/\delta T$ over the bands is weighted towards lower frequencies; an effect that becomes dramatically more pronounced at shorter wavelengths.   Ultimately, to convert SPIRE maps in $\rm Jy\, sr^{-1}$ to $\rm \micro K_{\rm CMB}$ they must be multiplied by factors of $3.664\times10^{-7}$, $1.897\times10^{-8}$, and $2.652\times10^{-10}$  

To compare the 350\rmicron\ band directly to the same band in BLAST requires a slight color correction, as their passbands are not quite the same.  This correction, from BLAST to SPIRE, is 0.968 in the maps, or 0.937 in the power spectra.  At 250 and 500\rmicron\ those conversions are respectively 0.994 and 0.996  in the maps, or 0.989 and 0.992 in the power spectra, i.e., negligible.  

To compare the 350\rmicron\ and 500\rmicron\ bands to the $857$ and $545$\,GHz (or 350 and 550\rmicron) \emph{Planck} bands  also requires color corrections due to shifts in the band centers.  Those conversions, from \emph{Planck}/HFI to SPIRE at 350 and 500\rmicron, are 0.99 and 1.14 in the maps, or 0.99 and 1.30 in the power spectra, respectively.  
  
Lastly, the CMB power spectrum is conventionally reported versus multipole $\ell$, while in the (sub)millimeter the convention is to report it versus angular wavenumber, $k_{\theta}=1/ \lambda$, which is also known as $\sigma$ in the literature, and is typically expressed in $\rm arcmin^{-1}$.  In the small-angle approximation the two are related by $\ell = 2\pi k_{\theta}$.

\section{D. Alternative Masking Spectra}
\label{sec:alt}
\begin{figure}[t!]
\centering
\includegraphics[width=0.7\textwidth]{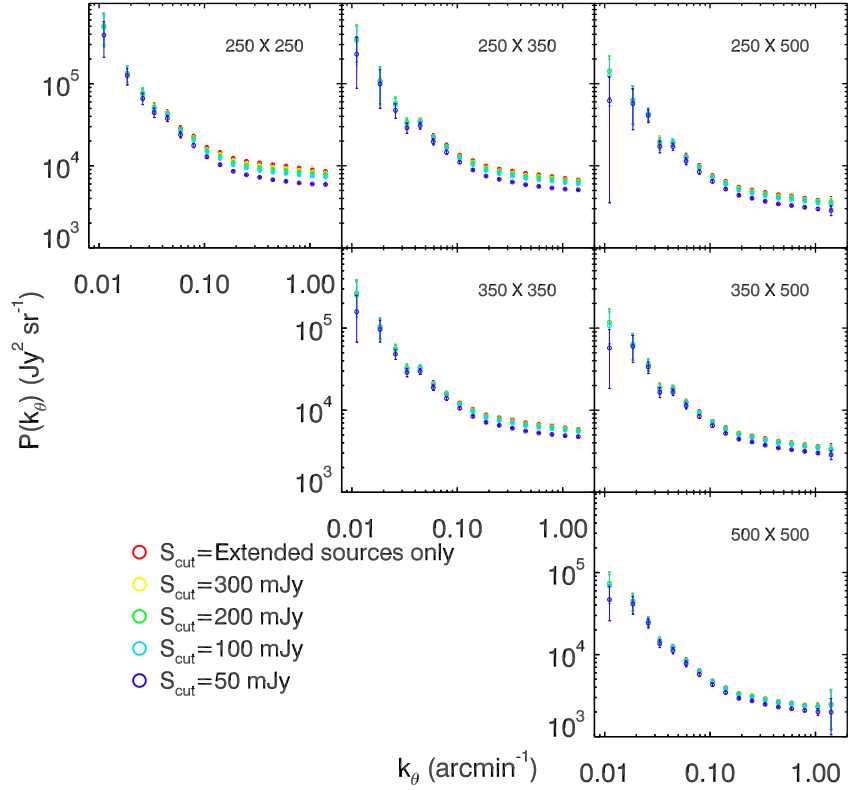}

\caption{Combined spectra vs.\@ flux cut of masked sources, plotted as circles with error bars.  The best estimates of the cirrus spectra in each field (\S~\ref{sec:cirrus}) are removed before combining.   } 
\label{fig:ps_vs_flux_cut_250mask}
\end{figure}
As described in \S~\ref{sec:mask}, an alternative set of spectra are calculated on sets of maps which have all been masked similarly.  I.e., rather than only mask those sources above the flux density cut in that band, all sources identified at 250\rmicron\ are masked at 350 and 500\rmicron\ as well. Note that as before, sources are masked with circles whose sizes are $1.1\cdot \rm FWHM$.  I.e., the locations of all masked sources are the same band to band, but the sizes of the masks are not.

The combined spectra are presented in Figure~\ref{fig:ps_vs_flux_cut_250mask}.  We expect the impact of the alternative masking scheme to be most noticeable in the Poisson and 1-halo terms at 350 and 500\rmicron, where previously very few sources were masked.  Poisson level estimates are given in Table~\ref{tab:poisson_250mask}.  We find a decline in the Poisson level for the 50\,mJy flux cut is 5.5 and 8.8\% at 350 and 500\rmicron.  
\section{E. Tables}
\label{sec:tables}

\begin{table*}
 \centering
 \begin{tabular}{c|ccccc}
 & $S > 50\, \rm mJy$ & $S > 100\, \rm mJy$ & $S > 200\, \rm mJy$ & $S > 300\, \rm mJy$ & Extended \\ 
 \hline
 $ 250 \times 250 $ & $ (6.0 \pm 0.1)\times 10^{3} $ & $ (7.4 \pm 0.1)\times 10^{3} $ & $ (8.1 \pm 0.1)\times 10^{3} $ & $ (8.4 \pm 0.1)\times 10^{3} $ & $ (9.0 \pm 0.1)\times 10^{3} $  \\ 
 $ 250 \times 350 $ & $ (5.1 \pm 0.1)\times 10^{3} $ & $ (6.2 \pm 0.1)\times 10^{3} $ & $ (6.6 \pm 0.1)\times 10^{3} $ & $ (6.7 \pm 0.1)\times 10^{3} $ & $ (7.0 \pm 0.1)\times 10^{3} $  \\ 
 $ 250 \times 500 $ & $ (3.0 \pm 0.0)\times 10^{3} $ & $ (3.6 \pm 0.1)\times 10^{3} $ & $ (3.8 \pm 0.1)\times 10^{3} $ & $ (3.9 \pm 0.1)\times 10^{3} $ & $ (4.0 \pm 0.1)\times 10^{3} $  \\ 
 $ 350 \times 350 $ & $ (5.2 \pm 0.0)\times 10^{3} $ & $ (5.9 \pm 0.1)\times 10^{3} $ & $ (6.1 \pm 0.1)\times 10^{3} $ & $ (6.1 \pm 0.1)\times 10^{3} $ & $ (6.2 \pm 0.1)\times 10^{3} $  \\ 
 $ 350 \times 500 $ & $ (3.1 \pm 0.0)\times 10^{3} $ & $ (3.6 \pm 0.0)\times 10^{3} $ & $ (3.7 \pm 0.0)\times 10^{3} $ & $ (3.7 \pm 0.0)\times 10^{3} $ & $ (3.8 \pm 0.0)\times 10^{3} $  \\ 
 $ 500 \times 500 $ & $ (2.2 \pm 0.0)\times 10^{3} $ & $ (2.4 \pm 0.0)\times 10^{3} $ & $ (2.4 \pm 0.0)\times 10^{3} $ & $ (2.4 \pm 0.0)\times 10^{3} $ & $ (2.4 \pm 0.0)\times 10^{3} $  \\ 
\hline
\end{tabular}
\caption{Best-fit Poisson levels as shown in Figure~\ref{fig:shot}. }
\label{tab:poisson}
\end{table*}
\begin{table*}
 \centering
 \begin{tabular}{c|ccccc}
 & $S > 50\, \rm mJy$ & $S > 100\, \rm mJy$ & $S > 200\, \rm mJy$ & $S > 300\, \rm mJy$ & Extended \\ 
 \hline
 $ 250 \times 250 $ & $ (5.8 \pm 0.1)\times 10^{3} $ & $ (7.3 \pm 0.1)\times 10^{3} $ & $ (7.9 \pm 0.1)\times 10^{3} $ & $ (8.2 \pm 0.1)\times 10^{3} $ & $ (8.6 \pm 0.2)\times 10^{3} $  \\ 
 $ 250 \times 350 $ & $ (5.0 \pm 0.1)\times 10^{3} $ & $ (6.0 \pm 0.1)\times 10^{3} $ & $ (6.4 \pm 0.1)\times 10^{3} $ & $ (6.5 \pm 0.1)\times 10^{3} $ & $ (6.7 \pm 0.1)\times 10^{3} $  \\ 
 $ 250 \times 500 $ & $ (2.9 \pm 0.1)\times 10^{3} $ & $ (3.5 \pm 0.1)\times 10^{3} $ & $ (3.6 \pm 0.1)\times 10^{3} $ & $ (3.7 \pm 0.1)\times 10^{3} $ & $ (3.8 \pm 0.1)\times 10^{3} $  \\ 
 $ 350 \times 350 $ & $ (4.7 \pm 0.1)\times 10^{3} $ & $ (5.5 \pm 0.1)\times 10^{3} $ & $ (5.7 \pm 0.1)\times 10^{3} $ & $ (5.8 \pm 0.1)\times 10^{3} $ & $ (5.9 \pm 0.1)\times 10^{3} $  \\ 
 $ 350 \times 500 $ & $ (2.9 \pm 0.1)\times 10^{3} $ & $ (3.4 \pm 0.1)\times 10^{3} $ & $ (3.5 \pm 0.1)\times 10^{3} $ & $ (3.5 \pm 0.1)\times 10^{3} $ & $ (3.5 \pm 0.1)\times 10^{3} $  \\ 
 $ 500 \times 500 $ & $ (1.9 \pm 0.1)\times 10^{3} $ & $ (2.2 \pm 0.1)\times 10^{3} $ & $ (2.3 \pm 0.1)\times 10^{3} $ & $ (2.3 \pm 0.1)\times 10^{3} $ & $ (2.3 \pm 0.1)\times 10^{3} $  \\ 
\hline
\end{tabular}
\caption{Best-fit Poisson levels as shown in Figure~\ref{fig:shot}. }
\label{tab:poisson_250mask}
\end{table*}
\newpage

%
%
\begin{center}
\begin{longtable*}{c|cccccc}
\caption[]
{Combined power spectra for all levels of masking.  At each wavelength, only sources above the flux cut are masked. } \label{tab:combined_spectra} \\

\endfirsthead

\multicolumn{7}{c}{{ \tablename\ \thetable{} -- Continued from previous page}} \\ \hline
\endhead

\multicolumn{7}{r}{{Continued on next page\ldots}} \\ 
\endfoot

\endlastfoot

 \hline
 & & & & & &  \\ [0.25mm]
$k_{\theta} $ & \multicolumn{6}{c}{Only Extended Sources Masked} \\ [0.5mm] 
$ [\rm arcmin^{-1}] $ & $ 250 \times250 $ & $ 250 \times350 $ & $ 250 \times500 $ & $ 350 \times350 $ & $ 350 \times500 $ & $ 500 \times500 $  \\ 
 \hline
 $ 0.011 $ &  $ (5.15 \pm 3.02)\times 10^{5} $ &  $ (3.59 \pm 2.45)\times 10^{5} $ &  $ (1.46 \pm 1.08)\times 10^{5} $ &  $ (2.70 \pm 1.69)\times 10^{5} $ &  $ (1.19 \pm 0.76)\times 10^{5} $ &  $ (7.41 \pm 3.87)\times 10^{4} $ \\ 
 $ 0.019 $ &  $ (1.34 \pm 0.43)\times 10^{5} $ &  $ (1.09 \pm 0.73)\times 10^{5} $ &  $ (6.28 \pm 4.37)\times 10^{4} $ &  $ (1.04 \pm 0.42)\times 10^{5} $ &  $ (6.31 \pm 3.18)\times 10^{4} $ &  $ (4.45 \pm 1.51)\times 10^{4} $ \\ 
 $ 0.026 $ &  $ (7.72 \pm 1.70)\times 10^{4} $ &  $ (5.80 \pm 1.57)\times 10^{4} $ &  $ (4.27 \pm 1.08)\times 10^{4} $ &  $ (5.60 \pm 1.07)\times 10^{4} $ &  $ (3.62 \pm 0.86)\times 10^{4} $ &  $ (2.55 \pm 0.46)\times 10^{4} $ \\ 
 $ 0.033 $ &  $ (5.18 \pm 0.94)\times 10^{4} $ &  $ (3.39 \pm 0.67)\times 10^{4} $ &  $ (1.96 \pm 0.46)\times 10^{4} $ &  $ (3.27 \pm 0.53)\times 10^{4} $ &  $ (1.87 \pm 0.37)\times 10^{4} $ &  $ (1.49 \pm 0.22)\times 10^{4} $ \\ 
 $ 0.044 $ &  $ (4.40 \pm 0.54)\times 10^{4} $ &  $ (3.55 \pm 0.46)\times 10^{4} $ &  $ (1.95 \pm 0.28)\times 10^{4} $ &  $ (3.34 \pm 0.37)\times 10^{4} $ &  $ (1.85 \pm 0.23)\times 10^{4} $ &  $ (1.23 \pm 0.15)\times 10^{4} $ \\ 
 $ 0.059 $ &  $ (2.84 \pm 0.32)\times 10^{4} $ &  $ (2.27 \pm 0.26)\times 10^{4} $ &  $ (1.30 \pm 0.17)\times 10^{4} $ &  $ (2.12 \pm 0.22)\times 10^{4} $ &  $ (1.26 \pm 0.14)\times 10^{4} $ &  $ (8.48 \pm 0.95)\times 10^{3} $ \\ 
 $ 0.079 $ &  $ (2.27 \pm 0.20)\times 10^{4} $ &  $ (1.78 \pm 0.14)\times 10^{4} $ &  $ (9.98 \pm 0.86)\times 10^{3} $ &  $ (1.61 \pm 0.12)\times 10^{4} $ &  $ (9.54 \pm 0.74)\times 10^{3} $ &  $ (6.36 \pm 0.52)\times 10^{3} $ \\ 
 $ 0.105 $ &  $ (1.67 \pm 0.11)\times 10^{4} $ &  $ (1.35 \pm 0.08)\times 10^{4} $ &  $ (7.60 \pm 0.44)\times 10^{3} $ &  $ (1.23 \pm 0.06)\times 10^{4} $ &  $ (7.35 \pm 0.38)\times 10^{3} $ &  $ (4.79 \pm 0.27)\times 10^{3} $ \\ 
 $ 0.141 $ &  $ (1.45 \pm 0.08)\times 10^{4} $ &  $ (1.16 \pm 0.05)\times 10^{4} $ &  $ (6.47 \pm 0.31)\times 10^{3} $ &  $ (1.03 \pm 0.04)\times 10^{4} $ &  $ (6.15 \pm 0.26)\times 10^{3} $ &  $ (3.95 \pm 0.19)\times 10^{3} $ \\ 
 $ 0.187 $ &  $ (1.23 \pm 0.06)\times 10^{4} $ &  $ (9.90 \pm 0.42)\times 10^{3} $ &  $ (5.48 \pm 0.23)\times 10^{3} $ &  $ (8.79 \pm 0.32)\times 10^{3} $ &  $ (5.26 \pm 0.18)\times 10^{3} $ &  $ (3.37 \pm 0.13)\times 10^{3} $ \\ 
 $ 0.250 $ &  $ (1.13 \pm 0.05)\times 10^{4} $ &  $ (9.10 \pm 0.36)\times 10^{3} $ &  $ (5.08 \pm 0.20)\times 10^{3} $ &  $ (8.09 \pm 0.27)\times 10^{3} $ &  $ (4.88 \pm 0.15)\times 10^{3} $ &  $ (3.15 \pm 0.11)\times 10^{3} $ \\ 
 $ 0.333 $ &  $ (1.08 \pm 0.05)\times 10^{4} $ &  $ (8.61 \pm 0.33)\times 10^{3} $ &  $ (4.74 \pm 0.17)\times 10^{3} $ &  $ (7.58 \pm 0.24)\times 10^{3} $ &  $ (4.53 \pm 0.13)\times 10^{3} $ &  $ (2.90 \pm 0.09)\times 10^{3} $ \\ 
 $ 0.445 $ &  $ (1.03 \pm 0.04)\times 10^{4} $ &  $ (8.09 \pm 0.28)\times 10^{3} $ &  $ (4.44 \pm 0.16)\times 10^{3} $ &  $ (7.05 \pm 0.21)\times 10^{3} $ &  $ (4.20 \pm 0.12)\times 10^{3} $ &  $ (2.68 \pm 0.09)\times 10^{3} $ \\ 
 $ 0.593 $ &  $ (9.93 \pm 0.39)\times 10^{3} $ &  $ (7.77 \pm 0.26)\times 10^{3} $ &  $ (4.29 \pm 0.16)\times 10^{3} $ &  $ (6.72 \pm 0.20)\times 10^{3} $ &  $ (4.03 \pm 0.11)\times 10^{3} $ &  $ (2.58 \pm 0.10)\times 10^{3} $ \\ 
 $ 0.790 $ &  $ (9.40 \pm 0.35)\times 10^{3} $ &  $ (7.38 \pm 0.23)\times 10^{3} $ &  $ (4.03 \pm 0.17)\times 10^{3} $ &  $ (6.42 \pm 0.18)\times 10^{3} $ &  $ (3.81 \pm 0.12)\times 10^{3} $ &  $ (2.43 \pm 0.13)\times 10^{3} $ \\ 
 $ 1.054 $ &  $ (8.92 \pm 0.32)\times 10^{3} $ &  $ (7.01 \pm 0.22)\times 10^{3} $ &  $ (3.80 \pm 0.25)\times 10^{3} $ &  $ (6.12 \pm 0.18)\times 10^{3} $ &  $ (3.62 \pm 0.18)\times 10^{3} $ &  $ (2.33 \pm 0.34)\times 10^{3} $ \\ 
 $ 1.406 $ &  $ (8.54 \pm 0.31)\times 10^{3} $ &  $ (6.71 \pm 0.23)\times 10^{3} $ &  $ (3.59 \pm 0.86)\times 10^{3} $ &  $ (5.84 \pm 0.24)\times 10^{3} $ &  $ (3.39 \pm 0.70)\times 10^{3} $ &  $ (2.50 \pm 1.79)\times 10^{3} $ \\ 
 \hline
 \hline
 & & & & & &  \\ [0.25mm]
$k_{\theta} $ & \multicolumn{6}{c}{Sources with $ S > 300\, \rm mJy$ Masked} \\ [0.5mm] 
$ [\rm arcmin^{-1}] $ & $ 250 \times250 $ & $ 250 \times350 $ & $ 250 \times500 $ & $ 350 \times350 $ & $ 350 \times500 $ & $ 500 \times500 $  \\ 
 \hline
 $ 0.011 $ &  $ (5.09 \pm 2.99)\times 10^{5} $ &  $ (3.57 \pm 2.44)\times 10^{5} $ &  $ (1.45 \pm 1.09)\times 10^{5} $ &  $ (2.70 \pm 1.69)\times 10^{5} $ &  $ (1.19 \pm 0.76)\times 10^{5} $ &  $ (7.41 \pm 3.88)\times 10^{4} $ \\ 
 $ 0.019 $ &  $ (1.34 \pm 0.43)\times 10^{5} $ &  $ (1.09 \pm 0.74)\times 10^{5} $ &  $ (6.36 \pm 4.41)\times 10^{4} $ &  $ (1.04 \pm 0.43)\times 10^{5} $ &  $ (6.35 \pm 3.19)\times 10^{4} $ &  $ (4.46 \pm 1.51)\times 10^{4} $ \\ 
 $ 0.026 $ &  $ (7.73 \pm 1.70)\times 10^{4} $ &  $ (5.83 \pm 1.59)\times 10^{4} $ &  $ (4.31 \pm 1.09)\times 10^{4} $ &  $ (5.59 \pm 1.07)\times 10^{4} $ &  $ (3.63 \pm 0.86)\times 10^{4} $ &  $ (2.56 \pm 0.46)\times 10^{4} $ \\ 
 $ 0.033 $ &  $ (5.11 \pm 0.93)\times 10^{4} $ &  $ (3.39 \pm 0.67)\times 10^{4} $ &  $ (1.95 \pm 0.45)\times 10^{4} $ &  $ (3.29 \pm 0.53)\times 10^{4} $ &  $ (1.88 \pm 0.37)\times 10^{4} $ &  $ (1.49 \pm 0.22)\times 10^{4} $ \\ 
 $ 0.044 $ &  $ (4.32 \pm 0.52)\times 10^{4} $ &  $ (3.51 \pm 0.46)\times 10^{4} $ &  $ (1.94 \pm 0.28)\times 10^{4} $ &  $ (3.32 \pm 0.38)\times 10^{4} $ &  $ (1.85 \pm 0.23)\times 10^{4} $ &  $ (1.23 \pm 0.14)\times 10^{4} $ \\ 
 $ 0.059 $ &  $ (2.71 \pm 0.30)\times 10^{4} $ &  $ (2.20 \pm 0.26)\times 10^{4} $ &  $ (1.26 \pm 0.16)\times 10^{4} $ &  $ (2.11 \pm 0.23)\times 10^{4} $ &  $ (1.25 \pm 0.14)\times 10^{4} $ &  $ (8.42 \pm 0.94)\times 10^{3} $ \\ 
 $ 0.079 $ &  $ (2.17 \pm 0.17)\times 10^{4} $ &  $ (1.73 \pm 0.13)\times 10^{4} $ &  $ (9.77 \pm 0.84)\times 10^{3} $ &  $ (1.60 \pm 0.12)\times 10^{4} $ &  $ (9.46 \pm 0.73)\times 10^{3} $ &  $ (6.31 \pm 0.51)\times 10^{3} $ \\ 
 $ 0.105 $ &  $ (1.57 \pm 0.08)\times 10^{4} $ &  $ (1.30 \pm 0.07)\times 10^{4} $ &  $ (7.42 \pm 0.43)\times 10^{3} $ &  $ (1.22 \pm 0.06)\times 10^{4} $ &  $ (7.31 \pm 0.39)\times 10^{3} $ &  $ (4.77 \pm 0.28)\times 10^{3} $ \\ 
 $ 0.141 $ &  $ (1.36 \pm 0.06)\times 10^{4} $ &  $ (1.11 \pm 0.04)\times 10^{4} $ &  $ (6.30 \pm 0.28)\times 10^{3} $ &  $ (1.02 \pm 0.04)\times 10^{4} $ &  $ (6.10 \pm 0.25)\times 10^{3} $ &  $ (3.93 \pm 0.18)\times 10^{3} $ \\ 
 $ 0.187 $ &  $ (1.15 \pm 0.04)\times 10^{4} $ &  $ (9.48 \pm 0.31)\times 10^{3} $ &  $ (5.33 \pm 0.20)\times 10^{3} $ &  $ (8.68 \pm 0.26)\times 10^{3} $ &  $ (5.21 \pm 0.17)\times 10^{3} $ &  $ (3.35 \pm 0.13)\times 10^{3} $ \\ 
 $ 0.250 $ &  $ (1.06 \pm 0.04)\times 10^{4} $ &  $ (8.71 \pm 0.27)\times 10^{3} $ &  $ (4.92 \pm 0.17)\times 10^{3} $ &  $ (8.00 \pm 0.22)\times 10^{3} $ &  $ (4.84 \pm 0.14)\times 10^{3} $ &  $ (3.13 \pm 0.11)\times 10^{3} $ \\ 
 $ 0.333 $ &  $ (1.00 \pm 0.04)\times 10^{4} $ &  $ (8.19 \pm 0.24)\times 10^{3} $ &  $ (4.58 \pm 0.14)\times 10^{3} $ &  $ (7.49 \pm 0.19)\times 10^{3} $ &  $ (4.49 \pm 0.12)\times 10^{3} $ &  $ (2.88 \pm 0.09)\times 10^{3} $ \\ 
 $ 0.445 $ &  $ (9.53 \pm 0.32)\times 10^{3} $ &  $ (7.69 \pm 0.21)\times 10^{3} $ &  $ (4.29 \pm 0.13)\times 10^{3} $ &  $ (6.96 \pm 0.17)\times 10^{3} $ &  $ (4.16 \pm 0.11)\times 10^{3} $ &  $ (2.66 \pm 0.09)\times 10^{3} $ \\ 
 $ 0.593 $ &  $ (9.22 \pm 0.31)\times 10^{3} $ &  $ (7.39 \pm 0.20)\times 10^{3} $ &  $ (4.14 \pm 0.13)\times 10^{3} $ &  $ (6.64 \pm 0.16)\times 10^{3} $ &  $ (3.99 \pm 0.11)\times 10^{3} $ &  $ (2.56 \pm 0.10)\times 10^{3} $ \\ 
 $ 0.790 $ &  $ (8.81 \pm 0.29)\times 10^{3} $ &  $ (7.05 \pm 0.19)\times 10^{3} $ &  $ (3.91 \pm 0.14)\times 10^{3} $ &  $ (6.35 \pm 0.15)\times 10^{3} $ &  $ (3.78 \pm 0.11)\times 10^{3} $ &  $ (2.42 \pm 0.13)\times 10^{3} $ \\ 
 $ 1.054 $ &  $ (8.42 \pm 0.27)\times 10^{3} $ &  $ (6.74 \pm 0.18)\times 10^{3} $ &  $ (3.70 \pm 0.22)\times 10^{3} $ &  $ (6.06 \pm 0.16)\times 10^{3} $ &  $ (3.59 \pm 0.17)\times 10^{3} $ &  $ (2.32 \pm 0.34)\times 10^{3} $ \\ 
 $ 1.406 $ &  $ (8.15 \pm 0.27)\times 10^{3} $ &  $ (6.50 \pm 0.20)\times 10^{3} $ &  $ (3.51 \pm 0.75)\times 10^{3} $ &  $ (5.80 \pm 0.22)\times 10^{3} $ &  $ (3.37 \pm 0.67)\times 10^{3} $ &  $ (2.49 \pm 1.78)\times 10^{3} $ \\ 
 \hline
 \hline
 & & & & & &  \\ [0.25mm]
$k_{\theta} $ & \multicolumn{6}{c}{Sources with $ S > 200\, \rm mJy$ Masked} \\ [0.5mm] 
$ [\rm arcmin^{-1}] $ & $ 250 \times250 $ & $ 250 \times350 $ & $ 250 \times500 $ & $ 350 \times350 $ & $ 350 \times500 $ & $ 500 \times500 $  \\ 
 \hline
 $ 0.011 $ &  $ (5.05 \pm 2.96)\times 10^{5} $ &  $ (3.55 \pm 2.43)\times 10^{5} $ &  $ (1.44 \pm 1.08)\times 10^{5} $ &  $ (2.70 \pm 1.68)\times 10^{5} $ &  $ (1.19 \pm 0.77)\times 10^{5} $ &  $ (7.41 \pm 3.88)\times 10^{4} $ \\ 
 $ 0.019 $ &  $ (1.34 \pm 0.43)\times 10^{5} $ &  $ (1.09 \pm 0.74)\times 10^{5} $ &  $ (6.42 \pm 4.42)\times 10^{4} $ &  $ (1.05 \pm 0.43)\times 10^{5} $ &  $ (6.36 \pm 3.20)\times 10^{4} $ &  $ (4.46 \pm 1.51)\times 10^{4} $ \\ 
 $ 0.026 $ &  $ (7.69 \pm 1.69)\times 10^{4} $ &  $ (5.76 \pm 1.57)\times 10^{4} $ &  $ (4.32 \pm 1.09)\times 10^{4} $ &  $ (5.62 \pm 1.07)\times 10^{4} $ &  $ (3.63 \pm 0.86)\times 10^{4} $ &  $ (2.56 \pm 0.46)\times 10^{4} $ \\ 
 $ 0.033 $ &  $ (5.05 \pm 0.92)\times 10^{4} $ &  $ (3.34 \pm 0.66)\times 10^{4} $ &  $ (1.91 \pm 0.45)\times 10^{4} $ &  $ (3.26 \pm 0.52)\times 10^{4} $ &  $ (1.87 \pm 0.36)\times 10^{4} $ &  $ (1.49 \pm 0.22)\times 10^{4} $ \\ 
 $ 0.044 $ &  $ (4.26 \pm 0.52)\times 10^{4} $ &  $ (3.48 \pm 0.45)\times 10^{4} $ &  $ (1.93 \pm 0.27)\times 10^{4} $ &  $ (3.31 \pm 0.37)\times 10^{4} $ &  $ (1.84 \pm 0.23)\times 10^{4} $ &  $ (1.23 \pm 0.14)\times 10^{4} $ \\ 
 $ 0.059 $ &  $ (2.70 \pm 0.30)\times 10^{4} $ &  $ (2.19 \pm 0.25)\times 10^{4} $ &  $ (1.26 \pm 0.17)\times 10^{4} $ &  $ (2.09 \pm 0.22)\times 10^{4} $ &  $ (1.24 \pm 0.14)\times 10^{4} $ &  $ (8.42 \pm 0.94)\times 10^{3} $ \\ 
 $ 0.079 $ &  $ (2.14 \pm 0.17)\times 10^{4} $ &  $ (1.71 \pm 0.13)\times 10^{4} $ &  $ (9.73 \pm 0.85)\times 10^{3} $ &  $ (1.59 \pm 0.11)\times 10^{4} $ &  $ (9.42 \pm 0.74)\times 10^{3} $ &  $ (6.31 \pm 0.52)\times 10^{3} $ \\ 
 $ 0.105 $ &  $ (1.54 \pm 0.08)\times 10^{4} $ &  $ (1.29 \pm 0.06)\times 10^{4} $ &  $ (7.40 \pm 0.42)\times 10^{3} $ &  $ (1.21 \pm 0.06)\times 10^{4} $ &  $ (7.27 \pm 0.39)\times 10^{3} $ &  $ (4.77 \pm 0.28)\times 10^{3} $ \\ 
 $ 0.141 $ &  $ (1.32 \pm 0.05)\times 10^{4} $ &  $ (1.10 \pm 0.04)\times 10^{4} $ &  $ (6.23 \pm 0.28)\times 10^{3} $ &  $ (1.01 \pm 0.04)\times 10^{4} $ &  $ (6.08 \pm 0.25)\times 10^{3} $ &  $ (3.93 \pm 0.18)\times 10^{3} $ \\ 
 $ 0.187 $ &  $ (1.11 \pm 0.04)\times 10^{4} $ &  $ (9.30 \pm 0.29)\times 10^{3} $ &  $ (5.26 \pm 0.19)\times 10^{3} $ &  $ (8.60 \pm 0.26)\times 10^{3} $ &  $ (5.19 \pm 0.17)\times 10^{3} $ &  $ (3.35 \pm 0.13)\times 10^{3} $ \\ 
 $ 0.250 $ &  $ (1.02 \pm 0.03)\times 10^{4} $ &  $ (8.54 \pm 0.24)\times 10^{3} $ &  $ (4.87 \pm 0.16)\times 10^{3} $ &  $ (7.92 \pm 0.21)\times 10^{3} $ &  $ (4.81 \pm 0.14)\times 10^{3} $ &  $ (3.13 \pm 0.11)\times 10^{3} $ \\ 
 $ 0.333 $ &  $ (9.58 \pm 0.30)\times 10^{3} $ &  $ (7.99 \pm 0.21)\times 10^{3} $ &  $ (4.52 \pm 0.13)\times 10^{3} $ &  $ (7.41 \pm 0.18)\times 10^{3} $ &  $ (4.46 \pm 0.12)\times 10^{3} $ &  $ (2.88 \pm 0.09)\times 10^{3} $ \\ 
 $ 0.445 $ &  $ (9.12 \pm 0.28)\times 10^{3} $ &  $ (7.51 \pm 0.19)\times 10^{3} $ &  $ (4.23 \pm 0.12)\times 10^{3} $ &  $ (6.88 \pm 0.16)\times 10^{3} $ &  $ (4.13 \pm 0.10)\times 10^{3} $ &  $ (2.66 \pm 0.09)\times 10^{3} $ \\ 
 $ 0.593 $ &  $ (8.85 \pm 0.27)\times 10^{3} $ &  $ (7.22 \pm 0.18)\times 10^{3} $ &  $ (4.08 \pm 0.12)\times 10^{3} $ &  $ (6.57 \pm 0.16)\times 10^{3} $ &  $ (3.96 \pm 0.10)\times 10^{3} $ &  $ (2.56 \pm 0.10)\times 10^{3} $ \\ 
 $ 0.790 $ &  $ (8.46 \pm 0.25)\times 10^{3} $ &  $ (6.90 \pm 0.17)\times 10^{3} $ &  $ (3.85 \pm 0.13)\times 10^{3} $ &  $ (6.30 \pm 0.15)\times 10^{3} $ &  $ (3.75 \pm 0.11)\times 10^{3} $ &  $ (2.42 \pm 0.13)\times 10^{3} $ \\ 
 $ 1.054 $ &  $ (8.12 \pm 0.24)\times 10^{3} $ &  $ (6.60 \pm 0.17)\times 10^{3} $ &  $ (3.65 \pm 0.20)\times 10^{3} $ &  $ (6.02 \pm 0.15)\times 10^{3} $ &  $ (3.57 \pm 0.16)\times 10^{3} $ &  $ (2.32 \pm 0.33)\times 10^{3} $ \\ 
 $ 1.406 $ &  $ (7.89 \pm 0.25)\times 10^{3} $ &  $ (6.37 \pm 0.18)\times 10^{3} $ &  $ (3.46 \pm 0.71)\times 10^{3} $ &  $ (5.76 \pm 0.21)\times 10^{3} $ &  $ (3.36 \pm 0.65)\times 10^{3} $ &  $ (2.49 \pm 1.77)\times 10^{3} $ \\ 
 \hline
 \hline
 & & & & & &  \\ [0.25mm]
$k_{\theta} $ & \multicolumn{6}{c}{Sources with $ S > 100\, \rm mJy$ Masked} \\ [0.5mm] 
$ [\rm arcmin^{-1}] $ & $ 250 \times250 $ & $ 250 \times350 $ & $ 250 \times500 $ & $ 350 \times350 $ & $ 350 \times500 $ & $ 500 \times500 $  \\ 
 \hline
 $ 0.011 $ &  $ (4.88 \pm 2.91)\times 10^{5} $ &  $ (3.44 \pm 2.40)\times 10^{5} $ &  $ (1.37 \pm 1.06)\times 10^{5} $ &  $ (2.65 \pm 1.67)\times 10^{5} $ &  $ (1.17 \pm 0.76)\times 10^{5} $ &  $ (7.41 \pm 3.88)\times 10^{4} $ \\ 
 $ 0.019 $ &  $ (1.33 \pm 0.43)\times 10^{5} $ &  $ (1.07 \pm 0.73)\times 10^{5} $ &  $ (6.31 \pm 4.39)\times 10^{4} $ &  $ (1.04 \pm 0.42)\times 10^{5} $ &  $ (6.37 \pm 3.20)\times 10^{4} $ &  $ (4.45 \pm 1.51)\times 10^{4} $ \\ 
 $ 0.026 $ &  $ (7.34 \pm 1.63)\times 10^{4} $ &  $ (5.53 \pm 1.53)\times 10^{4} $ &  $ (4.31 \pm 1.09)\times 10^{4} $ &  $ (5.46 \pm 1.05)\times 10^{4} $ &  $ (3.60 \pm 0.85)\times 10^{4} $ &  $ (2.55 \pm 0.46)\times 10^{4} $ \\ 
 $ 0.033 $ &  $ (4.83 \pm 0.89)\times 10^{4} $ &  $ (3.23 \pm 0.65)\times 10^{4} $ &  $ (1.87 \pm 0.44)\times 10^{4} $ &  $ (3.20 \pm 0.51)\times 10^{4} $ &  $ (1.82 \pm 0.36)\times 10^{4} $ &  $ (1.47 \pm 0.22)\times 10^{4} $ \\ 
 $ 0.044 $ &  $ (4.20 \pm 0.51)\times 10^{4} $ &  $ (3.43 \pm 0.45)\times 10^{4} $ &  $ (1.91 \pm 0.27)\times 10^{4} $ &  $ (3.28 \pm 0.37)\times 10^{4} $ &  $ (1.83 \pm 0.23)\times 10^{4} $ &  $ (1.22 \pm 0.14)\times 10^{4} $ \\ 
 $ 0.059 $ &  $ (2.63 \pm 0.29)\times 10^{4} $ &  $ (2.15 \pm 0.25)\times 10^{4} $ &  $ (1.24 \pm 0.16)\times 10^{4} $ &  $ (2.06 \pm 0.22)\times 10^{4} $ &  $ (1.23 \pm 0.14)\times 10^{4} $ &  $ (8.36 \pm 0.94)\times 10^{3} $ \\ 
 $ 0.079 $ &  $ (2.04 \pm 0.17)\times 10^{4} $ &  $ (1.67 \pm 0.13)\times 10^{4} $ &  $ (9.54 \pm 0.84)\times 10^{3} $ &  $ (1.57 \pm 0.12)\times 10^{4} $ &  $ (9.36 \pm 0.74)\times 10^{3} $ &  $ (6.28 \pm 0.51)\times 10^{3} $ \\ 
 $ 0.105 $ &  $ (1.46 \pm 0.07)\times 10^{4} $ &  $ (1.25 \pm 0.06)\times 10^{4} $ &  $ (7.25 \pm 0.41)\times 10^{3} $ &  $ (1.19 \pm 0.06)\times 10^{4} $ &  $ (7.20 \pm 0.38)\times 10^{3} $ &  $ (4.74 \pm 0.27)\times 10^{3} $ \\ 
 $ 0.141 $ &  $ (1.23 \pm 0.04)\times 10^{4} $ &  $ (1.05 \pm 0.04)\times 10^{4} $ &  $ (6.04 \pm 0.27)\times 10^{3} $ &  $ (9.84 \pm 0.37)\times 10^{3} $ &  $ (5.99 \pm 0.25)\times 10^{3} $ &  $ (3.92 \pm 0.18)\times 10^{3} $ \\ 
 $ 0.187 $ &  $ (1.03 \pm 0.03)\times 10^{4} $ &  $ (8.84 \pm 0.26)\times 10^{3} $ &  $ (5.10 \pm 0.18)\times 10^{3} $ &  $ (8.35 \pm 0.24)\times 10^{3} $ &  $ (5.08 \pm 0.16)\times 10^{3} $ &  $ (3.32 \pm 0.12)\times 10^{3} $ \\ 
 $ 0.250 $ &  $ (9.39 \pm 0.27)\times 10^{3} $ &  $ (8.08 \pm 0.21)\times 10^{3} $ &  $ (4.69 \pm 0.14)\times 10^{3} $ &  $ (7.68 \pm 0.19)\times 10^{3} $ &  $ (4.71 \pm 0.13)\times 10^{3} $ &  $ (3.11 \pm 0.10)\times 10^{3} $ \\ 
 $ 0.333 $ &  $ (8.79 \pm 0.24)\times 10^{3} $ &  $ (7.52 \pm 0.18)\times 10^{3} $ &  $ (4.34 \pm 0.12)\times 10^{3} $ &  $ (7.14 \pm 0.17)\times 10^{3} $ &  $ (4.35 \pm 0.11)\times 10^{3} $ &  $ (2.85 \pm 0.09)\times 10^{3} $ \\ 
 $ 0.445 $ &  $ (8.33 \pm 0.22)\times 10^{3} $ &  $ (7.05 \pm 0.16)\times 10^{3} $ &  $ (4.05 \pm 0.11)\times 10^{3} $ &  $ (6.65 \pm 0.15)\times 10^{3} $ &  $ (4.03 \pm 0.10)\times 10^{3} $ &  $ (2.64 \pm 0.08)\times 10^{3} $ \\ 
 $ 0.593 $ &  $ (8.09 \pm 0.22)\times 10^{3} $ &  $ (6.78 \pm 0.16)\times 10^{3} $ &  $ (3.91 \pm 0.11)\times 10^{3} $ &  $ (6.37 \pm 0.14)\times 10^{3} $ &  $ (3.87 \pm 0.10)\times 10^{3} $ &  $ (2.55 \pm 0.09)\times 10^{3} $ \\ 
 $ 0.790 $ &  $ (7.74 \pm 0.21)\times 10^{3} $ &  $ (6.49 \pm 0.15)\times 10^{3} $ &  $ (3.69 \pm 0.12)\times 10^{3} $ &  $ (6.11 \pm 0.14)\times 10^{3} $ &  $ (3.68 \pm 0.10)\times 10^{3} $ &  $ (2.40 \pm 0.13)\times 10^{3} $ \\ 
 $ 1.054 $ &  $ (7.49 \pm 0.20)\times 10^{3} $ &  $ (6.23 \pm 0.14)\times 10^{3} $ &  $ (3.50 \pm 0.19)\times 10^{3} $ &  $ (5.86 \pm 0.14)\times 10^{3} $ &  $ (3.51 \pm 0.16)\times 10^{3} $ &  $ (2.31 \pm 0.33)\times 10^{3} $ \\ 
 $ 1.406 $ &  $ (7.31 \pm 0.21)\times 10^{3} $ &  $ (6.03 \pm 0.16)\times 10^{3} $ &  $ (3.31 \pm 0.67)\times 10^{3} $ &  $ (5.64 \pm 0.20)\times 10^{3} $ &  $ (3.30 \pm 0.63)\times 10^{3} $ &  $ (2.48 \pm 1.75)\times 10^{3} $ \\ 
 \hline
 \hline
 & & & & & &  \\ [0.25mm]
$k_{\theta} $ & \multicolumn{6}{c}{Sources with $ S > 50\, \rm mJy$ Masked} \\ [0.5mm] 
$ [\rm arcmin^{-1}] $ & $ 250 \times250 $ & $ 250 \times350 $ & $ 250 \times500 $ & $ 350 \times350 $ & $ 350 \times500 $ & $ 500 \times500 $  \\ 
 \hline
 $ 0.011 $ &  $ (4.00 \pm 2.62)\times 10^{5} $ &  $ (2.55 \pm 2.08)\times 10^{5} $ &  $ (1.09 \pm 0.95)\times 10^{5} $ &  $ (2.00 \pm 1.44)\times 10^{5} $ &  $ (9.52 \pm 6.76)\times 10^{4} $ &  $ (6.80 \pm 3.64)\times 10^{4} $ \\ 
 $ 0.019 $ &  $ (1.25 \pm 0.41)\times 10^{5} $ &  $ (9.88 \pm 6.98)\times 10^{4} $ &  $ (5.88 \pm 4.26)\times 10^{4} $ &  $ (9.39 \pm 3.96)\times 10^{4} $ &  $ (5.90 \pm 3.03)\times 10^{4} $ &  $ (4.22 \pm 1.45)\times 10^{4} $ \\ 
 $ 0.026 $ &  $ (6.64 \pm 1.50)\times 10^{4} $ &  $ (4.80 \pm 1.36)\times 10^{4} $ &  $ (4.17 \pm 1.05)\times 10^{4} $ &  $ (4.94 \pm 0.96)\times 10^{4} $ &  $ (3.43 \pm 0.81)\times 10^{4} $ &  $ (2.54 \pm 0.46)\times 10^{4} $ \\ 
 $ 0.033 $ &  $ (4.46 \pm 0.83)\times 10^{4} $ &  $ (2.87 \pm 0.58)\times 10^{4} $ &  $ (1.72 \pm 0.40)\times 10^{4} $ &  $ (2.89 \pm 0.47)\times 10^{4} $ &  $ (1.64 \pm 0.33)\times 10^{4} $ &  $ (1.38 \pm 0.21)\times 10^{4} $ \\ 
 $ 0.044 $ &  $ (3.78 \pm 0.46)\times 10^{4} $ &  $ (3.08 \pm 0.41)\times 10^{4} $ &  $ (1.75 \pm 0.25)\times 10^{4} $ &  $ (2.96 \pm 0.34)\times 10^{4} $ &  $ (1.68 \pm 0.21)\times 10^{4} $ &  $ (1.17 \pm 0.14)\times 10^{4} $ \\ 
 $ 0.059 $ &  $ (2.38 \pm 0.27)\times 10^{4} $ &  $ (1.95 \pm 0.23)\times 10^{4} $ &  $ (1.15 \pm 0.15)\times 10^{4} $ &  $ (1.88 \pm 0.20)\times 10^{4} $ &  $ (1.13 \pm 0.13)\times 10^{4} $ &  $ (7.91 \pm 0.90)\times 10^{3} $ \\ 
 $ 0.079 $ &  $ (1.77 \pm 0.14)\times 10^{4} $ &  $ (1.47 \pm 0.12)\times 10^{4} $ &  $ (8.62 \pm 0.77)\times 10^{3} $ &  $ (1.43 \pm 0.10)\times 10^{4} $ &  $ (8.57 \pm 0.68)\times 10^{3} $ &  $ (5.97 \pm 0.49)\times 10^{3} $ \\ 
 $ 0.105 $ &  $ (1.28 \pm 0.07)\times 10^{4} $ &  $ (1.10 \pm 0.06)\times 10^{4} $ &  $ (6.50 \pm 0.38)\times 10^{3} $ &  $ (1.07 \pm 0.05)\times 10^{4} $ &  $ (6.50 \pm 0.34)\times 10^{3} $ &  $ (4.43 \pm 0.26)\times 10^{3} $ \\ 
 $ 0.141 $ &  $ (1.03 \pm 0.04)\times 10^{4} $ &  $ (8.85 \pm 0.33)\times 10^{3} $ &  $ (5.30 \pm 0.23)\times 10^{3} $ &  $ (8.59 \pm 0.32)\times 10^{3} $ &  $ (5.31 \pm 0.21)\times 10^{3} $ &  $ (3.65 \pm 0.16)\times 10^{3} $ \\ 
 $ 0.187 $ &  $ (8.57 \pm 0.26)\times 10^{3} $ &  $ (7.48 \pm 0.21)\times 10^{3} $ &  $ (4.44 \pm 0.15)\times 10^{3} $ &  $ (7.40 \pm 0.21)\times 10^{3} $ &  $ (4.53 \pm 0.14)\times 10^{3} $ &  $ (3.11 \pm 0.11)\times 10^{3} $ \\ 
 $ 0.250 $ &  $ (7.76 \pm 0.22)\times 10^{3} $ &  $ (6.79 \pm 0.17)\times 10^{3} $ &  $ (4.05 \pm 0.12)\times 10^{3} $ &  $ (6.73 \pm 0.17)\times 10^{3} $ &  $ (4.14 \pm 0.11)\times 10^{3} $ &  $ (2.88 \pm 0.09)\times 10^{3} $ \\ 
 $ 0.333 $ &  $ (7.23 \pm 0.18)\times 10^{3} $ &  $ (6.25 \pm 0.14)\times 10^{3} $ &  $ (3.69 \pm 0.10)\times 10^{3} $ &  $ (6.24 \pm 0.14)\times 10^{3} $ &  $ (3.77 \pm 0.09)\times 10^{3} $ &  $ (2.63 \pm 0.08)\times 10^{3} $ \\ 
 $ 0.445 $ &  $ (6.76 \pm 0.17)\times 10^{3} $ &  $ (5.79 \pm 0.13)\times 10^{3} $ &  $ (3.41 \pm 0.09)\times 10^{3} $ &  $ (5.80 \pm 0.12)\times 10^{3} $ &  $ (3.49 \pm 0.08)\times 10^{3} $ &  $ (2.44 \pm 0.07)\times 10^{3} $ \\ 
 $ 0.593 $ &  $ (6.47 \pm 0.16)\times 10^{3} $ &  $ (5.50 \pm 0.12)\times 10^{3} $ &  $ (3.25 \pm 0.09)\times 10^{3} $ &  $ (5.50 \pm 0.12)\times 10^{3} $ &  $ (3.32 \pm 0.08)\times 10^{3} $ &  $ (2.36 \pm 0.08)\times 10^{3} $ \\ 
 $ 0.790 $ &  $ (6.18 \pm 0.15)\times 10^{3} $ &  $ (5.25 \pm 0.11)\times 10^{3} $ &  $ (3.09 \pm 0.10)\times 10^{3} $ &  $ (5.31 \pm 0.11)\times 10^{3} $ &  $ (3.19 \pm 0.08)\times 10^{3} $ &  $ (2.27 \pm 0.11)\times 10^{3} $ \\ 
 $ 1.054 $ &  $ (6.00 \pm 0.15)\times 10^{3} $ &  $ (5.08 \pm 0.11)\times 10^{3} $ &  $ (2.94 \pm 0.15)\times 10^{3} $ &  $ (5.18 \pm 0.12)\times 10^{3} $ &  $ (3.09 \pm 0.13)\times 10^{3} $ &  $ (2.20 \pm 0.29)\times 10^{3} $ \\ 
 $ 1.406 $ &  $ (5.93 \pm 0.16)\times 10^{3} $ &  $ (5.00 \pm 0.13)\times 10^{3} $ &  $ (2.80 \pm 0.53)\times 10^{3} $ &  $ (5.05 \pm 0.17)\times 10^{3} $ &  $ (2.96 \pm 0.53)\times 10^{3} $ &  $ (2.37 \pm 1.57)\times 10^{3} $ \\ 
 \hline
 \hline
\hline
\end{longtable*}
\end{center}

%
%

\end{document}